\documentclass[apj]{emulateapj}
\usepackage{natbib}
\def\xmms{{\it XMM-Newton} }
\def\chandras{{\it Chandra} }
\def\suzakus{{\it Suzaku} }
\def\swifts{{\it Swift} }
\def\saxs{{\it Beppo-SAX} }
\def\rxtes{{\it RXTE} }

\def\his{HIFLUGCS }
\def\xmm{{\it XMM-Newton}}

\def\suzaku{{\it Suzaku}}
\def\swift{{\it Swift}}
\def\sax{{\it Beppo-SAX}}
\def\rxte{{\it RXTE}}
\def\integral{{\it INTEGRAL}}
\def\hi{HIFLUGCS}

\shorttitle{}
\shortauthors{Wik et al.}

\begin{document}

\title{The \swifts BAT Perspective on Non-thermal Emission in
\his Galaxy Clusters}

\author{Daniel R. Wik\altaffilmark{1,6}, Craig L. Sarazin\altaffilmark{2},
Yu-Ying Zhang\altaffilmark{3}, Wayne H. Baumgartner\altaffilmark{1},
Richard F. Mushotzky\altaffilmark{4}, Jack Tueller\altaffilmark{1},
Takashi Okajima\altaffilmark{1}, and Tracy E. Clarke\altaffilmark{5}}

\altaffiltext{1}{Astrophysics Science Division, 
NASA/Goddard Space Flight Center,
Greenbelt, MD 20771, USA; daniel.r.wik@nasa.gov}
\altaffiltext{2}{Department of Astronomy, University of Virginia
P. O. Box 400325, Charlottesville, VA 22904-4325}
\altaffiltext{3}{Argelander-Institut f\"ur Astronomie,
Universit\"at Bonn, Auf dem H\"ugel 71, 53121 Bonn, Germany}
\altaffiltext{4}{Department of Astronomy, University of Maryland, 
College Park, MD 20742, USA}
\altaffiltext{5}{Naval Research Laboratory, 4555 Overlook Ave. SW, 
Code 7213, Washington, DC 20375, USA}
\altaffiltext{6}{NASA Postdoctoral Program Fellow}

\begin{abstract}
The search for diffuse non-thermal, inverse Compton (IC) emission
from galaxy clusters at hard X-ray energies has been underway for
many years, with most detections being either of low significance
or controversial.
In this work, we investigate 14--195 keV spectra from the \swifts BAT 
all sky survey for evidence of non-thermal excess emission above the 
exponentially decreasing tail of thermal emission in the flux-limited \his sample.
To account for the thermal flux contribution at BAT energies, \xmms EPIC spectra
are extracted from coincident spatial regions so that both the
thermal and non-thermal spectral components can be determined
simultaneously.
We find marginally significant IC components in 6 clusters, though
after closer inspection and consideration of systematic errors
we are unable to claim a clear detection in any of them.
The spectra of all clusters are also summed to enhance a cumulative
non-thermal signal not quite detectable in individual clusters.
After constructing a model based on single temperature fits to the
\xmms data alone, we see no significant excess emission above that
predicted by the thermal model determined at soft energies.
This result also holds for the summed spectra of various subgroups,
except for the subsample of clusters with diffuse radio emission.
For clusters hosting a diffuse radio halo, relic, or mini-halo, non-thermal emission is
initially detected at the 
$\sim5\sigma$ confidence level, but modeling and
systematic uncertainties ultimately degrade this significance.
This marginal detection is driven by the mini-halo subgroup, suggesting
low average magnetic field strengths ($B \sim 0.1 \mu$G) in the cores of these clusters.
\end{abstract}


\section{Introduction}
\label{sec:bathi:intro}

A number of observations, mainly at radio frequencies, have established
that relativistic particles and
magnetic fields are part of the
intracluster medium (ICM) of galaxy clusters
\citep[e.g.,][]{GF04}.
The large ($\sim$Mpc) scale, diffuse structures known as radio halos
and relics are produced by relativistic electrons spiraling around
$\sim$$\mu$G magnetic fields.
Because halos and relics are not detected in every cluster, but are
only found in clusters with ongoing major merger activity
\citep{Buo01, SBR+01},
mergers probably temporarily reaccelerate underlying relativistic
populations \citep[e.g.,][]{Sar99, BB05}.
It is important to fully characterize
the non-thermal phase if the dynamics and general state of the ICM
is to be understood;
the proportion of energy tied up in these relativistic components, if
significant, may bias inferred mass estimates necessary to use clusters
as cosmological probes 
\citep[e.g.,][]{MAE+08, Vik+09, Van+10}.
Unfortunately, synchrotron emission alone cannot separately determine
particle and magnetic field energy densities, and so the total energy in the
non-thermal phase remains relatively unconstrained.
However, the electron population can be independently observed through
inverse Compton (IC) emission due to scattering of the ubiquitous
Cosmic Microwave Background (CMB) photons, which are up-scattered to X-ray
energies and may be observable if the electron population is sufficiently
large \citep{Rep79}.
Detections of IC emission, therefore, have the potential to determine
whether the non-thermal phase is energetically negligible or, particularly
if the average magnetic field is large, it is sizable enough to
affect the dynamics and structure of the thermal gas.

Thermal emission clearly dominates at $\sim$keV energies, so searches
for excess emission due to an IC spectral component are more easily undertaken
at very soft or hard ($>$ 10 keV) energies.
The latter range is particularly promising, given the exponential
decline in the thermal spectrum and the lack of Galactic
and solar wind charge exchange foregrounds that can hamper searches at
soft energies \citep{KLK+09, THF+07, BLB09}.
In particular, the \swifts BAT all sky survey \citep{Tue+10} provides 
a deep map of
hard energy (14--195 keV) emission from which non-thermal
excesses can be identified.
Its uniform coverage and impressive sensitivity makes 
it the most complete dataset from which to study the brightest
objects in a given class \citep[e.g.,][]{WMR+09}.
Whereas previous searches have concentrated on long pointed observations
of individual clusters, this survey allows a larger, more uniform sample
to be searched, as similarly done by \citet{Aje+09, Aje+10} for detected
BAT clusters.
To take full advantage of this capability, we have chosen the 
flux-limited \his sample
 \citep{RB02},
 which contains the brightest
clusters in the sky outside the Galactic plane.
The selection of the brightest clusters may provide the greatest opportunity to
detect IC emission, as in most models the nearest and most luminous clusters are expected
to have the strongest IC signal.
Also, because these clusters are bright and contained within a well-defined
survey, there already exist good observations at lower X-ray energies,
which can be used to strongly constrain the thermal properties of the 
ICM -- an important prerequisite for the robust detection of an IC excess.
Finally, the fact that \his is a complete flux-limited survey allows one to 
discuss the
statistical properties of their hard excesses by stacking the individual 
cluster observations.

Because they are nearby and bright, many of the clusters in \his
have been targets of IC searches with other telescopes, including
A3667 \citep{FSN+10}, A3112 \citep{BNL07}, A3376 \citep{Kaw+09},
A2256 \citep{FLO05}, A1367 \citep{HM01}, A2199 \citep{KS00},
and A2163 \citep{RGA06}.
Most often clusters are targeted because they host a radio halo or
relic, as the IC flux then leads to a direct measure of the average
magnetic field strength.
A large fraction of \his clusters were also included in an analysis
of all long exposure \saxs observations \citep{NOB+04}, which found
marginal evidence for non-thermal excesses in individual clusters
but a substantial excess in a stacked spectrum.
In general, an IC component distinct from thermal emission
in the hard band has been difficult to clearly identify,
with perhaps the only counter example being an exceptionally deep
observation of the Ophiuchus cluster \citep{EPP+08}.
The cluster most thoroughly searched for non-thermal
emission, also in \hi, is the Coma cluster.
Controversial \citep{RM04} detections with 
\rxtes \citep{RG02} and \saxs \citep{FOB+04} have recently been
challenged with comparable \suzakus \citep{WSF+09} observations
and a detailed analysis of the \swifts BAT survey data \citep{WSF+11}.

To perform the deepest hard X-ray survey of
non-thermal emission in clusters
to date, we jointly fit high quality \xmms EPIC and \swifts
BAT spectra, extracted from identical regions and cross-calibrated
to make their absolute spectral responses as consistent as possible.
We describe the data and its calibration in Section~\ref{sec:bathi:obs}.
In Section~\ref{sec:bathi:separate}, the thermal and non-thermal
character of the spectra are separately analyzed, and in
Section~\ref{sec:bathi:joint} they are jointly fit for each individual
cluster.
We also search for a statistical hard excess in sets of
stacked spectra for the entire sample and for several subsamples
in Section~\ref{sec:bathi:stack}.
Lastly, the implications of our results are discussed in
Section~\ref{sec:bathi:disc}.
We assume a flat cosmology with $\Omega_M = 0.23$ and 
$H_0 = 70$ km s$^{-1}$ Mpc$^{-1}$.
Unless otherwise stated, all uncertainties are given at the 90\% 
confidence level.

\section{Observations and Data Preparation}
\label{sec:bathi:obs}

\subsection{\xmms EPIC Spectra}
\label{sec:bathi:obs:xmm}

For the lower energy BAT bands,
it is very useful to have X-ray spectra at lower energies to 
constrain the thermal emission;
this is particularly true given that the
\swifts BAT survey spectra are coarsely binned 
(8 channels spanning 14 keV $< E <$ 195 keV).
Also, any non-thermal component in the BAT spectra must be consistent
with the spectra at softer energies.
\xmms is the ideal observatory to provide such complementary spectra.
For one, its large field of view (FOV) allows a higher fraction of the
total emission, which can be quite extended given the low redshifts
of the sample, to be detected in a single pointing.
Additionally, the EPIC instruments are sensitive to 5--10 keV photons,
which make them more useful for constraining the highest temperature
gas, and the telescopes have good spatial resolution so that point sources
can be excluded from the spectra.
Last, but of no less importance, \xmms has observed all but one 
(Abell 2244) of the clusters in \hi.
Unfortunately, another 4 cluster observations
(Abell 401, Abell 478, Abell 1736, and Abell 2163)
are heavily contaminated
by background flares and consequently unusable 
\citep[for more details, see][]{ZAC+11}.
However, the data for the remaining 59 clusters are of sufficient
quality to help constrain potential non-thermal signals in the
BAT energy bands.

We extract \xmms spectra for each cluster from the largest circular
region that either covers the FOV or extends to the point where
cosmic X-ray background (CXB) emission begins to dominate,
by summing the annular spectra from \citet{ZRF+09}.
To ensure near Gaussian statistics for $\chi^2$ fitting,
adjacent channels are grouped until each new bin contains at least
30 counts.
The centers and radii of the circular regions, along with
each pointing's observation ID, are listed in 
Table~\ref{tab:bathi:basic}.
Source spectra are extracted in concentric annuli within the region;
corresponding particle background spectra are derived from CLOSED
mode calibration data, which are renormalized based on 
3-10 keV events out of the FOV and outside of a
$15^{\prime}.4$ radius from the detector center
\citep[for details see Section 2.4 of][]{ZRF+09}.
The full background treatment is described in \citet{ZRF+09}.
As an additional step, we readjust the normalization of the particle
background spectra by hand to ensure the 7--12 keV continuum of the
cluster spectra have a more physical shape.
We define ``more physical'' as the background normalization that 
minimizes the $\chi^2$ statistic for a single temperature (1T)
(using the {\tt APEC} plasma emission 
model\footnote{http://cxc.harvard.edu/atomdb/sources\_apec.html}) 
individually fit to the
EPIC-pn (2 $< E <$ 12 keV) and MOS1 and MOS2 (2 $< E <$ 10 keV) spectra.
The new best-fit temperatures, after these initial renormalizations 
of the background, are compared to each other
and to previous measurements \citep[primarily][]{RB02}.
While this method may bias the background level,
especially if a single temperature model is a poor description of
a given spectrum, repeating this procedure with two temperature (2T)
and single temperature plus power law (T+NT) models yield comparable
or inferior results, usually favoring obvious under-subtractions
of the background that produce systematic patterns in the residuals.
We favor normalizations that leave the background slightly
under-subtracted, in order to avoid removing a real non-thermal
signature.
For the most part, the overall spectrum is only mildly affected
since much of the emission is at lower energies where the background
is a smaller fraction of the total.
One consequence is that instrumental lines, which are typically between 7.5
and 9.5 keV and mainly are a problem in the EPIC-pn spectra and which can vary in
intensity relative to the background continuum, can be under- or over-subtracted.
No resolved ICM lines exist in this range, so we simply ignore this
energy range when poor line subtractions occur, as in \cite{WSF+09}.
Based on the change in $\chi^2$ as the background normalization is
varied, a typical 90\% level uncertainty in the normalization is
$\sim 3$\%.

We choose to model, instead of subtract, one further background
component:  the CXB due to extragalactic sources.
\citet{LWP+02}, using \xmms sky fields, find that this component of 
the CXB is well fit by a power law with photon index of 1.42
in the hard band (2--10 keV).
Their results are in good agreement with other work in this band
\citep[e.g.,][]{MCL+03,DM04}.
We adopt their normalization at 1 keV of 
8.44 photons cm$^{-2}$ s$^{-1}$ keV$^{-1}$ sr$^{-1}$, which is scaled
to match the extraction area for each cluster.
The impact of cosmic variance, or the field-to-field variation
in CXB flux resulting from large scale
structure and source population selection, 
is not included as a systematic uncertainty in the following
analysis due to its small effect.
While cosmic variance increases with decreasing solid
angle, the high sensitivity of \xmms allows most of the sources
responsible for a higher variance to be removed, so for one of our
typical regions the 90\% uncertainty is only $\sim$10\% of the CXB flux.
Note that \citet{LWP+02} remove detected point sources as is done
here, so their spectrum can be directly applied as is.
The Galactic component of the CXB is also not considered, as it only
contributes below 1 keV,
and we restrict our fits to the 2--12 keV range.
\begin{deluxetable}{lccccc}
\tablewidth{0pt}
\tabletypesize{\small}
\tablecaption{\xmms Observations, Regions, and Cluster Classes
\label{tab:bathi:basic}}
\tablehead{
 & & $\alpha$(J2000) & $\delta$(J2000) & Radius &  \\
Name & ObsID & (deg) & (deg) & (arcmin) & Class\tablenotemark{a}
}
\startdata
A0085 & 0065140101 & \phn 10.45957 & \phn -9.30303 & 11.6667 & SCC \\
A0119 & 0505211001 & \phn 14.07130 & \phn -1.25327 & \phn 9.3333 & NCC \\
A0133 & 0144310101 & \phn 15.67971 & -21.87968 & \phn 6.7000 & SCC \\
NGC507 & 0080540101 & \phn 20.91068 & \phn 33.25063 & \phn 9.4667 & SCC \\
A0262 & 0109980101 & \phn 28.19002 & \phn 36.15114 & 13.5333 & SCC \\
A0400 & 0404010101 & \phn 44.42226 & \phn\phn  6.02696 & 12.6667 & NCC \\
A0399 & 0112260101 & \phn 44.46513 & \phn 13.04713 & 10.4000 & NCC,R \\
A3112 & 0105660101 & \phn 49.49456 & -44.23562 & \phn 6.9667 & SCC \\
Fornax & 0400620101 & \phn 54.61989 & -35.45122 & 10.6333 & SCC \\
2A0335 & 0109870101 & \phn 54.66787 & \phn\phn  9.96803 & \phn 8.9667 & SCC,R \\
IIIZw54 & 0505230401 & \phn 55.32801 & \phn 15.40390 & \phn 6.8667 & WCC \\
A3158 & 0300211301 & \phn 55.72316 & -53.63099 & \phn 9.1333 & NCC \\
NGC1550 & 0152150101 & \phn 64.90839 & \phn\phn  2.40929 & 11.6667 & SCC \\
EXO0422 & 0300210401 & \phn 66.46339 & \phn -8.56118 & \phn 7.1333 & SCC \\
A3266 & 0105260901 & \phn 67.81198 & -61.44835 & 12.0000 & WCC \\
A0496 & 0135120201 & \phn 68.40753 & -13.26069 & 10.1667 & SCC \\
A3376 & 0151900101 & \phn 90.54203 & -39.95994 & \phn 6.0000 & NCC,R \\
A3391 & 0505210401 & \phn 96.60081 & -53.69002 & \phn 6.7333 & NCC \\
A3395s & 0400010301 & \phn 96.69188 & -54.54530 & \phn 4.2000 & NCC \\
R1504 & 0401040101 & 106.37174 & -12.93125 & \phn 8.1992 & SCC,R \\
A0576 & 0205070301 & 110.35886 & \phn 55.75948 & \phn 9.0000 & WCC \\
A0754 & 0136740101 & 137.32574 & \phn -9.68781 & 10.1667 & NCC,R \\
HydraA & 0109980301 & 139.52491 & -12.09342 & \phn 5.0000 & NCC \\
A1060 & 0206230101 & 159.17853 & -27.52841 & \phn 8.4667 & WCC \\
A1367 & 0061740101 & 176.18539 & \phn 19.73211 & 10.0000 & NCC,R \\
MKW4 & 0093060101 & 181.11522 & \phn\phn  1.89480 & \phn 8.3333 & SCC \\
ZwCl1215 & 0300211401 & 184.41928 & \phn\phn  3.65818 & \phn 6.2333 & NCC \\
NGC4636 & 0111190701 & 190.70940 & \phn\phn  2.69179 & \phn 9.8333 & SCC \\
A3526 & 0406200101 & 192.21101 & -41.30430 & 12.9333 & SCC \\
A1644 & 0010420201 & 194.29469 & -17.40291 & 14.7333 & SCC \\
A1650 & 0093200101 & 194.67448 & \phn -1.75920 & \phn 5.1667 & WCC \\
A1651 & 0203020101 & 194.84310 & \phn -4.19633 & \phn 7.5000 & WCC \\
Coma & 0124711401 & 194.93888 & \phn 27.95150 & 14.6667 & NCC,R \\
NGC5044 & 0037950101 & 198.84908 & -16.38664 & 11.5000 & SCC \\
A3558 & 0107260101 & 202.00169 & -31.50027 & 10.2333 & WCC \\
A3562 & 0105261801 & 203.40201 & -31.67382 & \phn 6.1667 & WCC,R \\
A3571 & 0086950201 & 206.86609 & -32.86052 & \phn 8.5000 & WCC \\
A1795 & 0097820101 & 207.21991 & \phn 26.59282 & \phn 8.0000 & SCC \\
A3581 & 0205990101 & 211.87760 & -27.01320 & 11.0667 & SCC \\
MKW8 & 0300210701 & 220.17560 & \phn\phn  3.47159 & \phn 7.5667 & NCC \\
A2029 & 0111270201 & 227.73326 & \phn\phn  5.74264 & \phn 6.5000 & SCC,R \\
A2052 & 0109920101 & 229.18501 & \phn\phn  7.02012 & \phn 7.0667 & SCC \\
MKW3S & 0109930101 & 230.45945 & \phn\phn  7.70323 & \phn 9.6667 & SCC \\
A2065 & 0112240201 & 230.62112 & \phn 27.72063 & \phn 6.6667 & WCC \\
A2063 & 0550360101 & 230.77401 & \phn\phn  8.60701 & \phn 7.1667 & WCC \\
A2142 & 0111870301 & 239.56451 & \phn 27.25178 & \phn 6.6667 & WCC,R \\
A2147 & 0505210601 & 240.56789 & \phn 15.97177 & 11.3333 & NCC \\
A2199 & 0008030201 & 247.15461 & \phn 39.54811 & 12.3333 & SCC \\
A2204 & 0112230301 & 248.19604 & \phn\phn  5.57554 & \phn 6.1333 & SCC,R \\
A2256 & 0141380201 & 255.96829 & \phn 78.67197 & \phn 8.0000 & NCC,R \\
A2255 & 0112260801 & 258.22709 & \phn 64.06428 & \phn 8.1667 & NCC,R \\
A3667 & 0206850101 & 303.16966 & -56.84081 & 13.0000 & WCC,R \\
S1101 & 0123900101 & 348.49294 & -42.72664 & \phn 6.0333 & SCC \\
A2589 & 0204180101 & 350.98652 & \phn 16.77595 & \phn 5.0000 & WCC \\
A2597 & 0147330101 & 351.33334 & -12.12416 & \phn 6.5667 & SCC \\
A2634 & 0002960101 & 354.62099 & \phn 27.03107 & 11.0000 & WCC \\
A2657 & 0402190301 & 356.23640 & \phn\phn  9.19810 & \phn 5.6667 & WCC \\
A4038 & 0204460101 & 356.93602 & -28.14506 & 12.3333 & WCC \\
A4059 & 0109950201 & 359.25704 & -34.75803 & \phn 9.1333 & SCC

\enddata
\tablenotetext{a}{From \citet{HMR+10}: SCC = ``strong cool core cluster,''
WCC = ``weak cool core cluster,'' and NCC = ``non-cool core cluster'';
clusters that host a radio halo and/or relic are labeled with ``R''}
\end{deluxetable}

\subsection{\swifts BAT 58-month Survey Spectra}
\label{sec:bathi:obs:bat}

The \swifts mission and the properties of the survey are described in 
detail in
\citet[Section 2.2]{WSF+11} and in \citet{Tue+10}.
Similarly, we refer to that Section and the appendices for details
on the extraction and calibration of sources from survey image
data.
To briefly summarize, the flux calibration is tied to the Crab spectrum,
which we define to have the same spectrum as that observed by
\xmms for $E>2$ keV, extrapolated to BAT energies via an adopted
model based on \suzakus observations.
In this way, both the cross-normalization and spectral shape of the
\xmms and \swifts spectra will match, and continuous models can be
jointly fit to them simultaneously.
Unfortunately, independent measurements by various instruments,
including the \swifts BAT,
have recently demonstrated that the hard
X-ray spectrum of the Crab is in fact variable on yearly timescales
 \citep{Wil+11}.
At 14--50 keV energies, where the only appreciable amount of flux
is detected from clusters, the variation spans about 10\% over the
last 5 years, with a consistent decline only over the last 2 years
\citep[see Fig.~5 of ][]{Wil+11}.
Our flux calibration of BAT sources depends on an adopted model 
for the Crab spectrum, which is taken from \suzakus XIS and HXD-PIN
observations that took place in August 2005; this time occurs during
one of the higher flux periods.
Since the BAT survey spectrum of the Crab spans the following 5 years of
observations and averages over these fluctuations, the normalization
of our adopted model is only 2-3\% higher than the actual flux emitted.
The effect this has on our derived fluxes is to make them 2-3\% higher
than they actually are; this amount is equivalent to the 1$\sigma$ error
on the 14--20 keV flux of Coma, which is the highest signal-to-noise
flux considered here by a factor of 2.
Also, \citet{Wil+11} show that the recent decline in flux is more dramatic
for higher energy bands.
Both of these behaviors -- the overall decline in flux and the steepening
of the spectrum -- bias our derived BAT fluxes high, which could lead to
a higher chance of false non-thermal detections.
However, since we find no convincing evidence for non-thermal excesses
even given this probable effect, and since we allow for a 10\% cross-calibration
uncertainty between the BAT and EPIC spectra, which easily encompasses
this level of variability, it is clear that our choice of flux calibrator does not
strongly impact the following analysis, except to make our upper limits
slightly more conservative than they otherwise would be.

\begin{figure*}
\plotone{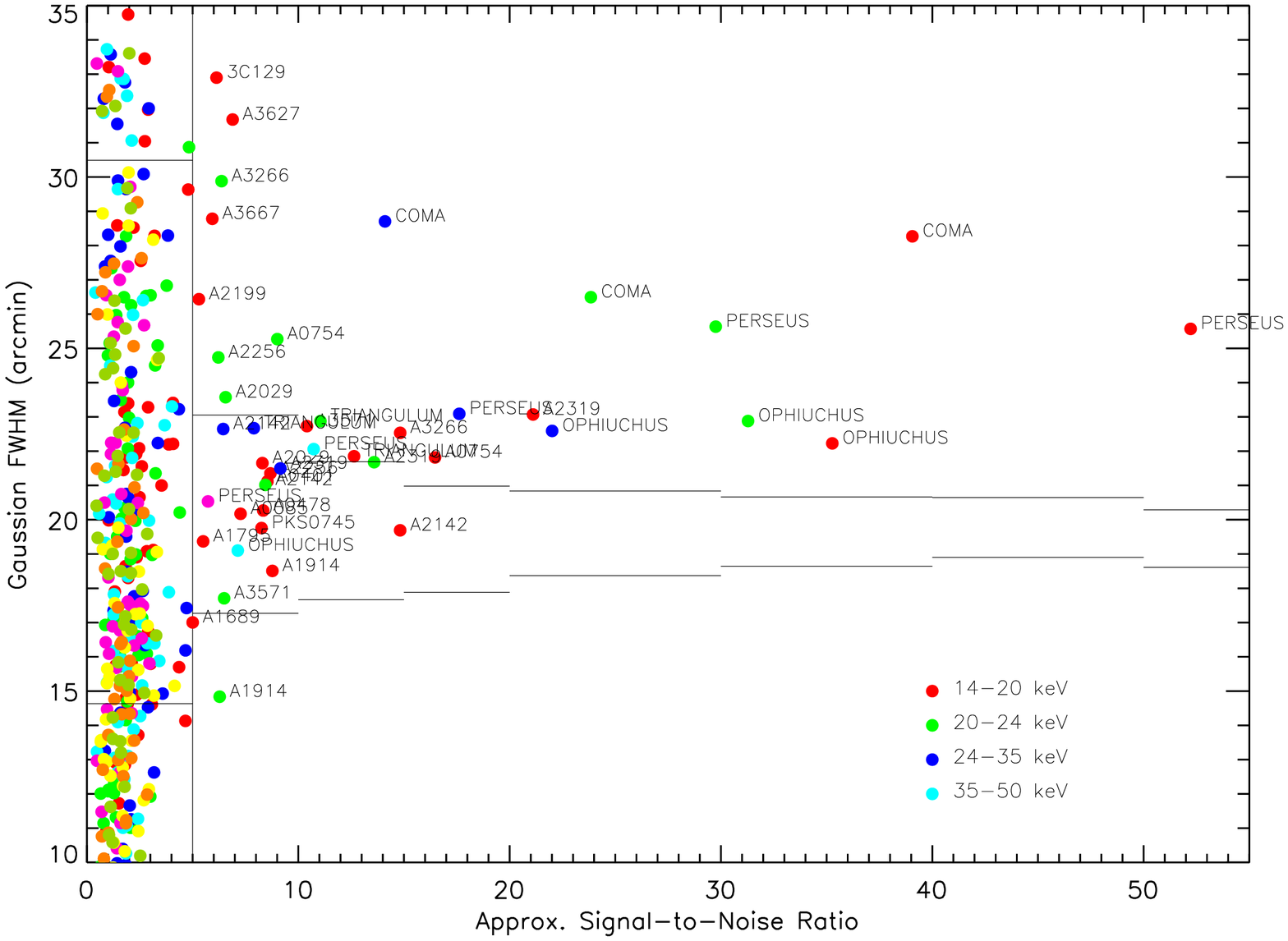}
\caption{Source extent as a function of approximate signal-to-noise
ratio (maximum pixel flux divided by local blank sky fluctuations)
in the 4 lowest energy BAT bands.
Non-cluster sources are shown as small dots (omitted in 
astro-ph version) and horizontal
lines mark the standard deviation of best-fit FWHM
values for the non-cluster sources in each signal-to-noise bin.
Galaxy clusters (colored circles) clearly trend above the mean
FWHM of $\sim$ 20\arcmin; clusters are labeled for S/N values
above 5; the only cluster detected at $E >$ 50 keV is Perseus, but
that emission is primarily due to the AGN in the center of NGC 1275.
Many bright galaxy clusters are at least somewhat resolved by the BAT,
which should be accounted for when extracting fluxes from the BAT survey.
However, FWHM estimates for sources below a S/N $\sim$ 10 can be 
particularly
contaminated by background fluctuations and may not be representative of
their true spatial extent; the 1$\sigma$ error on the FWHM estimates is
approximately given by the horizontal lines bounding the spread in
point source FWHM.
\label{fig:bathi:psf}}
\end{figure*}
While the standard processing of coded mask imaging data is designed
to extract the fluxes of point sources, it is also possible to extract
the flux of a mildly extended source, albeit with somewhat greater
uncertainty \citep{RGL+06,WSF+11}.
The large effective PSF
(full width at half maximum FWHM $\sim20$\arcmin)
for point sources in the survey means that even nearby
clusters of galaxies will appear only slightly extended; the FWHM
of the Coma cluster -- the most extended, reliably detected source 
in the survey -- is only 28\farcm5.
Note that while 4 clusters (Fornax, NGC 4636, A3526, and A1060)
have larger angular extents than Coma (based on angular $R_{500}$
estimates, \citet{EMP11}), they are all cooler, less massive systems and
thus either not detected
or only marginally detected by the BAT at 14--20 keV.
From Figure~\ref{fig:bathi:psf}, it is clear that detected clusters
(colored circles) are typically extended, relative to other sources.
The horizontal lines mark the standard deviation of best-fit FWHM
values for the non-cluster sources in each signal-to-noise bin;
they also represent the approximate error on FWHM estimates
for the clusters in each bin.
Individual clusters are labeled in the 4 lowest energy BAT bands 
when they are detected at a signal-to-noise ratio greater than 5.
We follow the procedure outlined in \citet{WSF+11} to extract fluxes for
diffuse sources, which requires the spatial distribution of the
emission to be known.
Because clusters are comparable in size to the effective spatial
resolution of the survey, detailed spatial models are not necessary
to extract accurate fluxes.
We consider generic $\beta$-model surface brightness profiles, which 
well represent the radial profiles at softer energies.
Taking a representative value for $\beta$ of 0.75, we find that 
all $> 3\sigma$ 
detected clusters (in a given band) can be well fit with 
core radii $r_c$ of either
4\arcmin, 6\arcmin, 8\arcmin, or 10\arcmin.
Profiles with $r_c<$ 4\arcmin\ are hard to distinguish from point
source profiles, so for any cluster emission that is too narrow to
be fit with the $r_c=4$\arcmin\ model is treated as a point source. 
The true spatial distribution may differ from these fiducial models,
but our aim is only to extract accurate fluxes, not describe the
distribution of hard X-ray emission.
For Coma, a $\beta$-model fit in the first BAT band (E1: 14--20 keV)
yields a total flux 9\% lower than that derived from a more
detailed model of its spatial distribution derived from an
\xmms temperature map \citep[see][]{WSF+11}, which accounts for the
NE-SW non-axisymmetric elongation of the emission \citep{ENC+07}.
While 9\% is a significant difference, Coma is one of the most 
significantly detected and is the most
extended cluster in the survey, so this deviation, which amounts to
a factor of only 1.6 times the 1-$\sigma$ error on the flux,
is the largest we would expect using this set of extended models.
Also, note that no energy dependence in FWHM values is detected;
e.g., Coma shows some variation with energy band, but these
measurements are all consistent within their uncertainties.

We also investigated the use of diffuse models for all the clusters, 
irrespective
of their observed extent, to account for the possibility that
we are missing low surface brightness emission obscured by noise.
Since the spatial distribution of $E > 10$ keV emission is unknown,
we assume $\beta$-model profiles derived from ROSAT images
\citep{RB02}.
For clusters with a clearly extended BAT profile, these models
reasonably, but usually not perfectly, follow the emission;
however, these profiles cannot be reliably distinguished from 
those at lower energies given that background fluctuations can
still distort the profile due to the low signal-to-noise ratios.
Spectral fits using these fluxes produce similar results to those
we present in this work, but because their associated errors are
larger, these spectra are generally less sensitive, so any
additional flux captured --- which is not significant --- is also diluted.
Therefore, these spectra are not considered further.

For clusters with modeled extended emission, we do not want to
include the portion of flux that falls outside the \xmms extraction
region during joint fits of the data, since the complementary softer
flux in the \xmms band spectra is not present.
Therefore, only the fraction of the flux that resides within the
\xmms region is included in the spectra derived here.
One uncertainty, particularly when emission is detected at lower
significance, is where the emission is actually coming from, given
the positional accuracy of the survey (a $5\sigma$ source detected
in a given band has a 90\% error circle of radius 6\arcmin).
Since the E1 band-derived positions are near the center of the
extraction region, within their respective error circles, we
assume the center of the hard band distribution is coincident with
the center of the \xmms extraction region except for A754, A3266,
and A2256.
For these detected clusters, their BAT positions are somewhat offset 
from the surface brightness peak due to an anisotropic temperature 
distribution produced by mergers 
\citep[see, e.g.,][]{HB95, FHM+06, SMM+02}.
Following this procedure,
we will not underestimate the coincident flux, although
overestimates may result that could lead to incorrect hard excesses.
However, since we are unable to significantly detect non-thermal
emission individually in any of the clusters, this procedure can only
cause us to be biased in favor of more conservative upper limits.


\section{Separate Fits to Individual \xmms EPIC and \swifts BAT Spectra}
\label{sec:bathi:separate}

Before combining the \swifts and \xmms  datasets, we characterize
each telescope's spectra separately.
The goal is to identify any problems with the data or our methodology that
might lead to biased results when the spectra are fit jointly.

\subsection{Single Temperature Fits to the EPIC Spectra}
\label{sec:bathi:separate:xmm}

The motivation for including \xmms spectra in the analysis is to
fully characterize the thermal properties of the hottest gas in the
ICM, which will contribute flux to the BAT energy bands.
Similarly, these lower energy spectra must be consistent with any
indication of a non-thermal component in the BAT spectra; for example,
a steep power law may best describe the BAT data but at lower energies
result in a poor description of the spectrum.
Since our purpose is not to fully characterize the total emission 
detectable by \xmm, but only capture the state of the hottest gas,
we ignore all events with energies below 2 keV.
Cool ($\la 1$ keV)  gas is completely unimportant at BAT
energies, and it will not overly bias $E > 2$ keV data.
We therefore initially consider EPIC spectra in the 2--12 keV range
for the pn and 2--10 keV range for the MOS detectors;
including photons down to 2 keV provides additional leverage during
spectral fitting, since most of the detected photons, regardless of temperature,
are at lower energies.

However, the lower end of this energy range presents two issues.
First, bright $\sim 1$ keV gas can significantly contribute to
the emission between 2 and 3 keV, which certainly exists in some of
the cool core clusters in \hi.
In single temperature fits, the average temperature will then be biased
low to accommodate this component, which could lead to thermal emission
being interpreted as a non-thermal excess.
Multi-temperature fits would alleviate this problem,
but most of the \xmms  data are not of sufficient quality to strongly constrain 
more than one
temperature component in this energy range.
Including $E < 2$ keV data to better constrain multi-temperature fits
would also require a more complicated analysis that will involve more
free parameters and, because the highest signal-to-noise ratios are
in the $\sim 1$ keV channels, fits would be driven by this data,
possibly resulting in biased high temperature components.
The second issue relates to the imperfectly calibrated gold edge
at 2.2 keV, where the response drops somewhat abruptly.
While on its own this feature does not strongly impact spectral fits,
because it lies near the edge of our energy range where the
signal-to-noise ratio is largest, secondary model components can be
``co-opted'' into better fitting this edge.
For instance, in a spectrum truly described by a gas at a
single temperature, the addition of a second temperature or non-thermal
component to the fit will cause the second component to ``fix''
any deviations at 
this edge, typically resulting in a low temperature or steep photon
index that has no real physical counterpart.

In practice, both of these effects can conspire to produce the appearance of a more
significant non-thermal spectral component than is warranted by the rest
of the data.
To counter both issues, we also perform fits to data with energies
$E > 3$ keV, which exclude the gold edge and any sizable emission
from $\la 1$ keV gas.
These spectra have lower signal-to-noise due to excluding the 2--3 keV emission,
but the high fluxes of clusters in our sample reduce this issue's
importance.
Single temperature fits in both the 2--12 keV and 3--12 keV ranges,
jointly fit to all three EPIC spectra (except for A3526, for which
the MOS-1 spectrum is ignored, and for A2142 and A2147, for which the
MOS-2 spectra are ignored), are given in Table~\ref{tab:bathi:xmmonly}.
The pn and MOS instrument cross-normalization is left as a free
parameter, which allows for a typical ($10 \pm 10$)\% difference
between their calibration \citep[e.g.,][]{Sno02}.
This cross-normalization factor is used and kept fixed during
all subsequent joint EPIC-BAT fits.
The change in the best-fit temperature from the $E > 2$ keV to
$E > 3$ keV fits is only $\sim 0.3$ keV on average, indicating that
the temperature is generally robust to the choice of the energy range,
but that higher energy photons come preferentially from higher
temperature gas, assuming the true temperature structure is not
isothermal but contains a continuous spectrum with gas at many
temperatures due to substructure and/or radial gradients
\citep{CDV+08, SMK+08}.

\begin{deluxetable*}{lcccccccccc}
\tablewidth{0pt}
\tablecaption{EPIC-only, Single Temperature Fit Parameters
\label{tab:bathi:xmmonly}}
\tablehead{
 & \multicolumn{4}{c}{Fits (2--12 keV)} & &
\multicolumn{4}{c}{Fits (3--12 keV)} \\
\cline{2-5}
\cline{7-10}
 & $kT$ & abund & Norm.\tablenotemark{a} & $\chi^2$/dof &
 & $kT$ & abund & Norm.\tablenotemark{a} & $\chi^2$/dof \\
Name & (keV) & (Z$_\odot$) & (cm$^{-5}$) & & &
(keV) & (Z$_\odot$) & (cm$^{-5}$) &
}\startdata
A0085 &  6.53$^{+ 0.20}_{- 0.19}$ & 
0.355$^{+0.031}_{-0.030}$ & 
0.0773$^{+0.0012}_{-0.0012}$ &  649.48/803 &  & 
 7.30$^{+ 0.37}_{- 0.36}$ & 
0.373$^{+0.036}_{-0.035}$ & 
0.0728$^{+0.0019}_{-0.0019}$ &  395.69/525 \\
A0119 &  5.73$^{+ 0.48}_{- 0.47}$ & 
0.227$^{+0.069}_{-0.068}$ & 
0.0314$^{+0.0012}_{-0.0011}$ &  226.39/270 &  & 
 7.18$^{+ 1.19}_{- 1.03}$ & 
0.248$^{+0.090}_{-0.084}$ & 
0.0280$^{+0.0024}_{-0.0019}$ &  110.39/153 \\
A0133 &  3.79$^{+ 0.14}_{- 0.13}$ & 
0.446$^{+0.050}_{-0.048}$ & 
0.0236$^{+0.0006}_{-0.0006}$ &  300.33/413 &  & 
 4.30$^{+ 0.36}_{- 0.28}$ & 
0.445$^{+0.053}_{-0.051}$ & 
0.0207$^{+0.0013}_{-0.0012}$ &  129.80/228 \\
NGC507 &  1.50$^{+ 0.08}_{- 0.08}$ & 
0.821$^{+0.239}_{-0.189}$ & 
0.0101$^{+0.0015}_{-0.0014}$ &  132.82/183 &  & 
 1.92$^{+ 0.37}_{- 0.27}$ & 
0.777$^{+0.660}_{-0.393}$ & 
0.0071$^{+0.0025}_{-0.0018}$ &   48.66/90 \\
A0262 &  2.23$^{+ 0.04}_{- 0.04}$ & 
0.485$^{+0.046}_{-0.044}$ & 
0.0549$^{+0.0015}_{-0.0014}$ &  584.83/668 &  & 
 2.37$^{+ 0.10}_{- 0.09}$ & 
0.395$^{+0.056}_{-0.053}$ & 
0.0548$^{+0.0030}_{-0.0029}$ &  284.19/370 \\
A0400 &  2.24$^{+ 0.12}_{- 0.12}$ & 
0.374$^{+0.108}_{-0.098}$ & 
0.0206$^{+0.0014}_{-0.0013}$ &  247.76/288 &  & 
 2.46$^{+ 0.33}_{- 0.26}$ & 
0.394$^{+0.181}_{-0.152}$ & 
0.0167$^{+0.0029}_{-0.0026}$ &  100.98/150 \\
A0399 &  7.44$^{+ 0.50}_{- 0.49}$ & 
0.224$^{+0.053}_{-0.053}$ & 
0.0356$^{+0.0010}_{-0.0010}$ &  269.07/368 &  & 
 8.10$^{+ 1.17}_{- 0.77}$ & 
0.237$^{+0.062}_{-0.059}$ & 
0.0343$^{+0.0018}_{-0.0018}$ &  143.88/220 \\
A3112 &  4.85$^{+ 0.13}_{- 0.13}$ & 
0.445$^{+0.029}_{-0.029}$ & 
0.0365$^{+0.0006}_{-0.0006}$ &  636.64/715 &  & 
 5.28$^{+ 0.27}_{- 0.22}$ & 
0.455$^{+0.032}_{-0.031}$ & 
0.0341$^{+0.0012}_{-0.0012}$ &  355.09/439 \\
Fornax &  1.66$^{+ 0.03}_{- 0.03}$ & 
0.743$^{+0.070}_{-0.065}$ & 
0.0191$^{+0.0009}_{-0.0009}$ &  867.36/801 &  & 
 2.06$^{+ 0.16}_{- 0.14}$ & 
0.233$^{+0.090}_{-0.080}$ & 
0.0196$^{+0.0023}_{-0.0020}$ &  464.04/494 \\
2A0335 &  3.03$^{+ 0.06}_{- 0.06}$ & 
0.423$^{+0.034}_{-0.033}$ & 
0.1014$^{+0.0019}_{-0.0019}$ &  482.66/649 &  & 
 3.22$^{+ 0.13}_{- 0.12}$ & 
0.395$^{+0.036}_{-0.035}$ & 
0.0966$^{+0.0041}_{-0.0041}$ &  234.21/372 \\
IIIZw54 &  2.63$^{+ 0.11}_{- 0.10}$ & 
0.297$^{+0.062}_{-0.058}$ & 
0.0198$^{+0.0008}_{-0.0007}$ &  302.14/404 &  & 
 3.07$^{+ 0.28}_{- 0.24}$ & 
0.238$^{+0.067}_{-0.062}$ & 
0.0171$^{+0.0016}_{-0.0015}$ &  154.89/219 \\
A3158 &  5.99$^{+ 0.37}_{- 0.35}$ & 
0.332$^{+0.057}_{-0.056}$ & 
0.0407$^{+0.0012}_{-0.0012}$ &  263.15/351 &  & 
 6.67$^{+ 0.72}_{- 0.60}$ & 
0.351$^{+0.066}_{-0.064}$ & 
0.0376$^{+0.0023}_{-0.0020}$ &  146.23/208 \\
NGC1550 &  1.42$^{+ 0.05}_{- 0.04}$ & 
0.522$^{+0.090}_{-0.079}$ & 
0.0285$^{+0.0023}_{-0.0022}$ &  198.84/263 &  & 
 1.60$^{+ 0.19}_{- 0.15}$ & 
0.282$^{+0.205}_{-0.158}$ & 
0.0292$^{+0.0076}_{-0.0060}$ &   78.01/119 \\
EXO0422 &  3.06$^{+ 0.07}_{- 0.07}$ & 
0.357$^{+0.033}_{-0.032}$ & 
0.0304$^{+0.0006}_{-0.0006}$ &  597.13/744 &  & 
 3.23$^{+ 0.15}_{- 0.13}$ & 
0.337$^{+0.036}_{-0.034}$ & 
0.0284$^{+0.0013}_{-0.0013}$ &  318.60/437 \\
A3266 &  8.34$^{+ 0.30}_{- 0.28}$ & 
0.196$^{+0.030}_{-0.030}$ & 
0.0797$^{+0.0011}_{-0.0011}$ &  831.15/1051 &  & 
 8.59$^{+ 0.90}_{- 0.44}$ & 
0.197$^{+0.035}_{-0.032}$ & 
0.0788$^{+0.0021}_{-0.0024}$ &  559.63/721 \\
A0496 &  4.36$^{+ 0.08}_{- 0.10}$ & 
0.394$^{+0.021}_{-0.021}$ & 
0.0835$^{+0.0012}_{-0.0010}$ & 1003.00/1083 &  & 
 4.68$^{+ 0.14}_{- 0.14}$ & 
0.388$^{+0.022}_{-0.022}$ & 
0.0790$^{+0.0019}_{-0.0019}$ &  610.07/757 \\
A3376 &  4.00$^{+ 0.29}_{- 0.27}$ & 
0.498$^{+0.126}_{-0.118}$ & 
0.0108$^{+0.0005}_{-0.0005}$ &  129.58/167 &  & 
 5.76$^{+ 1.15}_{- 0.94}$ & 
0.454$^{+0.146}_{-0.130}$ & 
0.0085$^{+0.0010}_{-0.0008}$ &   52.36/75 \\
A3391 &  6.45$^{+ 0.33}_{- 0.31}$ & 
0.312$^{+0.050}_{-0.049}$ & 
0.0207$^{+0.0005}_{-0.0005}$ &  371.79/482 &  & 
 6.85$^{+ 0.58}_{- 0.49}$ & 
0.315$^{+0.054}_{-0.052}$ & 
0.0200$^{+0.0009}_{-0.0008}$ &  199.48/294 \\
A3395s &  5.76$^{+ 0.66}_{- 0.66}$ & 
0.248$^{+0.102}_{-0.099}$ & 
0.0077$^{+0.0004}_{-0.0004}$ &  113.23/205 &  & 
 5.95$^{+ 1.30}_{- 1.13}$ & 
0.246$^{+0.112}_{-0.104}$ & 
0.0075$^{+0.0011}_{-0.0008}$ &   52.46/115 \\
R1504 &  8.54$^{+ 0.61}_{- 0.38}$ & 
0.412$^{+0.045}_{-0.042}$ & 
0.0492$^{+0.0011}_{-0.0011}$ & 1629.30/1341 &  & 
 8.59$^{+ 0.76}_{- 0.51}$ & 
0.416$^{+0.053}_{-0.047}$ & 
0.0490$^{+0.0019}_{-0.0017}$ & 1283.21/1010 \\
A0576 &  4.06$^{+ 0.28}_{- 0.26}$ & 
0.377$^{+0.087}_{-0.083}$ & 
0.0245$^{+0.0012}_{-0.0011}$ &  167.36/217 &  & 
 4.29$^{+ 0.68}_{- 0.47}$ & 
0.378$^{+0.094}_{-0.087}$ & 
0.0228$^{+0.0025}_{-0.0023}$ &   81.90/120 \\
A0754 &  9.16$^{+ 0.38}_{- 0.37}$ & 
0.281$^{+0.032}_{-0.032}$ & 
0.0703$^{+0.0008}_{-0.0008}$ &  780.81/951 &  & 
 9.43$^{+ 0.55}_{- 0.54}$ & 
0.285$^{+0.034}_{-0.033}$ & 
0.0697$^{+0.0015}_{-0.0014}$ &  523.22/636 \\
HydraA &  3.98$^{+ 0.09}_{- 0.09}$ & 
0.286$^{+0.026}_{-0.025}$ & 
0.0452$^{+0.0008}_{-0.0008}$ &  607.64/709 &  & 
 4.39$^{+ 0.19}_{- 0.18}$ & 
0.282$^{+0.026}_{-0.026}$ & 
0.0412$^{+0.0015}_{-0.0014}$ &  329.19/434 \\
A1060 &  3.20$^{+ 0.05}_{- 0.05}$ & 
0.406$^{+0.024}_{-0.023}$ & 
0.0592$^{+0.0008}_{-0.0008}$ &  853.00/963 &  & 
 3.44$^{+ 0.10}_{- 0.09}$ & 
0.384$^{+0.024}_{-0.024}$ & 
0.0544$^{+0.0015}_{-0.0014}$ &  498.30/632 \\
A1367 &  3.79$^{+ 0.12}_{- 0.12}$ & 
0.297$^{+0.037}_{-0.036}$ & 
0.0327$^{+0.0007}_{-0.0007}$ &  472.97/594 &  & 
 4.18$^{+ 0.25}_{- 0.23}$ & 
0.292$^{+0.039}_{-0.038}$ & 
0.0302$^{+0.0016}_{-0.0014}$ &  250.18/335 \\
MKW4 &  1.69$^{+ 0.12}_{- 0.11}$ & 
0.660$^{+0.248}_{-0.190}$ & 
0.0145$^{+0.0024}_{-0.0021}$ &   46.08/97 &  & 
 1.76$^{+ 0.29}_{- 0.26}$ & 
0.924$^{+1.095}_{-0.535}$ & 
0.0119$^{+0.0063}_{-0.0041}$ &   18.41/39 \\
ZwCl1215 &  7.15$^{+ 0.35}_{- 0.34}$ & 
0.283$^{+0.038}_{-0.037}$ & 
0.0257$^{+0.0005}_{-0.0005}$ &  455.11/603 &  & 
 7.66$^{+ 0.53}_{- 0.52}$ & 
0.300$^{+0.044}_{-0.042}$ & 
0.0248$^{+0.0009}_{-0.0008}$ &  269.38/363 \\
NGC4636 &  0.95$^{+ 0.11}_{- 0.08}$ & 
0.848$^{+0.478}_{-0.255}$ & 
0.0060$^{+0.0016}_{-0.0015}$ &  227.15/354 &  & 
 3.44$^{+ 4.08}_{- 1.88}$ & 
0.000$^{+1.498}_{-0.000}$ & 
0.0019$^{+0.0008}_{-0.0010}$ &   95.54/145 \\
A3526 &  3.95$^{+ 0.04}_{- 0.02}$ & 
0.544$^{+0.010}_{-0.007}$ & 
0.1080$^{+0.0005}_{-0.0016}$ & 3533.06/2257 &  & 
 4.02$^{+ 0.06}_{- 0.05}$ & 
0.522$^{+0.011}_{-0.010}$ & 
0.1085$^{+0.0019}_{-0.0028}$ & 2409.76/1925 \\
A1644 &  5.12$^{+ 0.24}_{- 0.23}$ & 
0.294$^{+0.046}_{-0.045}$ & 
0.0443$^{+0.0012}_{-0.0012}$ &  389.51/525 &  & 
 5.74$^{+ 0.52}_{- 0.56}$ & 
0.306$^{+0.052}_{-0.050}$ & 
0.0412$^{+0.0027}_{-0.0021}$ &  227.60/296 \\
A1650 &  5.96$^{+ 0.17}_{- 0.17}$ & 
0.393$^{+0.026}_{-0.026}$ & 
0.0275$^{+0.0004}_{-0.0004}$ &  748.85/910 &  & 
 6.13$^{+ 0.26}_{- 0.25}$ & 
0.396$^{+0.028}_{-0.028}$ & 
0.0271$^{+0.0007}_{-0.0007}$ &  449.63/594 \\
A1651 &  6.43$^{+ 0.37}_{- 0.35}$ & 
0.389$^{+0.057}_{-0.056}$ & 
0.0348$^{+0.0011}_{-0.0011}$ &  197.37/326 &  & 
 6.82$^{+ 0.75}_{- 0.56}$ & 
0.405$^{+0.070}_{-0.061}$ & 
0.0338$^{+0.0021}_{-0.0019}$ &  118.80/190 \\
Coma &  8.53$^{+ 0.19}_{- 0.13}$ & 
0.248$^{+0.015}_{-0.015}$ & 
0.2443$^{+0.0016}_{-0.0016}$ & 1787.94/2158 &  & 
 8.65$^{+ 0.26}_{- 0.22}$ & 
0.249$^{+0.015}_{-0.015}$ & 
0.2439$^{+0.0030}_{-0.0025}$ & 1445.12/1826 \\
NGC5044 &  1.21$^{+ 0.04}_{- 0.04}$ & 
0.797$^{+0.148}_{-0.124}$ & 
0.0247$^{+0.0029}_{-0.0027}$ &  393.86/488 &  & 
 1.48$^{+ 0.19}_{- 0.14}$ & 
0.627$^{+0.464}_{-0.294}$ & 
0.0192$^{+0.0061}_{-0.0047}$ &  176.46/229 \\
A3558 &  5.92$^{+ 0.10}_{- 0.10}$ & 
0.323$^{+0.015}_{-0.015}$ & 
0.0665$^{+0.0005}_{-0.0005}$ & 1271.71/1456 &  & 
 6.25$^{+ 0.15}_{- 0.15}$ & 
0.334$^{+0.016}_{-0.016}$ & 
0.0641$^{+0.0010}_{-0.0010}$ &  904.58/1124 \\
A3562 &  5.09$^{+ 0.65}_{- 0.55}$ & 
0.417$^{+0.155}_{-0.146}$ & 
0.0175$^{+0.0013}_{-0.0012}$ &   54.45/125 &  & 
 5.69$^{+ 1.52}_{- 1.09}$ & 
0.416$^{+0.173}_{-0.153}$ & 
0.0163$^{+0.0028}_{-0.0022}$ &   24.75/69 \\
A3571 &  7.24$^{+ 0.15}_{- 0.15}$ & 
0.372$^{+0.019}_{-0.019}$ & 
0.1104$^{+0.0011}_{-0.0011}$ & 1610.57/1874 &  & 
 7.57$^{+ 0.21}_{- 0.21}$ & 
0.385$^{+0.021}_{-0.021}$ & 
0.1074$^{+0.0017}_{-0.0017}$ & 1260.22/1542 \\
A1795 &  5.67$^{+ 0.08}_{- 0.08}$ & 
0.369$^{+0.013}_{-0.013}$ & 
0.0797$^{+0.0006}_{-0.0006}$ & 1673.53/1907 &  & 
 5.89$^{+ 0.12}_{- 0.12}$ & 
0.375$^{+0.014}_{-0.014}$ & 
0.0781$^{+0.0011}_{-0.0011}$ & 1337.50/1575 \\
A3581 &  1.91$^{+ 0.04}_{- 0.04}$ & 
0.556$^{+0.059}_{-0.055}$ & 
0.0271$^{+0.0010}_{-0.0010}$ &  431.36/546 &  & 
 2.04$^{+ 0.12}_{- 0.10}$ & 
0.404$^{+0.082}_{-0.074}$ & 
0.0275$^{+0.0024}_{-0.0023}$ &  198.57/278 \\
MKW8 &  3.36$^{+ 0.30}_{- 0.21}$ & 
0.350$^{+0.099}_{-0.092}$ & 
0.0135$^{+0.0007}_{-0.0007}$ &  150.85/221 &  & 
 3.88$^{+ 0.63}_{- 0.48}$ & 
0.321$^{+0.104}_{-0.097}$ & 
0.0119$^{+0.0014}_{-0.0014}$ &   69.64/120 \\
A2029 &  7.97$^{+ 0.22}_{- 0.22}$ & 
0.428$^{+0.029}_{-0.029}$ & 
0.0782$^{+0.0010}_{-0.0010}$ &  864.77/943 &  & 
 8.46$^{+ 0.39}_{- 0.30}$ & 
0.453$^{+0.034}_{-0.033}$ & 
0.0754$^{+0.0016}_{-0.0016}$ &  539.03/632 \\
A2052 &  3.01$^{+ 0.05}_{- 0.05}$ & 
0.500$^{+0.029}_{-0.029}$ & 
0.0480$^{+0.0007}_{-0.0007}$ &  717.07/849 &  & 
 3.22$^{+ 0.10}_{- 0.09}$ & 
0.471$^{+0.031}_{-0.031}$ & 
0.0454$^{+0.0015}_{-0.0015}$ &  426.04/523 \\
MKW3S &  3.36$^{+ 0.06}_{- 0.06}$ & 
0.388$^{+0.027}_{-0.026}$ & 
0.0392$^{+0.0006}_{-0.0006}$ &  693.59/838 &  & 
 3.65$^{+ 0.13}_{- 0.12}$ & 
0.385$^{+0.028}_{-0.028}$ & 
0.0361$^{+0.0012}_{-0.0012}$ &  398.37/515 \\
A2065 &  6.51$^{+ 0.60}_{- 0.49}$ & 
0.261$^{+0.078}_{-0.077}$ & 
0.0290$^{+0.0018}_{-0.0018}$ &  161.22/249 &  & 
 6.76$^{+ 1.14}_{- 0.83}$ & 
0.260$^{+0.082}_{-0.080}$ & 
0.0282$^{+0.0034}_{-0.0031}$ &   95.15/156 \\
A2063 &  4.34$^{+ 0.14}_{- 0.13}$ & 
0.345$^{+0.034}_{-0.032}$ & 
0.0371$^{+0.0008}_{-0.0007}$ &  640.80/774 &  & 
 4.55$^{+ 0.23}_{- 0.22}$ & 
0.344$^{+0.034}_{-0.033}$ & 
0.0356$^{+0.0014}_{-0.0013}$ &  419.72/510 \\
A2142 &  9.64$^{+ 2.83}_{- 1.88}$ & 
0.280$^{+0.217}_{-0.222}$ & 
0.0638$^{+0.0052}_{-0.0051}$ &  273.45/157 &  & 
 8.00$^{+ 4.37}_{- 1.96}$ & 
0.256$^{+0.189}_{-0.172}$ & 
0.0678$^{+0.0117}_{-0.0103}$ &  131.34/93 \\
A2147 &  5.17$^{+ 0.58}_{- 0.43}$ & 
0.238$^{+0.100}_{-0.096}$ & 
0.0410$^{+0.0024}_{-0.0023}$ &  164.44/220 &  & 
 6.46$^{+ 1.48}_{- 1.06}$ & 
0.249$^{+0.120}_{-0.113}$ & 
0.0341$^{+0.0041}_{-0.0037}$ &   82.91/129 \\
A2199 &  4.45$^{+ 0.09}_{- 0.09}$ & 
0.363$^{+0.021}_{-0.020}$ & 
0.1021$^{+0.0012}_{-0.0012}$ &  910.01/1069 &  & 
 4.59$^{+ 0.14}_{- 0.14}$ & 
0.366$^{+0.022}_{-0.021}$ & 
0.0999$^{+0.0023}_{-0.0023}$ &  565.55/737 \\
A2204 &  7.11$^{+ 0.24}_{- 0.23}$ & 
0.397$^{+0.029}_{-0.028}$ & 
0.0468$^{+0.0007}_{-0.0007}$ &  618.61/772 &  & 
 7.46$^{+ 0.33}_{- 0.32}$ & 
0.413$^{+0.033}_{-0.032}$ & 
0.0456$^{+0.0012}_{-0.0012}$ &  365.71/498 \\
A2256 &  6.97$^{+ 0.40}_{- 0.39}$ & 
0.299$^{+0.044}_{-0.043}$ & 
0.0530$^{+0.0013}_{-0.0011}$ &  324.79/434 &  & 
 8.07$^{+ 0.67}_{- 0.59}$ & 
0.338$^{+0.056}_{-0.053}$ & 
0.0491$^{+0.0019}_{-0.0018}$ &  176.12/253 \\
A2255 &  7.81$^{+ 0.95}_{- 0.87}$ & 
0.267$^{+0.110}_{-0.107}$ & 
0.0237$^{+0.0012}_{-0.0012}$ &   96.21/184 &  & 
 8.10$^{+ 1.69}_{- 1.35}$ & 
0.255$^{+0.118}_{-0.107}$ & 
0.0235$^{+0.0023}_{-0.0020}$ &   48.15/110 \\
A3667 &  6.62$^{+ 0.11}_{- 0.11}$ & 
0.266$^{+0.015}_{-0.015}$ & 
0.0761$^{+0.0006}_{-0.0006}$ & 1495.08/1643 &  & 
 7.20$^{+ 0.21}_{- 0.21}$ & 
0.277$^{+0.017}_{-0.017}$ & 
0.0728$^{+0.0010}_{-0.0010}$ & 1141.52/1311 \\
S1101 &  2.65$^{+ 0.06}_{- 0.06}$ & 
0.337$^{+0.038}_{-0.037}$ & 
0.0259$^{+0.0007}_{-0.0007}$ &  412.70/525 &  & 
 2.86$^{+ 0.14}_{- 0.13}$ & 
0.336$^{+0.044}_{-0.042}$ & 
0.0235$^{+0.0013}_{-0.0013}$ &  208.02/274 \\
A2589 &  3.69$^{+ 0.13}_{- 0.12}$ & 
0.542$^{+0.052}_{-0.050}$ & 
0.0205$^{+0.0005}_{-0.0005}$ &  326.99/437 &  & 
 3.87$^{+ 0.23}_{- 0.21}$ & 
0.545$^{+0.055}_{-0.053}$ & 
0.0197$^{+0.0010}_{-0.0010}$ &  155.20/243 \\
A2597 &  3.34$^{+ 0.07}_{- 0.06}$ & 
0.334$^{+0.025}_{-0.024}$ & 
0.0273$^{+0.0005}_{-0.0005}$ &  610.98/712 &  & 
 3.91$^{+ 0.17}_{- 0.16}$ & 
0.314$^{+0.025}_{-0.025}$ & 
0.0236$^{+0.0009}_{-0.0009}$ &  300.35/398 \\
A2634 &  4.55$^{+ 0.57}_{- 0.48}$ & 
0.271$^{+0.134}_{-0.127}$ & 
0.0184$^{+0.0013}_{-0.0012}$ &   96.47/131 &  & 
 4.89$^{+ 1.36}_{- 0.94}$ & 
0.271$^{+0.145}_{-0.136}$ & 
0.0176$^{+0.0032}_{-0.0026}$ &   51.98/70 \\
A2657 &  5.16$^{+ 0.32}_{- 0.29}$ & 
0.283$^{+0.065}_{-0.063}$ & 
0.0256$^{+0.0015}_{-0.0015}$ &  268.87/347 &  & 
 5.88$^{+ 0.69}_{- 0.65}$ & 
0.251$^{+0.069}_{-0.067}$ & 
0.0233$^{+0.0025}_{-0.0024}$ &  170.68/226 \\
A4038 &  3.20$^{+ 0.05}_{- 0.05}$ & 
0.365$^{+0.024}_{-0.024}$ & 
0.0596$^{+0.0009}_{-0.0009}$ &  870.70/1049 &  & 
 3.42$^{+ 0.12}_{- 0.11}$ & 
0.343$^{+0.026}_{-0.025}$ & 
0.0558$^{+0.0020}_{-0.0017}$ &  577.41/717 \\
A4059 &  4.24$^{+ 0.14}_{- 0.12}$ & 
0.425$^{+0.036}_{-0.035}$ & 
0.0342$^{+0.0007}_{-0.0007}$ &  471.59/685 &  & 
 4.48$^{+ 0.23}_{- 0.22}$ & 
0.425$^{+0.037}_{-0.036}$ & 
0.0329$^{+0.0013}_{-0.0012}$ &  251.18/418

\enddata
\tablenotetext{a}{Normalization of the {\tt APEC} thermal spectrum,
which is given by $\{ 10^{-14} / [ 4 \pi (1+z)^2 D_A^2 ] \} \, \int n_e n_H
\, dV$, where $z$ is the redshift, $D_A$ is the angular diameter distance,
$n_e$ is the electron density, $n_H$ is the ionized hydrogen density,
and $V$ is the volume of the cluster.}
\end{deluxetable*}
\subsection{Non-thermal Fits to the BAT Spectra}
\label{sec:bathi:separate:bat}

Our goal is to detect a non-thermal spectral component at hard energies,
but because the statistical weight of the BAT channels is so much less
than the EPIC channels (lower S/N and fewer of them, at least by an 
order of magnitude), we have to be careful not to let the \xmms data
unfairly drive the spectral fits.
To assess the sensitivity of our BAT spectra, we extract 10,000 blank
sky spectra from
uniformly distributed, random
positions at least 40\arcmin\ from
any known sources and greater than 20\arcdeg\ from the Galactic plane,
to mimic the selection function in \hi.
We then fit these spectra with a fiducial power law model of photon
index $\Gamma$ fixed at a value of 2, roughly the appropriate slope for IC
emission inferred from radio halos, relics, and mini-halos.
While the spectral index determined from the radio is typically steeper than
this (2.2-2.4), the electrons producing the radio emission at $\nu > 100$ MHz
have higher energies than those producing IC at $E < 50$ keV for
$B \lesssim 0.5 \mu$G, so a simple extrapolation may not be appropriate.
A clear flattening of the radio spectrum at low frequencies is apparent in
some cases, e.g., Coma \citep{TKW03} and A3562 \citep{Gia+05}, although this is not
universally found as in A2256 \citep{Bre08} and A2255 \citep{PB09}.
Since the BAT data are not particularly sensitive to the precise value of the
index, we choose a flatter slope to avoid poorly fitting the data at
$\sim$ keV energies where the power law distribution of relativistic
electrons is most likely to turn over in a steady state-like injection model
\citep[e.g.,][]{Sar99}.
\begin{figure}
\plotone{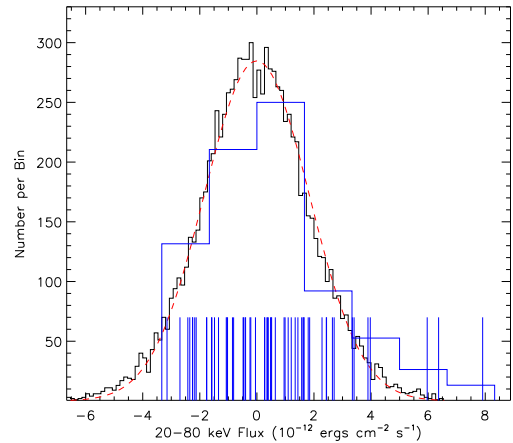}
\caption{The distribution of power law normalizations (with a fixed
photon index $\Gamma=2$) fit to 10,000 blank sky spectra extracted
from the BAT survey (narrowly-binned histogram).
The best-fit Gaussian distribution is overlaid as the smooth, dashed
line (red).
Similar best-fit normalizations are shown for the 59
\his clusters (see text for details), with individual normalizations
represented as vertical lines (blue).
The cluster histogram (wide bins) has been scaled up to show its agreement with
the blank sky spectra.
In general, the cluster BAT spectra lack any clear evidence for
a non-thermal component, except in a few cases comprising the positive
tail of the blue histogram.
\label{fig:bathi:bathist}}
\end{figure}

The distribution of best-fit normalizations from these power law fits
are presented in the narrow histogram in Figure~\ref{fig:bathi:bathist}.
They are well fit by a symmetric Gaussian (dashed smooth line) and
indicate a $1\sigma$ sensitivity threshold of 
$\sim2\times10^{-12}$ erg cm$^{-2}$ s$^{-1}$ (20--80 keV).
Similarly, the formal $3\sigma$ detection level is 
$5.8\times10^{-12}$ erg cm$^{-2}$ s$^{-1}$.
In principle, the BAT survey is sensitive enough to confirm or 
reject previous detections of hard excesses with fluxes 
$\sim10^{-11}$ erg cm$^{-2}$ s$^{-1}$
\citep[e.g.,][]{RG02, MDG02, FOB+04}.

Now we wish to compare our cluster spectra with this distribution,
but first we have to account for any thermal emission in the lower
energy bands.
The single temperature models derived with \xmms (2--12 keV) are
included as a second component along with the power law model,
with only its normalization left as a free parameter.
The resulting non-thermal normalizations are also given in
Figure~\ref{fig:bathi:bathist} as both the wider histogram
(scaled up) and as the vertical lines (showing individual values).
While the majority of cluster non-thermal components are consistent 
with the blank sky fits, there is a tail at positive normalizations
possibly indicative of a non-thermal excess.
However, the thermal contribution is not well determined in this
method and may be underestimated.
Intriguingly, the three clusters with the most significant non-thermal
component (A2029, A1367, and A1651) have positive fluxes, although 
marginally detected, in all 8 BAT bands; this rarely occurs for the
blank sky spectra.
We discuss these clusters in more detail later.
The main result from this analysis is that the BAT cluster spectra
have probably not reached a sensitivity level sufficient to detect
hard, non-thermal excesses, if they exist, in the brightest clusters.


\section{Joint Fits to the EPIC-BAT Spectra}
\label{sec:bathi:joint}

BAT fluxes are calibrated to match both the normalization and
the spectral shape of sources as detected by the \xmms EPIC-pn
instrument \citep{WSF+11}, and they are extracted from regions
identical to the \xmms extraction regions.
As such, continuous spectral models can be used over the
full 2--195 keV energy range to simultaneously fit both the
\xmms and \swifts spectra.
However, in individual cases the cross-normalization factor, $f_{CN}$,
may stray from a value of 1 as it does between the pn and MOS
instruments (see Section~\ref{sec:bathi:separate:xmm}).
We therefore adopt, along with a 3\% uncertainty in the \xmms
background normalizations, a conservative 10\% systematic uncertainty
for $f_{CN}$.
Because no compelling evidence for non-thermal emission is found in
the nominally calibrated spectra (see analysis below), we only consider
these uncertainties when deriving 90\% confidence interval upper
limits.

\begin{figure}
\plotone{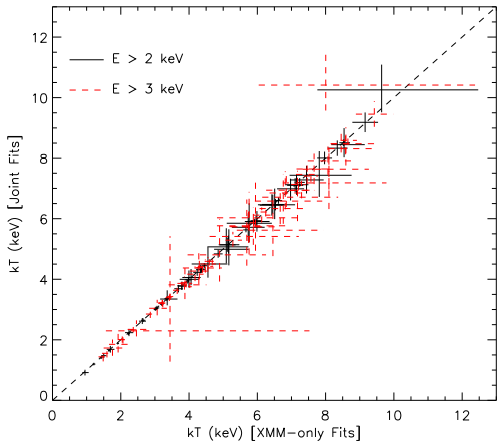}
\caption{A comparison of best-fit temperature values in 1T fits
to only the \xmms spectra (x-axis)
and to the EPIC and
BAT spectra simultaneously (y-axis).
Solid lines indicate the 90\% error interval for $E > 2$ keV fits,
dashed (red) lines for $E > 3$ keV, and the dashed diagonal line
represents equality between the two temperature determinations.
Jointly fitting both datasets yields consistent temperatures to those
derived only in the \xmms band.
Fitting over a slightly higher
energy range ($E > 3$ keV), while increasing the average temperature
by $\sim 0.3$ keV in the EPIC bandpass (see text), does not increase
the joint fit temperatures as much; note how the dashed points fall
slightly below equality for moderately hot clusters.
\label{fig:bathi:xjkts}}
\end{figure}
\subsection{General Properties from the Joint Analysis}
\label{sec:bathi:joint:general}

For each cluster, 3 simple spectral models are employed to describe the
emission covering 2 orders of magnitude in energy: a single temperature
thermal model (1T), a two temperature model (2T), and a thermal plus
non-thermal model (T$+$IC).
Due to the limited sensitivity of the \swifts data,
more complicated models cannot be constrained; for example,
the separate temperature
components in the 2T model are generally poorly constrained in our
analysis.
Above 50 keV, the {\tt APEC} emission model is replaced with {\tt MeKa}
because {\tt APEC} is not defined above 50 keV in the
implementation of {\tt XSpec} used here (Version 12.6.0k).
Note that the {\tt MeKaL} emission model could also be used continuously
across this energy range, if the look-up table switch is turned off.
For the thermal component, the temperature, abundance, redshift, and
normalization are all varied.
The individual abundances and redshifts
in the 2T model are tied together.
The non-thermal photon index is initially fixed at $\Gamma=2$, 
typical of radio halos, and the
normalization is allowed to vary;
when the photon index is fit for,
it is always fixed to the best-fit value before errors for other
parameters are derived.
In general, the photon index is poorly constrained, allowing for
a wide range of normalizations, which are then less straightforward
to evaluate.
The purpose of fitting for the photon index is to make sure that
we are not biased against detectable IC components with indices that 
differ from the fiducial value.

\begin{figure}
\plotone{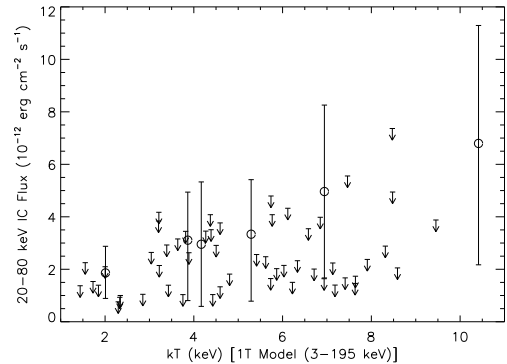}
\caption{Upper limits and measurements of the non-thermal spectral
component in the 3--195 keV joint fits as a function of cluster
temperature.
Limits and error bars indicate the 90\% confidence interval without
considering the impact of systematic uncertainties.
In general, an excess attributable to IC emission is not observed,
and the few detections, discussed individually in the text,
have marginal statistical significance.
\label{fig:bathi:uls}}
\end{figure}
Because of complications arising at energies between 2 and 3 keV
(see Section~\ref{sec:bathi:separate:xmm}),
we perform these fits for both the 2--195 keV 
(Table~\ref{tab:bathi:joint1}) and the 3--195 keV
(Table~\ref{tab:bathi:joint2}) spectral ranges.
The $E > 2$ keV fits, at first glance, suggest that there may be
evidence for a non-thermal component in a majority of \his clusters.
Many of the clusters with some evidence, at least at the 90\% level,
of a non-thermal excess are, unexpectedly, low temperature clusters
without significant detections at BAT energies.
In these cases, the non-thermal component is serving to ``adjust''
a problem at lower energies -- due to either incompletely modeled low
temperature components, an imperfectly calibrated response at the gold
edge, or both.
The significance of these instances will disappear from fits 
within a slightly higher energy range, while real non-thermal emission
will become a higher proportion of the total flux and so this component
should not greatly diminish in significance.
A drastic reduction in the number of marginally detected non-thermal 
excesses is seen when comparing Tables~\ref{tab:bathi:joint1} and
\ref{tab:bathi:joint2}; only 6 clusters are detected to have such
emission at the 90\% confidence level (statistical).
These clusters will be discussed individually in 
Section~\ref{sec:bathi:joint:individual}.

While the 3--12 keV band avoids some possible systematic uncertainties 
with the
\xmms response and complications from cooler gas, the narrower range may 
reduce
our ability to strongly constrain multi-temperature components in the 
spectra.
One concern is that a weak non-thermal emission component might be 
indistinguishable from
a purely thermal model with a slightly elevated temperature.
Note, however, that the 3--12 keV band temperatures in 
Section~\ref{sec:bathi:separate:xmm} are typically only $\sim0.3$ keV 
higher than
the 2--12 keV temperatures.
Therefore, the 1T model temperatures should agree for the joint fits
over both energy ranges, which is found to be the case in
Figure~\ref{fig:bathi:xjkts}.
Temperatures derived from joint fits are consistent with those found
using only the \xmms spectra, for both energy ranges.
For the most part, temperatures from the joint fit 3--195 keV fits are 
in good agreement with 
or slightly lower than the 3--12 keV temperatures.
The contribution of the BAT data in this case is to somewhat {\it lower}
the best-fit temperature, contrary to the expectation if a detectable
non-thermal excess were present.
The 3--195 keV non-thermal flux limits and possible detections 
(90\%, statistical) are shown
in Figure~\ref{fig:bathi:uls}.

\subsection{Individual Cases}
\label{sec:bathi:joint:individual}

Six clusters have a formal detection of non-thermal emission in the 
3--195 keV band.
Two of these 6 clusters are also in the top 3 of candidates for emission
based on their BAT-only fits: A1651 and A2142.
The other cluster in this top 3 -- with the largest non-thermal
normalization of all the clusters -- is A2029, so we will
include this cluster with the 6 ``detected'' clusters as worth some
brief discussion.
The clusters are listed in order of decreasing non-thermal flux.

\begin{figure}
\plotone{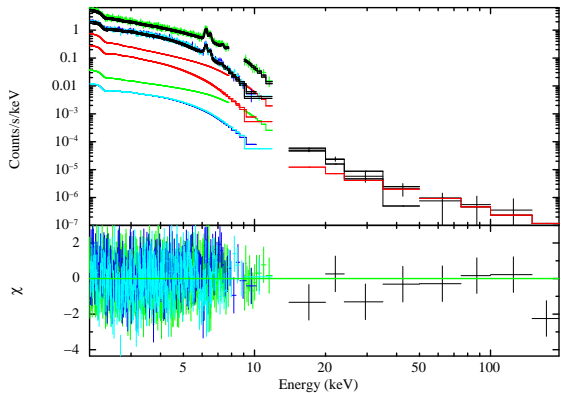}
\caption{Abell 2029: The T$+$IC model simultaneous fit to the
EPIC ($E < 12$ keV) and BAT ($E > 14$ keV, black data points and residuals) spectra.
The EPIC-pn spectrum and residuals are in green, and the MOS 1 and 2
spectra/residuals are in dark and light blue, respectively.
The like-colored lines below these spectra show the CXB model
contribution.
The total model fit and thermal contribution is represented by the
black histogram, and the red lines represent the non-thermal ($\Gamma=2$)
spectral component.
\label{fig:bathi:2029}}
\end{figure}
\begin{figure}
\plotone{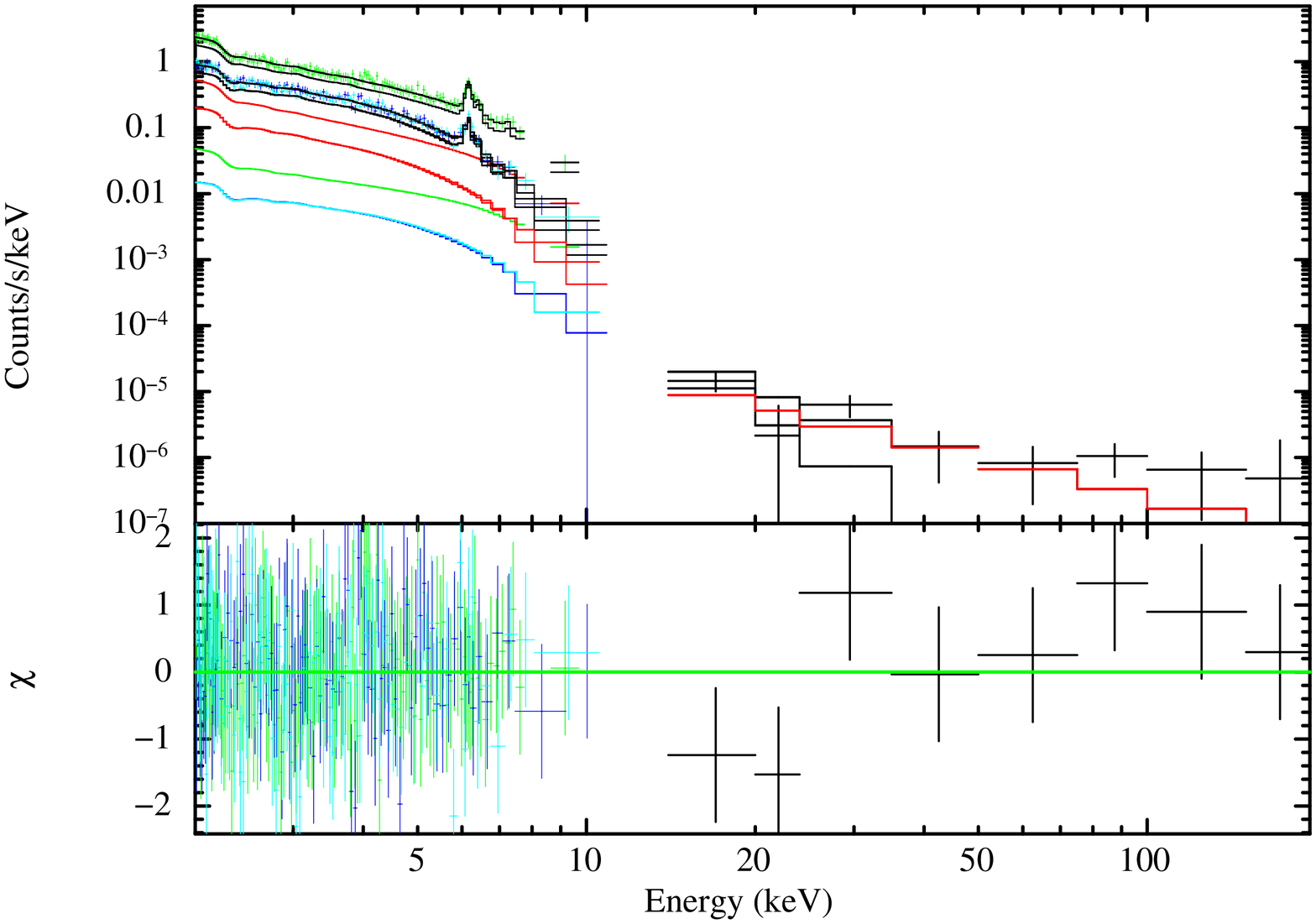}
\caption{Abell 1651: The T$+$IC model simultaneous fit to the
EPIC ($E < 12$ keV) and BAT ($E > 14$ keV) spectra.
The notation is identical to Figure~\ref{fig:bathi:2029}.
\label{fig:bathi:1651}}
\end{figure}
\begin{figure}
\plotone{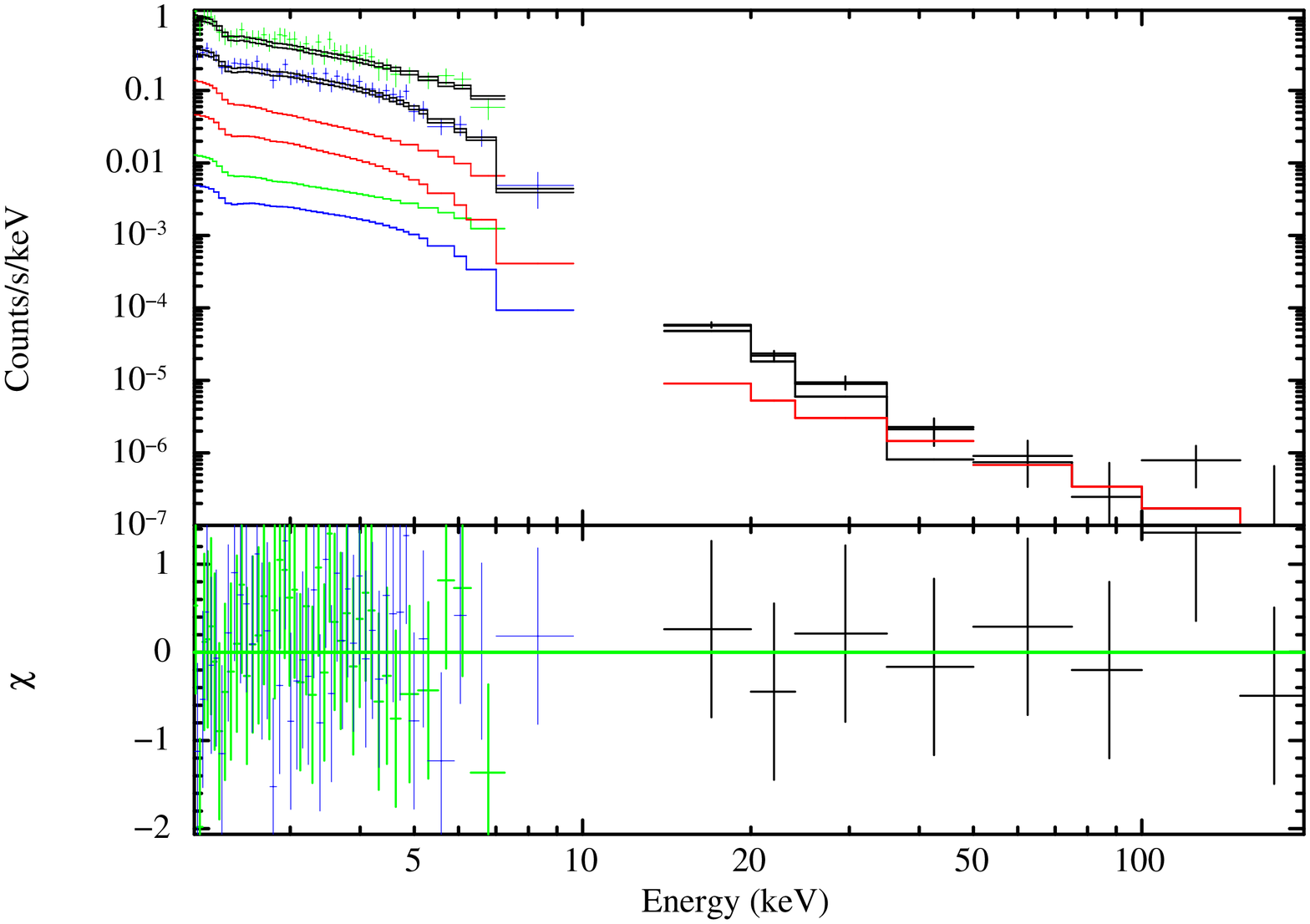}
\caption{Abell 2142: The T$+$IC model simultaneous fit to the
EPIC ($E < 12$ keV) and BAT ($E > 14$ keV) spectra.
The notation is identical to Figure~\ref{fig:bathi:2029}.
\label{fig:bathi:2142}}
\end{figure}
\begin{figure}
\plotone{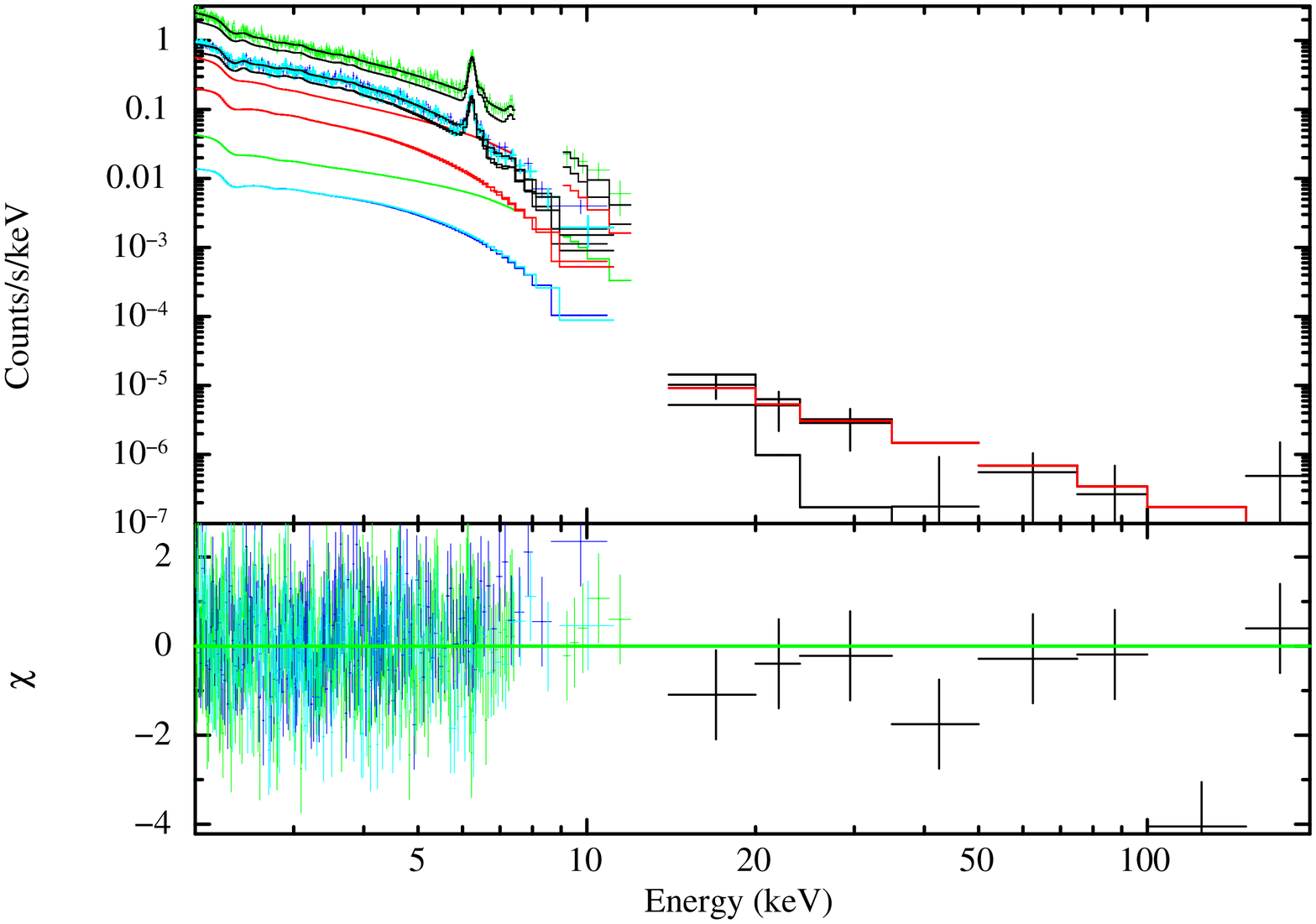}
\caption{Abell 3112: The T$+$IC model simultaneous fit to the
EPIC ($E < 12$ keV) and BAT ($E > 14$ keV) spectra.
The notation is identical to Figure~\ref{fig:bathi:2029}.
\label{fig:bathi:3112}}
\end{figure}
\begin{figure}
\plotone{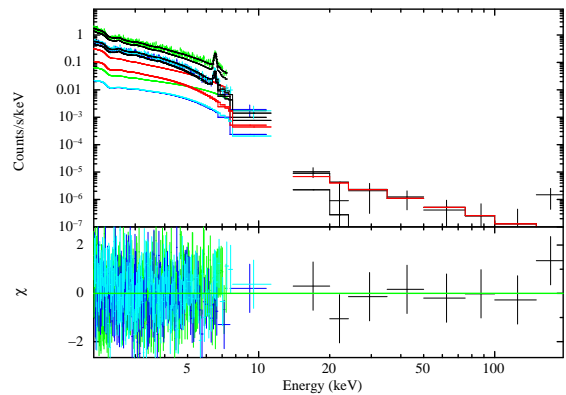}
\caption{Abell 1367: The T$+$IC model simultaneous fit to the
EPIC ($E < 12$ keV) and BAT ($E > 14$ keV) spectra.
The notation is identical to Figure~\ref{fig:bathi:2029}.
\label{fig:bathi:1367}}
\end{figure}
{\it A2029} (Fig.~\ref{fig:bathi:2029}): 
This hot ($\sim8$ keV), cool core cluster has been
studied in detail with \chandras \citep{CBS04}, who explore the interaction
between cool gas and the radio AGN in the cluster center.
The cluster is elongated but relatively regular; no evidence exists
for major merger activity; however, a minor merger may be producing
the spiral surface brightness enhancement in the center.
Also, no evidence for an X-ray counterpart of the AGN is visible in
the \chandras data.
In addition to the radio jets, the core of the cluster is also host
to an extended radio minihalo \citep{MGM+09}.
As with radio halos and relics, IC emission may be detectable from the
minihalo if the magnetic field is small; \citet{TBG94} measured a
lower limit of $B\ga$ 0.11-0.19 $\mu$G with Faraday RM observations
of the jet.
The implied magnetic field strength, if we take as the IC flux that
found with the 2--195 keV fit, is $B \sim 0.08$ $\mu$G, roughly
consistent with their field strength.

But have we really detected IC from the cluster core?
The significance of the non-thermal component completely disappears
in the 3--195 keV fit; all three model combinations match the data
equally well.
Also, the 2T model formally provides a better fit to the 2--195 keV spectrum where
the non-thermal component is detected.
The second temperature component, $\sim 0.3$ keV, is consistent with
a low temperature component of 0.11 keV observed by \citet{CBS04}.
Given these results, it is more likely that the non-thermal component is trying
to mimic the low $kT$ cool core component in the 2--3 keV range, since its
significance disappears if this energy range is ignored.
However, it is worth noting that the BAT data do generally support
hard emission at higher energies, although at low signal-to-noise.
Such hard emission could be due, on the other hand, to heavily
obscured emission from a background AGN within the FOV.
The spatial distribution of BAT emission is consistent with that from a 
point source in all bands.

{\it A1651} (Fig.~\ref{fig:bathi:1651}): 
This cluster has a weak cool core, which means that
while there is no significant temperature gradient in the center,
the cooling time of the gas in the center is short \citep{HMR+10}.
Note that in a bimodal classification A1651 would not be considered
to have a cool core given its high central entropy of 90 keV cm$^2$
\citep{CDV+09}.
Given the similarity between its BAT data and that of A2029,
an obscured AGN of similar flux could be responsible for the
marginally detected positive flux in the higher energy bands.
However, in this case the T$+$IC model is a significantly better
fit than is the 2T model; $\Delta \chi^2$ improves by 9 (2--195 keV) and
5 (3--195 keV) over the 1T and 2T models.
If there were no hard excess,
the probability that the 6 highest energy bands measure flux above
the thermal component, given that BAT fluctuations are Gaussian,
is $\left(\frac{1}{2}\right)^6$, or 1.6\%,
which is not impressive in a sample of 59 clusters.
The BAT spectrum is certainly suggestive, but
considering the excess is not significant at the $3\sigma$ level
for the 3--195 keV fit, and only just at this level in the 2--195 keV
fit -- without including systematic uncertainties -- we cannot claim 
to have detected a non-thermal component in this cluster.
However, the evidence is perhaps strongest in this case, which is
contrary to the expectation that such an excess is most likely in
a merging cluster, particularly one with a radio halo or relic.

{\it A2142} (Fig.~\ref{fig:bathi:2142}): 
As the hottest cluster in the sample, the BAT is easily able to
detect this cluster's high energy emission, which we might expect to exhibit
a non-thermal excess since it also hosts a radio halo
\citep{GF00}.
Both the T$+$IC and 2T models indicate that hard excess emission
may be present; in the latter case, the second temperature component
is unphysically high, acquiring the highest allowed temperature value.
However, \citet{NOB+04} estimate that 2 Seyfert galaxy nuclei within
17\arcmin\ of the cluster center contribute $\sim$30\% of the hard
band emission detected by \sax; a similar amount of contamination
would be expected in the BAT spectrum.
Unfortunately, the \xmms observation places this cluster right
on the edge of the FOV, so over half (55.6\%, based on a comparison
with a pointed {\it ROSAT} PSPC image) of the soft band emission is
missing from the EPIC spectra.
We rescale the \xmms spectra to correct for the lost flux;
the BAT source is equivalent to a point source, so it is not possible to 
correct the BAT emission for the \xmms FOV.
The correction to the \xmms flux could be off by a sizable factor
if the $E > 2$ keV emission is distributed differently than the
$E < 2$ keV emission where {\it ROSAT} is sensitive.
The significance of the non-thermal excess here is only at the
$2\sigma$ level, mainly due to the poor statistics at \xmms energies.
While inconclusive, the BAT spectrum warrants further analysis 
using better data below 12 keV.

{\it A3112} (Fig.~\ref{fig:bathi:3112}):
Using both \chandras and \xmms data, \citet{BNL07} have claimed to
see both a hard and soft excess that is consistent with a non-thermal origin.
If this is the correct interpretation of these spectra, the IC
excess would be clearly detectable in the BAT spectrum
given our sensitivity.
While a non-thermal component is detected in our joint fits, it has
well below the predicted flux of \citet{BNL07}; our $3\sigma$
upper limit on the non-thermal normalization, using a photon index
$\Gamma=1.8$ that matches their best-fit value, is 3 times lower
than their estimate.
The quality of our 1T model fits is significantly less than for
either the 2T or T$+$IC models; while those fits are of similar
quality, the 2T fit yields physically reasonable temperatures
and lower $\chi^2$ values ($\Delta\chi^2 \sim 3$) than the T$+$IC model
over both energy ranges.
A non-thermal excess may in fact exist in this cluster, but a perhaps more
likely scenario is that the ICM here is less isothermal than
is typical in clusters, requiring several temperature components to
adequately explain the cluster emission.
The analysis of the \chandras data by \citet{TSB+03} in fact demonstrates
the multi-temperature structure of this cluster, which may be exaggerated
by significant gas cooling outside the core.
In any case, the BAT data do not argue strongly in favor of an IC
interpretation for the excess emission above $\sim7$ keV observed in the
\xmms data; as can be seen in Figure~\ref{fig:bathi:3112}, the power
law component nearly ubiquitously overpredicts fluxes in the BAT spectrum.
A more detailed exploration of the spatial and thermal structure at 
$E < 12$ keV is certainly warranted.

{\it A1367} (Fig.~\ref{fig:bathi:1367}):
This cluster hosts a radio relic in its outskirts \citep{GT83}, 
and so IC emission is expected at some level in the radio relic
region; however, the \xmm/\swifts extraction region does not
contain the relic, so we are unable to address the magnetic field
strength.
Using \rxte, \citet{HM01} potentially detect a non-thermal component,
although a two temperature fit better describes their spectrum.
The marginally detected IC emission we see is consistent with
their non-thermal flux, whether we use a photon index of 2.0 or their
value (based on the spectrum of the radio relic) of 2.9.
Our 2T model fit, in the 2--195 keV band, is as good as the T$+$IC model fit,
and given the marginally detected fluxes in the BAT bands, a 2T
description of the ICM in this early stage, forming cluster
cannot be ruled out.
However, the positive BAT fluxes and the consistency of our 
non-thermal fit with the analysis of the \rxtes spectrum warrants future 
investigation of this cluster's hard X-ray emission.

{\it A2589 \& Fornax}: Neither of the BAT spectra of these clusters
show particular evidence that that they have detected emission
of any kind in any band.
The first 2 bands of A2589's spectrum are just inconsistent with
zero flux at the $1\sigma$ level, but a marginal detection in these
bands is consistent with the thermal component.
In both cases, the BAT spectrum is not sensitive enough to
exclude the non-thermal component driven by the \xmms data;
since the BAT data do not further constrain the non-thermal
component in these cases,
we will not discuss these clusters further.

\begin{deluxetable*}{lccccccccc}
\tablewidth{0pt}
\tabletypesize{\small}
\tablecaption{Upper Limits to 20--80 keV non-thermal Flux from 
EPIC and BAT Joint Fits
\label{tab:bathi:uls}}
\tablehead{
 & \multicolumn{4}{c}{Spectral Band: 2--195 keV} && 
 \multicolumn{4}{c}{Spectral Band: 3--195 keV} \\
\cline{2-5}
\cline{7-10}
Name & 90\%\tablenotemark{a} & $3\sigma_{\Gamma=2}$\tablenotemark{a} & 
$\Gamma$ & $3\sigma$\tablenotemark{a} & &
 90\%\tablenotemark{a} & $3\sigma_{\Gamma=2}$\tablenotemark{a} & $\Gamma$ & 
 $3\sigma$\tablenotemark{a}
}
\startdata
A0085 & \phn  0.729 & \phn  6.66 &  2.00 & \phn  6.66 &  & \phn  0.729 & \phn  6.66 &  1.05 & \phn  5.20 \\
A0119 & \phn  1.689 & \phn  6.58 &  2.12 & \phn  6.43 &  & \phn  1.689 & \phn  6.58 &  9.14 & 3.1$\times 10^{-7}$ \\
A0133 & \phn  3.759 & \phn  6.88 &  2.17 & \phn  6.40 &  & \phn  3.759 & \phn  6.88 &  2.14 & \phn  4.71 \\
NGC507 & \phn  1.725 & \phn  1.77 &  1.90 & \phn  2.13 &  & \phn  1.725 & \phn  1.77 &  1.97 & \phn  2.08 \\
A0262 & \phn  0.804 & \phn  4.05 &  2.00 & \phn  4.05 &  & \phn  0.804 & \phn  4.05 &  3.77 & \phn  0.15 \\
A0400 & \phn  0.645 & \phn  2.26 &  9.02 & 2.1$\times 10^{-8}$ &  & \phn  0.645 & \phn  2.26 &  8.80 & 6.9$\times 10^{-7}$ \\
A0399 & \phn  1.818 & \phn  4.04 &  2.07 & \phn  3.62 &  & \phn  0.032 & \phn  4.04 &  4.53 & \phn  0.01 \\
A3112 & \phn  5.420 & \phn  8.98 &  2.00 & \phn  8.98 &  & \phn  5.420 & \phn  8.98 &  2.18 & \phn  6.98 \\
Fornax & \phn  3.365 & \phn  3.13 &  2.00 & \phn  3.13 &  & \phn  3.365 & \phn  3.13 &  2.18 & \phn  2.67 \\
2A0335 & \phn  1.626 & \phn  6.80 &  2.00 & \phn  6.80 &  & \phn  1.626 & \phn  6.80 &  2.18 & \phn  4.31 \\
IIIZw54 & \phn  2.626 & \phn  4.32 &  2.14 & \phn  3.48 &  & \phn  2.626 & \phn  4.32 &  2.33 & \phn  2.63 \\
A3158 & \phn  2.453 & \phn  5.08 &  1.93 & \phn  5.24 &  & \phn  2.453 & \phn  5.08 &  1.98 & \phn  4.56 \\
NGC1550 & \phn  2.478 & \phn  1.66 &  2.06 & \phn  1.94 &  & \phn  2.478 & \phn  1.66 &  2.39 & \phn  1.60 \\
EXO0422 & \phn  4.171 & \phn  5.18 &  2.01 & \phn  5.14 &  & \phn  4.171 & \phn  5.18 &  2.03 & \phn  4.98 \\
A3266 & \phn  5.131 & \phn  7.25 &  1.98 & \phn  7.43 &  & \phn  5.131 & \phn  7.25 &  1.89 & \phn  7.87 \\
A0496 & \phn  0.733 & \phn  6.28 &  2.00 & \phn  6.28 &  & \phn  0.733 & \phn  6.28 &  2.11 & \phn  3.33 \\
A3376 & \phn  5.582 & \phn  6.01 &  2.10 & \phn  5.03 &  & \phn  5.582 & \phn  6.01 &  2.02 & \phn  5.77 \\
A3391 & \phn  4.609 & \phn  6.30 &  2.00 & \phn  6.25 &  & \phn  4.609 & \phn  6.30 &  2.00 & \phn  6.08 \\
A3395s & \phn  3.548 & \phn  3.52 &  2.00 & \phn  3.44 &  & \phn  3.548 & \phn  3.52 &  2.37 & \phn  2.48 \\
R1504 & \phn  9.934 & \phn  5.87 &  2.01 & \phn  5.72 &  & \phn  5.826 & \phn  5.87 &  2.62 & 31.90 \\
A0576 & \phn  3.671 & \phn  6.44 &  2.00 & \phn  6.44 &  & \phn  3.671 & \phn  6.44 &  2.02 & \phn  6.36 \\
A0754 & \phn  4.332 & \phn  7.61 &  2.00 & \phn  7.61 &  & \phn  4.332 & \phn  7.61 &  1.98 & 13.77 \\
HydraA & \phn  4.169 & \phn  8.72 &  2.08 & \phn  8.55 &  & \phn  4.169 & \phn  8.72 &  2.06 & \phn  5.78 \\
A1060 & \phn  0.469 & \phn  7.30 &  2.19 & \phn  7.11 &  & \phn  0.469 & \phn  7.30 &  2.61 & \phn  1.49 \\
A1367 & \phn  5.676 & \phn  8.24 &  2.00 & \phn  8.37 &  & \phn  5.676 & \phn  8.24 &  2.00 & \phn  7.30 \\
MKW4 & \phn  3.453 & \phn  1.98 &  1.93 & \phn  2.07 &  & \phn  3.453 & \phn  1.98 &  1.69 & \phn  3.80 \\
ZwCl1215 & \phn  1.585 & \phn  5.45 &  2.00 & \phn  7.27 &  & \phn  1.585 & \phn  5.45 &  2.26 & \phn  2.84 \\
NGC4636 & \phn  0.886 & \phn  0.61 &  2.00 & \phn  0.61 &  & \phn  0.886 & \phn  0.61 &  2.16 & \phn  0.63 \\
A3526 & \phn  3.709 & 11.23 &  2.00 & 11.59 &  & \phn  3.709 & 11.23 &  2.16 & \phn  2.03 \\
A1644 & \phn  4.915 & \phn  8.65 &  2.00 & \phn  8.65 &  & \phn  4.915 & \phn  8.65 &  2.08 & \phn  7.53 \\
A1650 & \phn  4.306 & \phn  6.63 &  2.00 & \phn  6.62 &  & \phn  4.306 & \phn  6.63 &  2.00 & \phn  6.71 \\
A1651 & \phn  8.497 & 10.82 &  2.00 & 10.83 &  & \phn  8.497 & 10.82 &  1.96 & 11.06 \\
Coma & \phn  1.509 & \phn  6.55 &  2.03 & \phn  6.48 &  & \phn  1.509 & \phn  6.55 &  0.47 & \phn  3.20 \\
NGC5044 & \phn  1.831 & \phn  1.92 &  2.14 & \phn  1.46 &  & \phn  1.831 & \phn  1.92 &  2.45 & \phn  0.89 \\
A3558 & \phn  0.229 & \phn  9.20 &  2.53 & \phn  1.85 &  & \phn  0.223 & \phn  9.20 &  2.58 & \phn  0.64 \\
A3562 & \phn  2.387 & \phn  5.52 &  4.98 & 1.3$\times 10^{-3}$ &  & \phn  2.311 & \phn  5.52 &  8.97 & 7.8$\times 10^{-7}$ \\
A3571 & \phn  0.442 & \phn  8.52 &  2.39 & \phn  3.51 &  & \phn  0.442 & \phn  8.52 &  9.46 & 1.3$\times 10^{-7}$ \\
A1795 & \phn  2.143 & \phn  7.22 &  2.68 & \phn  1.09 &  & \phn  2.143 & \phn  7.22 &  2.10 & \phn  2.99 \\
A3581 & \phn  2.351 & \phn  2.76 &  2.00 & \phn  2.75 &  & \phn  2.351 & \phn  2.76 &  2.34 & \phn  1.70 \\
MKW8 & \phn  3.508 & \phn  4.99 &  2.04 & \phn  4.80 &  & \phn  3.508 & \phn  4.99 &  2.05 & \phn  4.92 \\
A2029 & \phn  7.633 & 13.33 &  2.00 & 13.31 &  & \phn  7.633 & 13.33 &  1.82 & 11.50 \\
A2052 & \phn  3.430 & \phn  6.95 &  2.00 & \phn  6.95 &  & \phn  3.430 & \phn  6.95 &  2.14 & \phn  4.60 \\
MKW3S & \phn  3.280 & \phn  5.96 &  2.00 & \phn  5.96 &  & \phn  3.280 & \phn  5.96 &  1.95 & \phn  5.34 \\
A2065 & \phn  3.842 & \phn  5.59 &  2.00 & \phn  5.60 &  & \phn  3.842 & \phn  5.59 &  2.00 & \phn  5.95 \\
A2063 & \phn  2.573 & \phn  6.58 &  2.13 & \phn  5.81 &  & \phn  2.580 & \phn  6.58 &  2.16 & \phn  5.01 \\
A2142 & 11.494 & 13.51 &  1.95 & 13.48 &  & 11.494 & 13.51 &  1.98 & 14.65 \\
A2147 & \phn  2.203 & \phn  5.50 &  2.01 & \phn  5.52 &  & \phn  2.203 & \phn  5.50 &  8.62 & 2.1$\times 10^{-6}$ \\
A2199 & \phn  3.768 & \phn  7.53 &  2.20 & \phn  8.05 &  & \phn  3.768 & \phn  7.53 &  2.17 & \phn  5.89 \\
A2204 & \phn  5.809 & 11.75 &  2.09 & \phn  9.30 &  & \phn  5.809 & 11.75 &  2.05 & \phn  8.11 \\
A2256 & \phn  3.013 & \phn  7.38 &  2.28 & \phn  5.07 &  & \phn  3.013 & \phn  7.38 &  2.36 & \phn  2.98 \\
A2255 & \phn  1.321 & \phn  2.73 &  6.24 & 2.3$\times 10^{-5}$ &  & \phn  1.321 & \phn  2.73 &  9.41 & 2.5$\times 10^{-7}$ \\
A3667 & \phn  1.932 & 12.00 &  2.39 & \phn  4.01 &  & \phn  1.932 & 12.00 &  2.20 & \phn  2.79 \\
S1101 & \phn  1.009 & \phn  2.23 &  2.43 & \phn  1.65 &  & \phn  1.009 & \phn  2.23 &  2.08 & \phn  2.13 \\
A2589 & \phn  5.045 & \phn  5.82 &  2.00 & \phn  5.82 &  & \phn  5.045 & \phn  5.82 &  2.02 & \phn  6.42 \\
A2597 & \phn  2.620 & \phn  6.10 &  2.00 & \phn  6.09 &  & \phn  2.620 & \phn  6.10 &  2.41 & \phn  2.90 \\
A2634 & \phn  1.806 & \phn  4.34 &  7.28 & 3.3$\times 10^{-6}$ &  & \phn  1.806 & \phn  4.34 &  9.50 & 4.7$\times 10^{-7}$ \\
A2657 & \phn  0.613 & \phn  4.58 &  2.00 & \phn  4.58 &  & \phn  0.613 & \phn  4.58 &  7.28 & 2.6$\times 10^{-5}$ \\
A4038 & \phn  3.117 & \phn  7.43 &  2.00 & \phn  7.43 &  & \phn  3.117 & \phn  7.43 &  2.12 & \phn  4.73 \\
A4059 & \phn  1.981 & \phn  3.82 &  2.44 & \phn  2.44 &  & \phn  0.060 & \phn  3.82 &  9.85 & 6.1$\times 10^{-8}$

\enddata
\tablenotetext{a}{20--80 keV, 10$^{-12}$ erg cm$^{-2}$ s$^{-1}$}
\end{deluxetable*}
\subsection{Upper Limits}
\label{sec:bathi:joint:uls}

While some evidence for non-thermal emission is present in several
of the \his clusters, in none of these cases is a significant excess
indicated by both the BAT and EPIC spectra that could not plausibly
be explained by a multi-temperature state of the ICM.
In many cases, the BAT spectra simply lacked the signal-to-noise to
meaningfully constrain the existence of excess emission;
we therefore derive upper limits for a non-thermal component in our 
joint spectra.
Three limits are presented for each energy range (2--195 keV and 
3--195 keV) considered: a 90\% confidence level limit including
systematic uncertainties in $f_{\rm CN}$ and the EPIC backgrounds,
as described in Section~\ref{sec:bathi:obs:xmm}, and two $3\sigma$
limits, without systematic uncertainties included,
for our fiducial photon index of $\Gamma=2$ and for the
best-fit value of $\Gamma$.
After fitting for $\Gamma$, it is then fixed at that value when the 
upper limit is computed.
The systematic terms are included in the 90\% limits as described in 
\citet{WSF+09}.
Upper limits are reported as 20--80 keV fluxes in units of
$10^{-12}$ erg cm$^{-2}$ s$^{-1}$ in Table~\ref{tab:bathi:uls}.
Note that when $\Gamma$ is much steeper than 2, the power law
component is constrained only by the low energy spectrum and the
20--80 keV flux limits are not reflective of the sensitivity of the BAT survey.
In some instances, usually for lower temperature clusters, the 90\%
limit exceeds the $3\sigma$ limits; in these fits, the systematic uncertainties in 
$f_{\rm CN}$ and/or the EPIC background dominates over the
statistical uncertainty in the spectra.
For example, in a low temperature cluster lowering the EPIC backgrounds 
significantly hardens the spectra, while modifying $f_{\rm CN}$
such that already poorly constraining BAT fluxes are 10\% higher,
will allow a much larger IC-like component to fit the data than would
be allowed statistically.
In hotter clusters, adjusting the background has less of an effect
on their spectral shape, and because they are hot they tend to be
more significantly detected by the BAT, so that modifying $f_{\rm CN}$
cannot drastically affect the non-thermal component.


\section{Joint Fits to Stacked EPIC-BAT Spectra}
\label{sec:bathi:stack}

In some clusters, as noted above, hints of a non-thermal excess are
present, even if we cannot argue for their definite detection.
If the excess does exist in several clusters, but just below the
detection threshold, we may be able to increase the signal-to-noise
enough for a statistical detection by stacking the cluster spectra.
For simplicity, we stack only the EPIC-pn \xmms spectra, which have the
highest sensitivity especially at higher
energies.
Stacking the MOS spectra would be complicated by the variable pn/MOS
cross-calibration factor and the fact that 3 of the cluster MOS spectra
have been excluded from our analysis.
Both the pn and BAT spectra are straightforwardly summed, as are the
pn backgrounds, and their errors are propagated.
Because the same response matrix is used for all the BAT spectra,
we are able to use this unmodified file with the stacked spectrum.
To create an average response matrix for use with the stacked pn
spectrum, we first multiply the individual redistribution matrices
by their respective auxiliary response files, which contain the
effective area per incoming photon energy.
Then, a weighted average is performed on the new response files,
with weighting factors proportional to each spectrum's
2--7 keV count rate.
This procedure ensures that the final response matrix will best
represent the instrumental response for the majority of photons.
In any case, an unweighted response file was also created and no
significantly different results were produced when using it.
The CXB model normalizations were
summed and included in the spectral fits.

In all, we create 8 stacked spectra based on different
groupings of the 59 \his clusters for which we have \xmms data:
``All'' clusters, ``Hot'' ($kT > 7$ keV, from the 2--12 keV fits), 
``Cool'' ($kT < 7$ keV), 
``Radio'' clusters hosting either a large-scale radio halo and/or relic or a
smaller, central mini-halo, ``No Radio''
clusters that do not host any of these types of diffuse radio emission,
non-cool-core clusters (``NCC''),
strong cool core clusters (``SCC''),
and weak cool core clusters (``WCC''), as
defined by \citet{HMR+10} and listed in Table~\ref{tab:bathi:basic}.
These categories are designed to separate the sample into subgroups
which might have different average levels of non-thermal emission.
For example, IC emission must exist at some level in clusters with a 
radio halo or relic,
but may not be present in clusters more generally.
Thus, we might expect the
``Radio'' clusters to preferentially have non-thermal
excesses, which are enhanced when they are stacked together and not
diluted by the additional spectra from ``No Radio'' clusters that
have no such excess.

\begin{figure}
\plotone{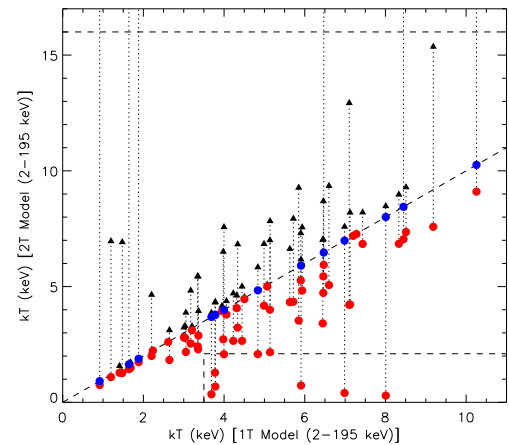}
\caption{The two temperature values in the 2T fits (2--195 keV)
with respect to the single temperature fit for each cluster over
the same range.
High (triangles, black) and low (circles, red) temperature values
in the 2T model for the same cluster are connected by dotted lines
for clarity.
Blue circles are 1T  fits to clusters with an unphysical best-fit 2T 
model, with either unrealistically high temperatures
($kT > 16$ keV) or a low temperature component that over-predicts
the emission below 2 keV.
(These have $kT \la 2.1$ keV for clusters with single temperatures of
3.5 keV or hotter.)
These excluded regions are indicated by dashed lines in the figure, and the
diagonal dashed line represents equality of 1T and 2T temperatures.
These temperature values are used to build the 2T$_{\rm J}$ model
used in fits to the stacked spectra.
\label{fig:bathi:tvs2t}}
\end{figure}

Because these clusters span a large range of temperatures and
redshifts, it is not appropriate to model the summed spectra
with a single or even several temperature model for the thermal
component.
Instead, we build multi-temperature models from the previous spectral
fits, for which we keep the spectral shape fixed and only allow the
overall normalization to vary during fits to the stacked spectra.
We consider the \xmm-only single temperature fits 
(Table~\ref{tab:bathi:xmmonly}) derived from 2--12 keV (1T$_{\rm X,>2}$)
and from 3--12 keV (1T$_{\rm X,>3}$), and the single (1T$_{\rm J}$)
and double (2T$_{\rm J}$) temperature fits derived from the 2--195 keV
joint spectra (Table~\ref{tab:bathi:joint1}).
To search for non-thermal emission in the stacked spectra, a power
law model is added to represent the IC component and the normalization
of the thermal model is allowed to vary.
Ideally, the shape of the thermal component would be able to adjust
to accommodate the IC signal, as it effectively does in the individual 
joint fits via the temperature parameter.
However, the non-thermal flux below 12 keV will
be small and should not cause the temperature to change in any
significant way.
For the 2T$_{\rm J}$ model, we want to avoid including unphysical
temperature components that may have been driven by calibration 
features at the edges of the spectral range in the individual 2T fits.
A low temperature ($\la 2$ keV) component's emission
measure may cause $<$ 2 keV emission to be significantly
overestimated in order to better fit the gold edge, for example.
Similarly, a slight under-subtraction of the \xmms background or positive
fluxes in the higher energy BAT bands may lead to unrealistically
high temperatures.
In Figure~\ref{fig:bathi:tvs2t}, we plot the temperature values
for this model relative to the 1T$_{\rm J}$ model temperatures.
We have removed unphysical temperature components from both the
2T$_{\rm J}$ model; the best-fit single temperature 
model is used in place of the 2T model for those clusters, which
are represented by blue circles in Figure~\ref{fig:bathi:tvs2t}.
Unphysical temperature components were found to have 
$kT > 16$ keV
and $kT < 2.1$ keV, if their 1T$_{\rm J}$ temperature is greater
than 3.5 keV.
In general, this latter cut eliminates temperature components 
that significantly over-predict the 0.5 keV $< E <$ 2 keV emission.

Thermal and thermal plus non-thermal fits to the stacked spectra
are given in Table~\ref{tab:bathi:stack}.
Considering only the fits to data with $E > 3$ keV, which excludes the
most problematic region of the spectra, we find no evidence
at the statistical 90\% level for
a non-thermal component in any of the stacked spectra
except in the case of ``Radio'' clusters.
In the table, the normalization of the thermal component in the 
``T$_{\rm Model}$-only'' fits is not shown,
only its $\chi^2$ value for comparison purposes.
For the ``T$_{\rm Model}+$IC'' fits, the photon index is fixed to
$\Gamma=2$ as was done previously for the joint fits.
The last 3 columns report the ``T$_{\rm Model}+$IC'' fits with $\Gamma$
as a free parameter; however, its value is fixed when errors are
computed.
In this case,
the photon index was initialized as $\Gamma=2$, so for spectra with no
particularly strong indication of non-thermal emission, the best-fit
normalization was set to zero and the photon index kept at or near
its initialized value; this explains why so many of the ``best-fit''
photon indices presented in the table are `2.00.'
In the case of large values of $\Gamma > 3$, the non-thermal component
is attempting to either represent incompletely modeled soft emission
from low temperature gas or correct an imperfectly calibrated gold
edge.
Even though these normalizations are large and quite significant, they
are so steep that the flux at hard energies is negligible and does not
represent an IC excess.
If $< 2$ keV emission were included in the fits, these large $\Gamma$
values would disappear as they would vastly over-predict the soft
emission.

\begin{figure}
\plotone{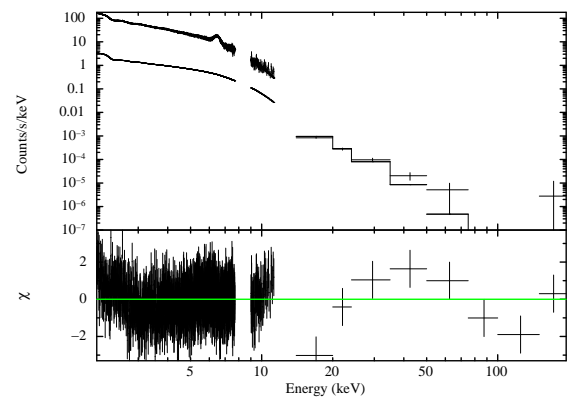}
\caption{The stacked spectrum of all 59 clusters with the combined
single temperature model fit (1T$_{\rm X,>2}$).
The EPIC-pn spectrum ($E < 12$ keV) and BAT spectrum ($E > 14$ keV)
are shown in the top panel, and their residuals in the lower panel.
The CXB contribution appears below the EPIC-pn data.
The problems between 2--3 keV (described in the text) clearly show up
in the residuals, as does a potential problem with low energy
BAT fluxes.
The combined single temperature model
determined from the 2--12 keV fits
is sufficient to explain the summed BAT spectrum; no non-thermal
excess is obvious.
\label{fig:bathi:stackall}}
\end{figure}
In Figure~\ref{fig:bathi:stackall}, the joint fit for the stacked spectra
of all 59 clusters is shown with the 1T$_{\rm X,>2}$ model.
The best-fit model normalization agrees with its expected value
to better
than 1\%, as do all the model fits without an IC component,
indicating that the average pn response is accurate.
Also, a difference in spectral shape appears below 3 keV, 
visible in the residuals, that highlights the problem with including
this emission in the fits.
The BAT data are well represented by this model, {\it even though
the temperature models were derived from fits to the \xmms spectra 
alone}.
The regular pattern in the BAT residuals is likely real, and is
apparent in most of the spectra of hot clusters such as Coma
\citep[see][]{WSF+11}.
When considering only one cluster, it seemed reasonable that this
residual pattern could simply be due to chance.
The pattern reappears in many of the individual joint
fits however, indicative of a systematic problem.
Because the BAT flux calibration is dominated by normalizing to the
Crab flux in each band, these fluxes are really only accurate for
objects with a spectral slope similar to the Crab's.
At these energies, cluster spectra are quite steep even for the
hottest temperatures, so some miscalibration would be expected.
Most likely, the first and possibly second energy bands have
underestimated fluxes, owing to the rapid rise of the instrumental
response with energy; clusters have proportionately more emission
at the lower energy part of the band than does the Crab, 
and so the internal band response is miscalibrated -- weighting the
higher energy part of the response more strongly than is appropriate
for thermal emission.
While this certainly affects our results, the only solution is to develop
a detailed response matrix model for the survey data.
Unfortunately, the detailed spectral response for the \swifts 
survey data currently has
much larger uncertainties than the Crab spectrum itself.

In general, the addition of a non-thermal component to these spectra
does not significantly improve the fits in Table~\ref{tab:bathi:stack}, 
{\it except} for the ``Radio'' -- and to a lesser degree the ``Hot" -- subsamples.
The ``All," ``Cool," ``No Radio,"  ``WCC," and ``SCC" stacks are found to lack a physically
plausible ($\Gamma \lesssim 3$) non-thermal component at the statistical-only
90\% level.
For the ``Hot" and ``NCC" sample fits, the IC component 
improves fits using the 1T$_{\rm X,>2}$ and 1T$_{\rm J}$ thermal models, but not
the 1T$_{\rm X,>3}$ and 2T$_{\rm J}$  models.
In contrast, the ``Radio" cluster sample
T$+$IC fits are not only clearly improved over the thermal-only fits, but the IC
component is significant at the 90\% level regardless of the thermal model considered.
Also, except in the case of the 2T$_{\rm J}$  model, allowing the IC index to vary shows
that the slope is consistent with expected indices (2-2.5).
The exact value of the index favored by the data should not be taken to represent the
true shape of the non-thermal component, however, since it is most strongly influenced by
the 2 keV $< E <$ 3 keV data and thus biased by gold edge calibration issues and
incompletely modeled cool gas.

\begin{figure}
\plotone{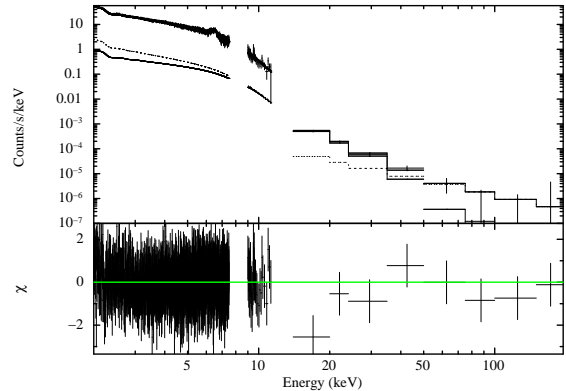}
\caption{The stacked spectrum of all clusters with large-scale,
diffuse radio halos or relics.
General features of the plot are the same as 
Figure~\ref{fig:bathi:stackall}.
The dotted line represents the best-fit non-thermal model with photon
index $\Gamma=2$, and the upper
solid line shows the 1T$_{\rm X,>2}$
thermal model.
\label{fig:bathi:stackradioic}}
\end{figure}

\begin{figure}
\plotone{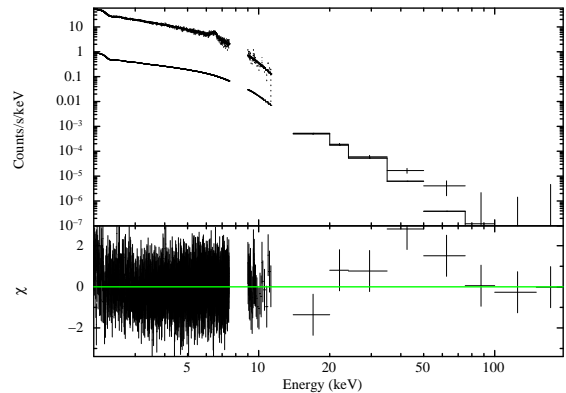}
\caption{The stacked spectrum of all clusters with large-scale,
diffuse radio halos or relics with the combined
single temperature model fit (1T$_{\rm X,>2}$).
General features of the plot are the same as 
Figure~\ref{fig:bathi:stackall}.
A slight excess is apparent in the BAT spectrum, due to either
a non-thermal spectral component (see Fig.~\ref{fig:bathi:stackradioic})
or a significant multi-temperature structure in the individual
clusters (see Fig.~\ref{fig:bathi:stackradio2t})
\label{fig:bathi:stackradio1t}}
\end{figure}

\begin{figure}
\plotone{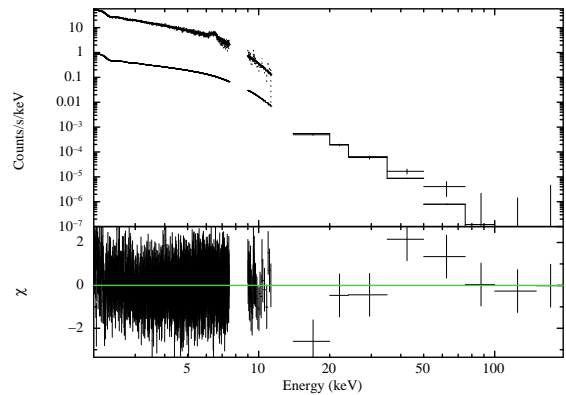}
\caption{The stacked spectrum of all clusters with large-scale,
diffuse radio halos or relics with the combined 
double temperature model fit (2T$_{\rm J}$).
General features of the plot are the same as 
Figure~\ref{fig:bathi:stackall}.
The combined 2T$_{\rm J}$ model can explain most of the 
slight excess seen when
the single temperature model (1T$_{\rm X,>2}$) is considered.
\label{fig:bathi:stackradio2t}}
\end{figure}

\begin{figure}
\plotone{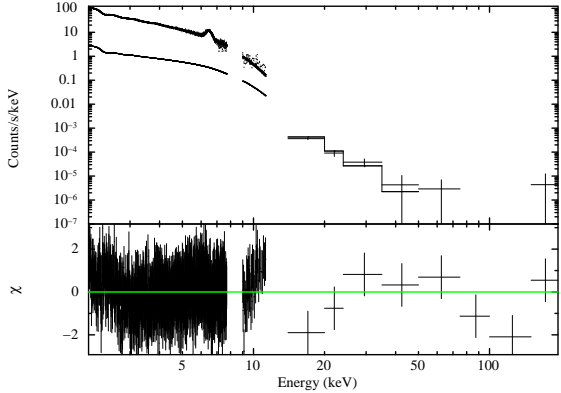}
\caption{The stacked spectrum of all clusters without diffuse
radio emission shown with the combined
single temperature model fit (1T$_{\rm X,>2}$).
General features of the plot are the same as 
Figure~\ref{fig:bathi:stackall}.
\label{fig:bathi:stacknotradio}}
\end{figure}

\begin{deluxetable*}{llcccccccccc}
\tablewidth{0pt}
\tablecaption{Fits to Stacked EPIC and BAT Spectra
\label{tab:bathi:stack}}
\tablehead{
 Sample & & T$_{\rm Model}$--only & \multicolumn{2}{c}{T$_{\rm Model}+$IC} & &
 \multicolumn{3}{c}{T$_{\rm Model}+$IC, $\Gamma_{\rm free}$} \\ 
 \cline{4-5}
 \cline{7-9}
(number) & T$_{\rm Model}$ & $\chi^2$/dof & Norm.\tablenotemark{a} & 
 $\chi^2$/dof && $\Gamma$ & Norm.\tablenotemark{a} & $\chi^2$/dof
}
\startdata
All & 1T$_{\rm X,>2}$ & 1218.34/1606 & $<$ 0.0165 & 1217.15/1605 &  &  6.94 & 1.0606$^{+0.1657}_{-0.1657}$ & 1107.44/1605 \\
(59) & 
1T$_{\rm X,>3}$ &  954.95/1406 & $<$ 0.0041 &  954.95/1405 &  &  2.00 & $<$ 0.0041 &  954.95/1405 \\
 & 1T$_{\rm J}$ & 1218.62/1606 & $<$ 0.0162 & 1217.56/1605 &  &  7.16 & 1.2145$^{+0.1926}_{-0.1926}$ & 1110.99/1605 \\
 & 2T$_{\rm J}$ & 1225.78/1606 & $<$ 0.0062 & 1225.78/1605 &  &  2.00 & $<$ 0.0062 & 1225.78/1605 \\
Hot & 1T$_{\rm X,>2}$ & 1037.57/1606 & 0.0093$^{+0.0049}_{-0.0049}$ & 1027.64/1605 &  &  2.21 & 0.0147$^{+0.0067}_{-0.0067}$ & 1024.50/1605 \\
(12) & 
1T$_{\rm X,>3}$ &  862.44/1406 & $<$ 0.0082 &  861.68/1405 &  &  0.83 & $<$ 9.3$\times 10^{-5}$ &  860.09/1405 \\
 & 1T$_{\rm J}$ & 1037.76/1606 & 0.0088$^{+0.0049}_{-0.0049}$ & 1028.96/1605 &  &  2.22 & 0.0140$^{+0.0067}_{-0.0067}$ & 1025.95/1605 \\
 & 2T$_{\rm J}$ & 1033.29/1606 & $<$ 0.0093 & 1031.20/1605 &  &  8.02 & 0.5347$^{+0.2086}_{-0.2086}$ & 1015.51/1605 \\
Cool & 1T$_{\rm X,>2}$ & 1101.49/1606 & $<$ 0.0069 & 1101.49/1605 &  &  2.00 & $<$ 0.0069 & 1101.49/1605 \\
(47) & 
1T$_{\rm X,>3}$ &  892.09/1406 & $<$ 0.0036 &  892.09/1405 &  &  2.00 & $<$ 0.0036 &  892.09/1405 \\
 & 1T$_{\rm J}$ & 1101.51/1606 & $<$ 0.0078 & 1101.51/1605 &  &  7.62 & 0.9748$^{+0.2172}_{-0.2172}$ & 1047.00/1605 \\
 & 2T$_{\rm J}$ & 1098.84/1606 & $<$ 0.0034 & 1098.84/1605 &  &  2.00 & $<$ 0.0034 & 1098.84/1605 \\
Radio & 1T$_{\rm X,>2}$ & 1007.22/1605 & 0.0137$^{+0.0055}_{-0.0055}$ &  990.54/1604 &  &  2.29 & 0.0265$^{+0.0080}_{-0.0080}$ &  977.31/1604 \\
(15) & 
1T$_{\rm X,>3}$ &  835.82/1405 & 0.0061$^{+0.0059}_{-0.0059}$ &  832.91/1404 &  &  2.01 & 0.0064$^{+0.0062}_{-0.0062}$ &  832.91/1404 \\
 & 1T$_{\rm J}$ & 1007.61/1605 & 0.0130$^{+0.0055}_{-0.0055}$ &  992.71/1604 &  &  2.32 & 0.0259$^{+0.0080}_{-0.0080}$ &  979.14/1604 \\
 & 2T$_{\rm J}$ & 1002.93/1605 & 0.0079$^{+0.0056}_{-0.0056}$ &  997.62/1604 &  &  6.80 & 0.3038$^{+0.0900}_{-0.0900}$ &  972.09/1604 \\
No Radio & 1T$_{\rm X,>2}$ & 1105.22/1606 & $<$ 0.0060 & 1105.22/1605 &  &  2.00 & $<$ 0.0060 & 1105.22/1605 \\
(44) & 
1T$_{\rm X,>3}$ &  895.38/1406 & $<$ 0.0029 &  895.38/1405 &  &  2.00 & $<$ 0.0029 &  895.38/1405 \\
 & 1T$_{\rm J}$ & 1104.03/1606 & $<$ 0.0067 & 1104.03/1605 &  &  7.47 & 0.8958$^{+0.1910}_{-0.1910}$ & 1044.52/1605 \\
 & 2T$_{\rm J}$ & 1105.19/1606 & $<$ 0.0033 & 1105.19/1605 &  &  2.00 & $<$ 0.0033 & 1105.19/1605 \\
NCC & 1T$_{\rm X,>2}$ &  894.14/1606 & 0.0054$^{+0.0055}_{-0.0049}$ &  890.82/1605 &  &  6.67 & 0.1830$^{+0.0770}_{-0.0770}$ &  878.85/1605 \\
(16) & 
1T$_{\rm X,>3}$ &  753.77/1406 & $<$ 0.0034 &  753.77/1405 &  &  2.00 & $<$ 0.0034 &  753.77/1405 \\
 & 1T$_{\rm J}$ &  892.13/1606 & 0.0052$^{+0.0055}_{-0.0049}$ &  889.10/1605 &  &  2.41 & 0.0122$^{+0.0076}_{-0.0076}$ &  885.20/1605 \\
 & 2T$_{\rm J}$ &  888.10/1606 & $<$ 0.0072 &  887.98/1605 &  &  7.19 & 0.2277$^{+0.1114}_{-0.1114}$ &  876.80/1605 \\
WCC & 1T$_{\rm X,>2}$ &  776.09/1606 & $<$ 0.0110 &  773.65/1605 &  &  4.18 & 0.0739$^{+0.0172}_{-0.0172}$ &  726.08/1605 \\
(17) & 
1T$_{\rm X,>3}$ &  619.71/1406 & $<$ 0.0057 &  619.71/1405 &  &  2.00 & $<$ 0.0057 &  619.71/1405 \\
 & 1T$_{\rm J}$ &  775.16/1606 & $<$ 0.0104 &  773.29/1605 &  &  4.30 & 0.0781$^{+0.0185}_{-0.0185}$ &  726.90/1605 \\
 & 2T$_{\rm J}$ &  776.97/1606 & $<$ 0.0085 &  776.33/1605 &  &  4.08 & 0.0681$^{+0.0162}_{-0.0162}$ &  729.05/1605 \\
SCC & 1T$_{\rm X,>2}$ & 1429.48/1606 & $<$ 0.0089 & 1428.53/1605 &  &  9.98 & 2.5281$^{+0.7309}_{-0.7309}$ & 1397.11/1605 \\
(26) & 
1T$_{\rm X,>3}$ & 1194.56/1406 & $<$ 0.0056 & 1194.56/1405 &  &  2.00 & $<$ 0.0056 & 1194.56/1405 \\
 & 1T$_{\rm J}$ & 1433.88/1606 & $<$ 0.0088 & 1432.82/1605 &  &  9.98 & 2.6175$^{+0.7326}_{-0.7326}$ & 1399.34/1605 \\
 & 2T$_{\rm J}$ & 1423.95/1606 & $<$ 0.0032 & 1423.95/1605 &  &  2.00 & $<$ 0.0032 & 1423.95/1605

\enddata
\tablenotetext{a}{At a photon energy of
1 keV in units of photons cm$^{-2}$ s$^{-1}$ keV$^{-1}$.}
\end{deluxetable*}
The best-fit non-thermal plus T$_{\rm X,>2}$ model for the Radio clusters is shown in
Figure~\ref{fig:bathi:stackradioic}.
For comparison, the T$_{\rm X,>2}$ and 2T$_{\rm J}$ fits with no IC
component are shown in Figure~\ref{fig:bathi:stackradio1t} and
Figure~\ref{fig:bathi:stackradio2t}, respectively.
The non-thermal component, plotted as a dotted line in the figure,
becomes competitive with the thermal emission in the 35--50 keV
band, where a somewhat significant excess is present in thermal-only 
model fits.
By contrast, the ``No Radio'' subsample shows no evidence
for an excess at hard energies (Fig.~\ref{fig:bathi:stacknotradio}).

Ignoring systematic uncertainties, the non-thermal signature is detected 
for the Radio clusters 
with $5.1\sigma$ confidence using the T$_{\rm X,>2}$ model and
$2.4\sigma$ with the T$_{\rm X,>3}$ model.
Including a conservative $f_{\rm CN}$ uncertainty of 
10\%, which assumes the average cross-calibration is incorrect by that much,
reduces the significances to $3.3\sigma$ and $0.7\sigma$, respectively.
While the detection is quite robust with the T$_{\rm X,>2}$ thermal model, 
both the normalization and significance of the detection degrades when 
using the T$_{\rm X,>3}$ model.
Since the temperatures making up the models are not very different
(see Fig.~\ref{fig:bathi:tvs2t}), the primary driver for this change must come from
removing the 2--3 keV data when fitting with the T$_{\rm X,>3}$ model.
The decreased significance should not necessarily be a concern
as the lowest energies statistically dominate $\chi^2$.
However, the factor of 2 drop in flux associated with the IC component
between the T$_{\rm X,>2}+$IC and  T$_{\rm X,>3}+$IC fits suggests that 
the power law is not driven by broadband IC emission but instead by 
features around 2--3 keV.
We cannot therefore reasonably claim an ensemble detection of excess hard 
emission at the $3\sigma$ level given the systematic and modeling uncertainties.


\section{Implications and Discussion}
\label{sec:bathi:disc}
In this work, we characterized the hard X-ray emission from \hi,
a sample of the brightest galaxy clusters outside the Galactic plane.
For the 59 out of 64 clusters with usable  \xmms data,
we searched for excesses over the thermal emission from gas in the
ICM in data from the 58-month \swifts BAT all-sky survey.
EPIC and BAT spectra were extracted from identical regions and 
carefully calibrated to allow straightforward joint fits that
simultaneously constrain the thermal and non-thermal emission in
both spectra.
We first considered fitting over an energy range of 2--195 keV but
found that low temperature gas and the gold edge in the \xmms spectra
could lead to false detections.
Ignoring the 2--3 keV data resolved this issue, although a somewhat weaker
constraint on the thermal component reduced our overall sensitivity.
From the 3--195 keV fits, six clusters were found to have marginal
evidence for a non-thermal excess, although none of these were deemed
significant enough to claim a detection, especially considering
systematic uncertainties in the EPIC background and
EPIC-BAT cross calibration normalizations.
We then stacked the spectra to look for a
significant statistical detection of non-thermal emission in the \his sample.
Unfortunately, the stacked spectra revealed no definitive excess.
Stacking subsamples of the \his clusters returned similar results,
except for a tantalizing but very marginal detection of a non-thermal component
in the stacked spectrum of all clusters that host radio halos and/or
relics or mini-halos --- the very clusters that are most expected to have detectable
IC emission.

\subsection{Comparison to Previous Studies}
\label{sec:bathi:disc:prevwork}

The lack of definitive hard X-ray excesses in our individual clusters is
consistent with the most recent searches with \suzaku, \integral, and \swift,
though somewhat less so with those of \rxtes and \sax.
Ignoring the Coma cluster, whose controversial hard energy emission is
discussed at length elsewhere \citep[e.g.,][]{WSF+11}, our analysis is not
clearly inconsistent with any previous observations, particularly given that
the possible existence of low-level, extended non-thermal emission has
not been considered in detail here \citep[as in][]{WSF+11}, 
which \rxtes and \saxs in particular would be sensitive to given their large FOVs.
For the clusters in our sample also observed by \rxte, A3667 \citep{RG04}
and A2256 \citep{RG03},
our upper limits agree with analyses of their data, at least considering the
two-temperature interpretation allowed for A2256, regardless of the
distribution of emission.
The recent \rxtes detection of non-thermal emission in NGC 5044 below 15 keV
by \citet{Hen11} lies below our detection threshold at higher energies.
For several of the clusters observed with \saxs and found to host non-thermal 
emission, such as A2256 \citep{FLO05}, A2199 \citep{KLM+99}, 
and A3526 \citep{MDG02},
our upper limits fall below their measured inverse Compton fluxes.
\citet{KLM+99} claim an extended non-thermal halo for A2199 between 0.5
and 1.5 Mpc, which is not inconsistent with its larger size at high energies 
(14--20 keV, see Fig.~\ref{fig:bathi:psf}); however, due to the low S/N of the
detection, this extent is also indistinguishable from that of a point source.
Upper limits from \suzakus for clusters A3667 \citep{Nak+09} and
A3376 \citep{Kaw+09} are obviously consistent with these results.

Similar studies of clusters detected by the BAT \citep{Aje+09, Aje+10}
have also failed to find definitive non-thermal excesses.
The only discrepancy is for A3667, for which both \citet{Aje+10} and
\citet{Nak+09} detect high temperature ($kT \sim 15$ keV) gas near the
center.
While we do not see strong evidence for a significant high temperature
component like this -- although our 2T, 2--195 keV fit does suggest a 
significant amount of hot gas ($kT \sim 9$ keV) --
the elongated shape caused by its ongoing merger
requires a more detailed analysis to more accurately extract its BAT fluxes
to properly assess this high temperature component.
In any case, a noteworthy difference between the methodology here and
in \citet{WSF+11} with that of \citet{Aje+09, Aje+10} is our use of the
technique developed by \citet{RGL+06} to recover extended source fluxes from
coded mask observations.
This procedure allows for a more direct spatial comparison between soft and
hard X-ray spectra such that no assumptions about the extent of hard band data
need to be made; however, the low relative extent and signal-to-noise generally
achieved makes this advantage critical only for the largest, brightest
clusters such as Perseus and Coma.

While some excesses in the stacked spectra are tantalizing,
equally good, and sometimes better, fits result when the 2T$_{\rm J}$
model is used.
Since only the normalization is allowed to vary in these fits, it
is hard to justify why the addition of an IC component really provides
a better description of the data, especially if the improvement in 
$\chi^2$ is minor.
Note that this comparison is only fair because the 2T models are all
physically reasonable descriptions of the ICM, otherwise we may be
inappropriately modeling non-thermal emission with an incorrect
thermal component.
The upper limits on non-thermal emission in the stacked spectra, when
applied on average to the clusters making up the stacked sample,
are more constraining than limits from individual fits.
The typical 90\% confidence level upper limit on the cumulative 
IC flux in the stacked spectra
is $10^{-11}$ erg cm$^{-2}$ s$^{-1}$ in the 20--80 keV
band, which translates to an average maximum flux per cluster 
4 to 8 times lower than this limit.

These results are in conflict with an analysis of a similar
sample of clusters observed by \saxs \citep{NOB+04}, which found 
systematic if marginal excesses for merging clusters.
Actually, these previous  IC flux estimates are not unlike
our results
in the 2--195 keV range, as are the temperatures of the thermal
component for clusters in both our and their samples.
However, over the 3--195 keV energy range, the 90\% error interval for
nearly all the excesses include zero.
This result is at least partly due to slightly higher best-fit
temperatures (see Section~\ref{sec:bathi:separate:xmm} for a more 
detailed discussion).
Since clusters are not isothermal, 
harder spectra such as those from the BAT will contain
proportionately more photons from higher temperature gas.
An example of this bias can be seen in the stacked
spectrum of \citet{NOB+04}; they observe a highly significant 
non-thermal excess, but the
steep IC component necessary to explain it would lead to detectable amounts
of non-thermal emission at softer energies, which is not seen.
The authors interpret this as evidence that the non-thermal emission is
significantly extended.
Our BAT data test this possibility, as extended emission is both 
detectable and not
detected by the BAT beyond that produced by the thermal gas.
Thus, it is unlikely that the non-thermal emission is very highly 
extended and strong.
In fact, the steep excess in their spectrum is exactly 
what would be expected for a strongly 
multi-temperature thermal structure -- which naturally results when
many clusters spanning a broad range in temperature are summed --
that is modeled as a single temperature component, which is what they do.
When we model our stacked BAT spectrum this way, we find a
temperature consistent with the average temperature of our clusters
and a very significant, steep ($\Gamma\sim2.8$) power law component,
identical to their best-fit photon index.
But the thermal component, {\it determined at hard energies}, will
be more highly weighted by hotter clusters, whose emission dominates.
If a single temperature component is used to model the thermal
emission for such a summed spectrum, then at the very least
the temperature needs to
be fixed to the weighted-average value in the band in which the
hard excess emission is expected to be found.
For example, in our sample, the count rate 
weighted-average temperature jumps
from 5.6 keV, when weighted by the 2--7 keV count rate, to
7.1 keV when the 14--50 keV count rate is used.
Here we have employed the temperatures determined from the 2--12 keV
fits.
Even so, the exact value of the temperature is less important than the
fact that the highly multi-temperature composite spectrum no longer
looks like a single temperature plasma.
The proper procedure is to use a truly
multi-temperature model based on the temperatures of the constituent
clusters, as we have done.
We suspect that, if the thermal component is similarly modeled for
the stacked spectrum of \citet{NOB+04}, the 
non-thermal excess will be reduced; however, it is
unlikely that all of their excess would disappear.

\subsection{Implied Magnetic Field Strengths}
\label{sec:bathi:disc:bfield}

Our most suggestive result from the various stacked subsamples,
that clusters hosting a radio halo or relic have the the most
significant indication of a non-thermal excess on average, is
also the least surprising of possible outcomes.
Because radio halos and relics are associated with mergers, which also
produce shocks and multi-temperature gas distributions, the more
appropriate thermal model to use might be the 2T$_{\rm J}$ model,
although even in this case a non-thermal component improves the fit.
Assuming the power law component represents true IC emission, we can determine
the average $B$ field in these clusters from the IC and synchrotron flux ratio as
outlined in \citet{WSF+09}.
Unfortunately, not all of the diffuse radio emission of these clusters falls within
the FOVs of the observations, so the measured IC normalization in 
Table~\ref{tab:bathi:stack} is not the best value we can use.
We therefore sum a new subsample of cluster emission, excluding those with radio
emission outside their XMM extraction regions, most notably the relics in
A3667, A1367, and A3376.
For the radio halo/relic clusters, we are left with radio emission from
Coma (4.4 Jy at 74 MHz \citep{HE80}, vignetted for the extraction region),
A754 (4 Jy at 74 MHz \citep{KCE+01}),
A3562 (220 mJy at 240 MHz \citep{Gia+05}),
A2256 (100 Jy at 22 MHz \citep{CBF72}, halo and relic emission combined),
A2255 (475 mJy at 150 MHz \citep{PB09}, halo emission only),
and A0399 (16 mJy at 1400 MHz \citep{MGF+10}).
Extrapolating these flux densities to a common 74 MHz given their individual
spectral indices, we find a total flux of 32 Jy.
Low frequency flux densities have been used where available to mitigate
the effect of spectral curvature on the IC/synchrotron flux ratios.
For the mini-halos, low frequency measurements are less common:
A2142 (18 mJy at 1400 MHz \citep{GF00}),
R1504 (121 mJy at 327 MHz \citep{GMB+11}),
A2029 (18.8 mJy at 1400 MHz \citep{MGM+09}),
2A0335 (22.6 mJy at 1500 MHz \citep{SBO95}).
The two remaining mini-halos, A2204 \citep{SFT09} and A2052 (Clarke, priv. comm.),
do not have published diffuse fluxes.
Assuming spectral indices of $\alpha=1.3$ ($F_\nu \propto \nu^{-\alpha}$) and
an average flux density of 18 mJy at 1400 MHz for all 6 clusters, a total flux of
$\sim 5$ Jy at 74 MHz is found.
Combining all these flux densities and using the power law normalization range 
in Table~\ref{tab:bathi:stack} 
(20--80 keV fluxes of $(1.3$--$3) \times 10^{-11}$ erg s$^{-1}$ cm$^{-2}$)
yields an average $B \sim 0.13$--$0.19 \mu$G.

Because mini-halos are confined to the cool cores of more relaxed clusters, completely
unlike large-scale halos and relics, it may be more appropriate to consider the two
classes separately.
Surprisingly, the fit to the new radio halo/relic stacked spectrum, 
made up of the clusters listed above, completely disfavors the addition of a non-thermal
component.
Taking the 90\% statistical-only upper limit to the IC flux of 
$5.7 \times 10^{-12}$ erg s$^{-1}$ cm$^{-2}$, we estimate $B > 0.26 \mu$G on average
in the central regions of these clusters.
In contrast, the best-fit IC flux to the stacked mini-halo spectrum is
$1.9 \times 10^{-11}$ erg s$^{-1}$ cm$^{-2}$, significant at nearly the $4\sigma$ level
(statistical-only), implying a magnetic field strength of $0.08 \mu$G on average.
For both cases the 1T$_{\rm J}$ thermal model is used over the 2--195 keV band.
Individually, the best-fit IC fluxes of A3562, A0399, and A2255 are quite low (implying
$B > 10 \mu$G) with 90\% (stat.$+$sys.) lower limits of 0.06 $\mu$G, 0.04 $\mu$G,
and 0.09 $\mu$G, respectively.
The remaining halo/relic clusters are more accommodating of a non-thermal component, with
best-fit/lower limit $B$ field strengths of 
0.21/0.11 $\mu$G (A0754), 0.54/0.21 $\mu$G (Coma), and 0.46/0.33 $\mu$G (A2256).
To compute these and the following $B$ field strengths, we use the average of the best-fit
normalizations in Tables~\ref{tab:bathi:joint1} and \ref{tab:bathi:joint2} and the larger
upper limit from Table~\ref{tab:bathi:uls}.
The mini-halo spectra typically favor the additional power law component, leading
to nominal best-fit IC fluxes not much lower than their upper limits and corresponding
 $B$ field strengths of 0.16/0.12 $\mu$G (2A0355), 0.06/0.04 $\mu$G (A2142),
 0.08/0.06 $\mu$G (A2204), 0.06/0.05 $\mu$G (A2029), 0.08/0.07 $\mu$G (A2052),
 and 0.10/0.05 $\mu$G (R1504), assuming a spectral index of 1.3 and radio
 flux densities of $\sim 18$ mJy at 1.4 GHz for A2052 and A2204.

These results suggest that mini-halos may involve lower $B$ fields, more easily
allowing the detection of IC emission as compared with larger halos and relics.
This general conclusion is consistent with the measurement of non-thermal emission
with {\it INTEGRAL} associated with the mini-halo in the core of the Ophiuchus
cluster that implies $B = 0.05$--$0.1 \mu$G \citep{NEK+09}.
While it is tempting to associate the IC flux with the entire mini-halo sample,
the vast majority of the BAT flux originates with A2029 and A2142 (three-quarters of
the 14-24 keV emission and $>90$\% of it at the higher energies).
These two cases are discussed individually in Section~\ref{sec:bathi:joint:individual}.
A2142 is particularly problematic, in that the \xmms extraction region only contains half
the cluster, which adds significant systematic uncertainty to the cross-calibration with the 
BAT spectrum, and in that the classification of the diffuse radio emission as a mini-halo,
as opposed to radio galaxy jets, is not ironclad.
We are therefore precluded from drawing strong conclusions from this result, which is
robustly driven by only a single cluster: A2029.
In any case, the non-thermal component in the mini-halo spectrum is consistent with and
somewhat driven by the BAT measurements, suggesting that mini-halos may be better
targets of future hard X-ray observations, especially given their smaller size, than
traditional halos and relics.



It may not be surprising that IC emission was not detected definitively
in the radio halo/relic clusters; direct measurements of cluster magnetic fields
through Faraday rotation measure (RM) studies typically find
line-of-sight $B$ fields on the order of several $\mu$G
\citep{GF04}.
Similar high values of $B$ are suggested by
the stability of cold fronts in merging clusters \citep{KMB+10}, although the flow may
locally amplify the fields in these regions, so they would not be representative of the 
global average field strength, even in mini-halos.
Also, RM magnetic field strengths could be biased high if stronger
fields are correlated with denser gas, since RM observations are
really measuring the electron density-weighted value of $B$ along
the line of sight \citep{Pet01}.
Such explanations, while entirely reasonable, were primarily developed
to explain the lower values of $B$ implied by earlier IC detections,
some of which have been more recently called into question
\citep[e.g., with \suzaku,][]{Nak+07, Nak+09, WSF+09}.
However, if our low significance evidence for IC emission associated with mini-halos
can be corroborated, then such low values of $B$ may in fact be common.
Our current sensitivity to IC emission with either pointed 
or survey observations can only detect non-thermal emission in 
clusters with radio halos if the magnetic fields are $\la$0.2 $\mu$G.
Note that it is possible to observe much fainter IC emission at
lower X-ray energies, and thus measure larger $B$ fields, in radio relics
that are significantly displaced from the X-ray luminous gas in cluster
centers \citep{FSN+10}.

\subsection{Dynamical Importance of the Non-thermal Electron Population}
\label{sec:bathi:disc:energetics}

Given that we have searched for and provided limits on IC emission from
relativistic electron populations in a complete sample of clusters, what does
that tell us about the general energetic importance of this population
relative to the thermal electron population of clusters in the low redshift universe?
Hydrostatic cluster mass estimates -- employed by studies using clusters as
cosmological probes -- typically assume a negligible amount of 
non-thermal pressure in central regions.
In the following, we attempt to confirm this presumption.
A power law IC spectrum is thought to originate from an energy distribution
of relativistic electrons proportional to $\gamma^{-p}$, where $\gamma$ is
the electron Lorentz factor and $p=2\alpha+1=2\Gamma-1$.
The total energy in these electrons depends on the range of $\gamma$s and
is particularly sensitive to the lower cut off energy where most of the electrons reside.
Following \citet{Mur+10}, a rare case in which rough observational estimates exist,
we take $\gamma_{\rm min} \sim 300$ and 
$\gamma_{\rm max} \sim 3 \times 10^4$.
The large uncertainty in these choices, which are impossible for us to estimate since 
2--200 keV energies correspond to $1500 \lesssim \gamma \lesssim 1.5 \times 10^4$,
limits the following discussion to order of magnitude estimates at best.

Our goal is to compare the typical non-thermal pressure, $P_{\rm NT} = u_{\rm NT}/3$,
where $ u_{\rm NT}$ is the energy density of the non-thermal population,
to the typical thermal pressure inside our extraction regions.
For simplicity, we adopt the universal pressure profile of \citet{Arn+10} from which
the thermal pressure inside the region (taken to be $\sim \frac{2}{3}$ of the extraction
radius) can be derived from the mass estimates at $R_{500}$ of \citet{ZAC+11}.
In individual clusters, the 90\% upper limits on the IC flux typically limit the
non-thermal electron to thermal electron pressure to less than a few tens of percent.
The limits are stricter for massive (hot) clusters ($\lesssim 10\%$), primarily
because the IC limits are all roughly comparable as a result of the uniformity of the
BAT sensitivity.
If we instead take the best-fit IC normalizations, the non-thermal pressure of nearly
all of the clusters is less than 10\% of thermal pressure with massive clusters
typically at less than a few percent.
Except for the least massive clusters in \hi, the comparable magnetic field pressure 
assuming $B = 1 \mu$G corresponds to $\sim 1\%$ of the thermal pressure.
Although we cannot measure $B$ fields this large in the radio clusters, as long
as $B \lesssim 3$--$7 \mu$G, it seems unlikely that non-thermal pressure support
will significantly contribute in cluster centers, especially for the more massive clusters
that are typically used for cosmological parameter estimation.

\subsection{Future Outlook}
\label{sec:bathi:disc:future}

Can the survey observations with the BAT be improved, beyond the 
increase in sensitivity which comes
with longer accumulating exposures?
Perhaps the clearest way forward is to better calibrate the spectral 
response of the BAT in narrower channel
so that the fluxes are more reliable for
steep thermal emission in the 14--24 keV energy range.
At present, we may be underestimating source fluxes in these bands.
If the first band is low by $\sim 2\sigma$ and the second by 
$\sim 1\sigma$, as suggested by the residuals in 
Figure~\ref{fig:bathi:stackall},
our non-thermal limits will increase by
about $1\sigma$ -- a small but non-negligible amount.
The most straightforward fix is to remake the survey using the BAT's
native 80 channels instead of binning them into 8 channels 
that are broad enough to be biased by the flux calibration with the Crab.
With such improved data, this study can be repeated with a sample
of all the known radio halo and relic clusters to definitively
detect the non-thermal excess hinted at in the stacked ``Radio''
subsample considered here, if it exists.

Ultimately, any IC detections, especially if marginal,
will have to be confirmed by the
upcoming missions with focussing hard X-ray telescopes,
namely {\it NuSTAR}\footnote{http://www.nustar.caltech.edu/} and
{\it Astro-H}\footnote{http://astro-h.isas.jaxa.jp/}.
By resolving both contaminating point sources and the location of
the hottest gas, these missions have the potential to achieve
higher sensitivities than have thus far been possible.
Also, these telescopes' narrow FOVs are well-suited to the smaller angular
extents of mini-halos, making them ideal targets even though the 
radio flux densities are typically lower.

\acknowledgments
We particularly owe the \swifts BAT team a hearty thanks for uniformly
processing the tens of thousands of individual pointings that make up
the BAT survey that allow it to be such an incredibly useful resource for 
studies like ours.
DRW and CLS were supported in part by NASA through Suzaku grants 
NNX08AZ99G, NNX09AH25G, and NNX09AH74G, and XMM-Newton grants 
NNX08AZ34G and NNX08AW83G.
YYZ acknowledges support from the German
BMBF through the Verbundforschung under grant No. 50 OR 1005.
The XMM-Newton project is an ESA Science Mission with instruments and
contributions directly funded by ESA Member States and the USA (NASA),
and it is supported by the Bundesministerium f\"ur Wirtschaft
und Technologie/Deutsches Zentrum f\"ur Luft- und Raumfahrt 
(BMWI/DLR, FKZ 50 OX 0001) and the Max-Planck Society. 
Basic research in radio astronomy at
the Naval Research Laboratory is supported by 6.1 Base funding.
This research was also supported in part by an appointment to the NASA 
Postdoctoral Program at the Goddard Space Flight Center, administered 
by Oak Ridge Associated Universities through a contract with NASA.


\LongTables

\begin{deluxetable}{lcccccccc}
\tablewidth{0pt}
\tablecaption{Joint Thermal and non-thermal Fits to the EPIC and 
BAT Spectra (2--195 keV)
\label{tab:bathi:joint1}}
\tablehead{
 & & \multicolumn{3}{c}{Component 1} && \multicolumn{2}{c}{Component 2} & \\
 \cline{3-5}
 \cline{7-8}
 & & $kT$ & abund & Norm.\tablenotemark{b} && $kT$ & 
Norm.\tablenotemark{c} & \\
Name & Model\tablenotemark{a} & (keV) & Z$_\odot$ & (cm$^{-5}$) & & (keV) &  & 
$\chi^2$/dof
}
\startdata
A0085 & 1T &  6.46$^{+ 0.19}_{- 0.18}$ & 
0.358$^{+0.031}_{-0.030}$ & 
0.0775$^{+0.0010}_{-0.0010}$ &  & 
 &  &  654.76/812 \\
 & 2T &      5.44 & 
    0.368 & 
   0.0497 &  & 
     8.69 &    0.0283 &    653.27/810 \\
 & T$+$IC &      6.46$^{+     0.19}_{-     0.19}$ & 
    0.365$^{+    0.036}_{-    0.035}$ & 
   0.0760$^{+   0.0025}_{-   0.0042}$ &  & 
 & 
$<$    0.0019 &    654.35/811 \\
A0119 & 1T &  5.72$^{+ 0.46}_{- 0.45}$ & 
0.223$^{+0.069}_{-0.068}$ & 
0.0316$^{+0.0011}_{-0.0010}$ &  & 
 &  &  231.53/279 \\
 & 2T &      4.34 & 
    0.240 & 
   0.0183 &  & 
     7.94 &    0.0138 &    229.46/277 \\
 & T$+$IC &      5.66$^{+     0.49}_{-     0.48}$ & 
    0.243$^{+    0.078}_{-    0.076}$ & 
   0.0295$^{+   0.0029}_{-   0.0040}$ &  & 
 & 
$<$    0.0019 &    230.33/278 \\
A0133 & 1T &  3.78$^{+ 0.14}_{- 0.13}$ & 
0.446$^{+0.050}_{-0.048}$ & 
0.0236$^{+0.0006}_{-0.0006}$ &  & 
 &  &  302.63/422 \\
 & 2T &      0.68 & 
    0.452 & 
   0.0133 &  & 
     4.34 &    0.0206 &    285.74/420 \\
 & T$+$IC &      3.36$^{+     0.40}_{-     0.20}$ & 
    0.640$^{+    0.147}_{-    0.178}$ & 
   0.0175$^{+   0.0047}_{-   0.0024}$ &  & 
 & 
   0.0018$^{+   0.0007}_{-   0.0014}$ &    297.02/421 \\
NGC507 & 1T &  1.48$^{+ 0.08}_{- 0.08}$ & 
0.827$^{+0.248}_{-0.193}$ & 
0.0100$^{+0.0015}_{-0.0014}$ &  & 
 &  &  144.17/192 \\
 & 2T &      1.27 & 
    0.876 & 
   0.0095 &  & 
     6.92 &    0.0008 &    136.45/190 \\
 & T$+$IC &      1.28$^{+     0.14}_{-     0.12}$ & 
    0.990$^{+    0.474}_{-    0.275}$ & 
   0.0084$^{+   0.0020}_{-   0.0021}$ &  & 
 & 
   0.0004$^{+   0.0002}_{-   0.0002}$ &    137.27/191 \\
A0262 & 1T &  2.20$^{+ 0.04}_{- 0.04}$ & 
0.485$^{+0.046}_{-0.044}$ & 
0.0545$^{+0.0014}_{-0.0014}$ &  & 
 &  &  591.81/677 \\
 & 2T &      2.00 & 
    0.482 & 
   0.0508 &  & 
     4.65 &    0.0052 &    588.00/675 \\
 & T$+$IC &      2.12$^{+     0.09}_{-     0.06}$ & 
    0.520$^{+    0.057}_{-    0.052}$ & 
   0.0519$^{+   0.0025}_{-   0.0025}$ &  & 
 & 
   0.0008$^{+   0.0006}_{-   0.0007}$ &    588.14/676 \\
A0400 & 1T &  2.24$^{+ 0.12}_{- 0.11}$ & 
0.368$^{+0.108}_{-0.098}$ & 
0.0207$^{+0.0014}_{-0.0013}$ &  & 
 &  &  265.48/297 \\
 & 2T &      2.23 & 
    0.366 & 
   0.0095 &  & 
     2.24 &    0.0112 &    265.30/295 \\
 & T$+$IC &      2.23$^{+     0.13}_{-     0.13}$ & 
    0.359$^{+    0.115}_{-    0.091}$ & 
   0.0208$^{+   0.0013}_{-   0.0020}$ &  & 
 & 
$<$    0.0005 &    265.30/296 \\
A0399 & 1T &  7.28$^{+ 0.47}_{- 0.45}$ & 
0.224$^{+0.052}_{-0.052}$ & 
0.0357$^{+0.0007}_{-0.0007}$ &  & 
 &  &  276.34/377 \\
 & 2T &      7.30 & 
    0.224 & 
   0.0108 &  & 
     7.27 &    0.0249 &    276.34/375 \\
 & T$+$IC &      7.29$^{+     0.46}_{-     0.45}$ & 
    0.223$^{+    0.053}_{-    0.051}$ & 
   0.0357$^{+   0.0007}_{-   0.0027}$ &  & 
 & 
$<$    0.0009 &    276.34/376 \\
A3112 & 1T &  4.84$^{+ 0.13}_{- 0.13}$ & 
0.447$^{+0.030}_{-0.029}$ & 
0.0365$^{+0.0006}_{-0.0005}$ &  & 
 &  &  655.48/724 \\
 & 2T &      2.08 & 
    0.495 & 
   0.0097 &  & 
     5.84 &    0.0284 &    635.35/722 \\
 & T$+$IC &      4.64$^{+     0.19}_{-     0.17}$ & 
    0.567$^{+    0.066}_{-    0.076}$ & 
   0.0284$^{+   0.0037}_{-   0.0026}$ &  & 
 & 
   0.0025$^{+   0.0008}_{-   0.0011}$ &    638.12/723 \\
Fornax & 1T &  1.64$^{+ 0.03}_{- 0.03}$ & 
0.748$^{+0.071}_{-0.066}$ & 
0.0188$^{+0.0009}_{-0.0009}$ &  & 
 &  &  847.93/810 \\
 & 2T &      1.43 & 
    0.796 & 
   0.0175 &  & 
    20.57 &    0.0016 &    777.35/808 \\
 & T$+$IC &      1.41$^{+     0.05}_{-     0.05}$ & 
    1.038$^{+    0.163}_{-    0.133}$ & 
   0.0134$^{+   0.0014}_{-   0.0014}$ &  & 
 & 
   0.0011$^{+   0.0002}_{-   0.0002}$ &    759.76/809 \\
2A0335 & 1T &  3.03$^{+ 0.06}_{- 0.06}$ & 
0.424$^{+0.034}_{-0.033}$ & 
0.0999$^{+0.0017}_{-0.0017}$ &  & 
 &  &  506.39/658 \\
 & 2T &      2.78 & 
    0.424 & 
   0.0524 &  & 
     3.31 &    0.0475 &    506.39/656 \\
 & T$+$IC &      2.99$^{+     0.09}_{-     0.09}$ & 
    0.438$^{+    0.044}_{-    0.042}$ & 
   0.0976$^{+   0.0037}_{-   0.0043}$ &  & 
 & 
$<$    0.0019 &    505.54/657 \\
IIIZw54 & 1T &  2.62$^{+ 0.10}_{- 0.09}$ & 
0.299$^{+0.062}_{-0.059}$ & 
0.0198$^{+0.0008}_{-0.0007}$ &  & 
 &  &  303.72/413 \\
 & 2T &      2.61 & 
    0.299 & 
   0.0103 &  & 
     2.63 &    0.0095 &    303.73/411 \\
 & T$+$IC &      2.20$^{+     0.31}_{-     0.17}$ & 
    0.407$^{+    0.092}_{-    0.114}$ & 
   0.0160$^{+   0.0028}_{-   0.0014}$ &  & 
 & 
   0.0012$^{+   0.0004}_{-   0.0008}$ &    298.06/412 \\
A3158 & 1T &  5.91$^{+ 0.35}_{- 0.33}$ & 
0.333$^{+0.057}_{-0.056}$ & 
0.0407$^{+0.0010}_{-0.0010}$ &  & 
 &  &  268.83/360 \\
 & 2T &      0.73 & 
    0.346 & 
   0.0066 &  & 
     6.17 &    0.0393 &    267.43/358 \\
 & T$+$IC &      5.85$^{+     0.35}_{-     0.35}$ & 
    0.343$^{+    0.065}_{-    0.062}$ & 
   0.0395$^{+   0.0020}_{-   0.0034}$ &  & 
 & 
$<$    0.0014 &    270.38/359 \\
NGC1550 & 1T &  1.41$^{+ 0.05}_{- 0.04}$ & 
0.519$^{+0.090}_{-0.079}$ & 
0.0286$^{+0.0023}_{-0.0022}$ &  & 
 &  &  199.27/272 \\
 & 2T &      1.28 & 
    0.532 & 
   0.0172 &  & 
     1.57 &    0.0116 &    198.50/270 \\
 & T$+$IC &      1.31$^{+     0.08}_{-     0.08}$ & 
    0.552$^{+    0.107}_{-    0.089}$ & 
   0.0274$^{+   0.0029}_{-   0.0027}$ &  & 
 & 
   0.0005$^{+   0.0003}_{-   0.0003}$ &    192.68/271 \\
EXO0422 & 1T &  3.05$^{+ 0.07}_{- 0.07}$ & 
0.359$^{+0.033}_{-0.032}$ & 
0.0303$^{+0.0005}_{-0.0005}$ &  & 
 &  &  604.98/753 \\
 & 2T &      2.18 & 
    0.363 & 
   0.0148 &  & 
     3.88 &    0.0164 &    600.49/751 \\
 & T$+$IC &      2.78$^{+     0.20}_{-     0.14}$ & 
    0.457$^{+    0.041}_{-    0.062}$ & 
   0.0258$^{+   0.0031}_{-   0.0021}$ &  & 
 & 
   0.0013$^{+   0.0006}_{-   0.0009}$ &    598.20/752 \\
A3266 & 1T &  8.33$^{+ 0.26}_{- 0.25}$ & 
0.193$^{+0.030}_{-0.030}$ & 
0.0797$^{+0.0009}_{-0.0009}$ &  & 
 &  &  836.80/1060 \\
 & 2T &      6.85 & 
    0.194 & 
   0.0203 &  & 
     8.98 &    0.0595 &    835.89/1058 \\
 & T$+$IC &      8.35$^{+     0.27}_{-     0.26}$ & 
    0.206$^{+    0.032}_{-    0.031}$ & 
   0.0761$^{+   0.0018}_{-   0.0035}$ &  & 
 & 
   0.0012$^{+   0.0011}_{-   0.0012}$ &    833.17/1059 \\
A0496 & 1T &  4.34$^{+ 0.08}_{- 0.10}$ & 
0.397$^{+0.023}_{-0.021}$ & 
0.0835$^{+0.0011}_{-0.0008}$ &  & 
 &  & 1011.89/1092 \\
 & 2T &      3.23 & 
    0.451 & 
   0.0556 &  & 
     6.83 &    0.0302 &    999.75/1090 \\
 & T$+$IC &      4.28$^{+     0.13}_{-     0.08}$ & 
    0.414$^{+    0.025}_{-    0.033}$ & 
   0.0810$^{+   0.0034}_{-   0.0032}$ &  & 
 & 
$<$    0.0019 &   1011.04/1091 \\
A3376 & 1T &  4.00$^{+ 0.29}_{- 0.27}$ & 
0.499$^{+0.126}_{-0.118}$ & 
0.0108$^{+0.0005}_{-0.0005}$ &  & 
 &  &  140.99/176 \\
 & 2T &      2.08 & 
    0.579 & 
   0.0061 &  & 
     7.57 &    0.0054 &    130.29/174 \\
 & T$+$IC &      3.06$^{+     0.59}_{-     0.46}$ & 
    1.329$^{+    1.706}_{-    0.586}$ & 
   0.0045$^{+   0.0026}_{-   0.0024}$ &  & 
 & 
   0.0018$^{+   0.0005}_{-   0.0008}$ &    130.39/175 \\
A3391 & 1T &  6.47$^{+ 0.32}_{- 0.31}$ & 
0.311$^{+0.050}_{-0.049}$ & 
0.0207$^{+0.0004}_{-0.0004}$ &  & 
 &  &  375.51/491 \\
 & 2T &      5.94 & 
    0.321 & 
   0.0195 &  & 
    47.56 &    0.0016 &    372.46/489 \\
 & T$+$IC &      6.56$^{+     0.39}_{-     0.38}$ & 
    0.379$^{+    0.064}_{-    0.087}$ & 
   0.0168$^{+   0.0034}_{-   0.0023}$ &  & 
 & 
   0.0012$^{+   0.0008}_{-   0.0011}$ &    371.91/490 \\
A3395s & 1T &  5.85$^{+ 0.66}_{- 0.64}$ & 
0.228$^{+0.095}_{-0.093}$ & 
0.0078$^{+0.0004}_{-0.0003}$ &  & 
 &  &  118.38/214 \\
 & 2T &      3.53 & 
    0.306 & 
   0.0041 &  & 
     9.27 &    0.0039 &    114.33/212 \\
 & T$+$IC &      5.79$^{+     0.75}_{-     0.74}$ & 
    0.282$^{+    0.163}_{-    0.114}$ & 
   0.0067$^{+   0.0014}_{-   0.0020}$ &  & 
 & 
$<$    0.0009 &    117.05/213 \\
R1504 & 1T &  8.45$^{+ 0.55}_{- 0.41}$ & 
0.416$^{+0.049}_{-0.047}$ & 
0.0491$^{+0.0011}_{-0.0010}$ &  & 
 &  & 1187.05/1053 \\
 & 2T &      7.05 & 
    0.425 & 
   0.0364 &  & 
    17.30 &    0.0135 &   1183.65/1051 \\
 & T$+$IC &     11.47$^{+     1.32}_{-     0.92}$ & 
    0.453$^{+    0.143}_{-    0.105}$ & 
   0.0437$^{+   0.0058}_{-   0.0058}$ &  & 
 & 
   0.0016$^{+   0.0015}_{-   0.0015}$ &   1272.90/1052 \\
A0576 & 1T &  4.06$^{+ 0.28}_{- 0.26}$ & 
0.376$^{+0.087}_{-0.083}$ & 
0.0244$^{+0.0010}_{-0.0010}$ &  & 
 &  &  175.63/226 \\
 & 2T &      3.80 & 
    0.377 & 
   0.0132 &  & 
     4.38 &    0.0112 &    175.62/224 \\
 & T$+$IC &      3.96$^{+     0.33}_{-     0.35}$ & 
    0.416$^{+    0.124}_{-    0.108}$ & 
   0.0220$^{+   0.0032}_{-   0.0038}$ &  & 
 & 
$<$    0.0019 &    174.70/225 \\
A0754 & 1T &  9.19$^{+ 0.33}_{- 0.32}$ & 
0.273$^{+0.032}_{-0.032}$ & 
0.0699$^{+0.0007}_{-0.0007}$ &  & 
 &  &  797.01/960 \\
 & 2T &      7.58 & 
    0.288 & 
   0.0504 &  & 
    15.37 &    0.0209 &    783.61/958 \\
 & T$+$IC &      9.26$^{+     0.34}_{-     0.34}$ & 
    0.292$^{+    0.034}_{-    0.036}$ & 
   0.0668$^{+   0.0035}_{-   0.0034}$ &  & 
 & 
$<$    0.0023 &    790.98/959 \\
HydraA & 1T &  3.98$^{+ 0.09}_{- 0.09}$ & 
0.286$^{+0.026}_{-0.025}$ & 
0.0453$^{+0.0007}_{-0.0007}$ &  & 
 &  &  607.52/718 \\
 & 2T &      2.72 & 
    0.325 & 
   0.0290 &  & 
     6.51 &    0.0181 &    590.83/716 \\
 & T$+$IC &      3.77$^{+     0.16}_{-     0.17}$ & 
    0.338$^{+    0.049}_{-    0.042}$ & 
   0.0390$^{+   0.0035}_{-   0.0034}$ &  & 
 & 
   0.0020$^{+   0.0010}_{-   0.0011}$ &    598.26/717 \\
A1060 & 1T &  3.20$^{+ 0.05}_{- 0.05}$ & 
0.404$^{+0.023}_{-0.023}$ & 
0.0589$^{+0.0007}_{-0.0007}$ &  & 
 &  &  858.66/972 \\
 & 2T &      3.29 & 
    0.404 & 
   0.0304 &  & 
     3.12 &    0.0285 &    858.63/970 \\
 & T$+$IC &      3.07$^{+     0.11}_{-     0.08}$ & 
    0.458$^{+    0.040}_{-    0.047}$ & 
   0.0539$^{+   0.0037}_{-   0.0027}$ &  & 
 & 
   0.0017$^{+   0.0008}_{-   0.0011}$ &    850.39/971 \\
A1367 & 1T &  3.77$^{+ 0.12}_{- 0.12}$ & 
0.299$^{+0.037}_{-0.037}$ & 
0.0326$^{+0.0007}_{-0.0006}$ &  & 
 &  &  481.37/603 \\
 & 2T &      1.28 & 
    0.307 & 
   0.0081 &  & 
     4.31 &    0.0273 &    472.10/601 \\
 & T$+$IC &      3.39$^{+     0.31}_{-     0.17}$ & 
    0.398$^{+    0.074}_{-    0.089}$ & 
   0.0254$^{+   0.0043}_{-   0.0025}$ &  & 
 & 
   0.0023$^{+   0.0008}_{-   0.0014}$ &    471.44/602 \\
MKW4 & 1T &  1.68$^{+ 0.11}_{- 0.11}$ & 
0.634$^{+0.239}_{-0.188}$ & 
0.0153$^{+0.0025}_{-0.0021}$ &  & 
 &  &   61.30/106 \\
 & 2T &      1.73 & 
    0.635 & 
   0.0123 &  & 
     1.49 &    0.0030 &     61.32/104 \\
 & T$+$IC &      1.68$^{+     0.11}_{-     0.17}$ & 
    0.668$^{+    0.255}_{-    0.192}$ & 
   0.0148$^{+   0.0024}_{-   0.0026}$ &  & 
 & 
$<$    0.0004 &     56.17/105 \\
ZwCl1215 & 1T &  7.12$^{+ 0.34}_{- 0.33}$ & 
0.284$^{+0.038}_{-0.037}$ & 
0.0255$^{+0.0004}_{-0.0004}$ &  & 
 &  &  465.57/612 \\
 & 2T &      4.23 & 
    0.304 & 
   0.0060 &  & 
     8.20 &    0.0199 &    464.24/610 \\
 & T$+$IC &      7.27$^{+     0.37}_{-     0.37}$ & 
    0.319$^{+    0.050}_{-    0.066}$ & 
   0.0229$^{+   0.0029}_{-   0.0026}$ &  & 
 & 
$<$    0.0017 &    463.19/611 \\
NGC4636 & 1T &  0.92$^{+ 0.09}_{- 0.09}$ & 
0.977$^{+0.630}_{-0.322}$ & 
0.0056$^{+0.0017}_{-0.0016}$ &  & 
 &  &  229.06/363 \\
 & 2T &     26.36 & 
    1.988 & 
   0.0003 &  & 
     0.75 &    0.0039 &    212.46/361 \\
 & T$+$IC &      0.75$^{+     0.06}_{-     0.06}$ & 
    4.985$^{+   -4.985}_{-    2.829}$ & 
   0.0016$^{+   0.0021}_{-   0.0003}$ &  & 
 & 
   0.0002$^{+   0.0001}_{-   0.0001}$ &    213.53/362 \\
A3526 & 1T &  3.95$^{+ 0.03}_{- 0.09}$ & 
0.544$^{+0.013}_{-0.015}$ & 
0.1109$^{+0.0025}_{-0.0037}$ &  & 
 &  & 2020.73/1762 \\
 & 2T &      3.95 & 
    0.544 & 
   0.0507 &  & 
     4.16 &    0.0513 &   2129.81/1760 \\
 & T$+$IC &      4.00$^{+     0.05}_{-     0.07}$ & 
    0.541$^{+    0.012}_{-    0.011}$ & 
   0.1021$^{+   0.0011}_{-   0.0047}$ &  & 
 & 
   0.0007$^{+   0.0016}_{-   0.0004}$ &   2089.03/1761 \\
A1644 & 1T &  5.14$^{+ 0.24}_{- 0.23}$ & 
0.294$^{+0.046}_{-0.045}$ & 
0.0442$^{+0.0011}_{-0.0011}$ &  & 
 &  &  394.74/534 \\
 & 2T &      4.00 & 
    0.312 & 
   0.0249 &  & 
     7.01 &    0.0198 &    390.98/532 \\
 & T$+$IC &      5.09$^{+     0.26}_{-     0.26}$ & 
    0.318$^{+    0.054}_{-    0.049}$ & 
   0.0405$^{+   0.0043}_{-   0.0048}$ &  & 
 & 
$<$    0.0027 &    392.98/533 \\
A1650 & 1T &  5.94$^{+ 0.17}_{- 0.16}$ & 
0.394$^{+0.026}_{-0.026}$ & 
0.0275$^{+0.0003}_{-0.0003}$ &  & 
 &  &  755.89/919 \\
 & 2T &      4.83 & 
    0.413 & 
   0.0151 &  & 
     7.57 &    0.0126 &    752.04/917 \\
 & T$+$IC &      5.93$^{+     0.19}_{-     0.19}$ & 
    0.451$^{+    0.075}_{-    0.059}$ & 
   0.0237$^{+   0.0032}_{-   0.0033}$ &  & 
 & 
   0.0011$^{+   0.0010}_{-   0.0010}$ &    749.07/918 \\
A1651 & 1T &  6.45$^{+ 0.36}_{- 0.35}$ & 
0.389$^{+0.057}_{-0.056}$ & 
0.0347$^{+0.0009}_{-0.0009}$ &  & 
 &  &  212.15/335 \\
 & 2T &      3.41 & 
    0.406 & 
   0.0049 &  & 
     7.02 &    0.0302 &    211.76/333 \\
 & T$+$IC &      6.46$^{+     0.44}_{-     0.42}$ & 
    0.501$^{+    0.123}_{-    0.096}$ & 
   0.0268$^{+   0.0043}_{-   0.0043}$ &  & 
 & 
   0.0025$^{+   0.0013}_{-   0.0013}$ &    202.64/334 \\
Coma & 1T &  8.51$^{+ 0.11}_{- 0.11}$ & 
0.248$^{+0.015}_{-0.015}$ & 
0.2434$^{+0.0013}_{-0.0013}$ &  & 
 &  & 1801.39/2167 \\
 & 2T &      7.36 & 
    0.248 & 
   0.0869 &  & 
     9.29 &    0.1573 &   1797.98/2165 \\
 & T$+$IC &      8.51$^{+     0.11}_{-     0.11}$ & 
    0.249$^{+    0.015}_{-    0.015}$ & 
   0.2429$^{+   0.0017}_{-   0.0038}$ &  & 
 & 
$<$    0.0015 &   1801.47/2166 \\
NGC5044 & 1T &  1.20$^{+ 0.04}_{- 0.04}$ & 
0.800$^{+0.151}_{-0.126}$ & 
0.0245$^{+0.0029}_{-0.0027}$ &  & 
 &  &  391.89/497 \\
 & 2T &      1.09 & 
    0.866 & 
   0.0237 &  & 
     6.96 &    0.0009 &    382.01/495 \\
 & T$+$IC &      1.10$^{+     0.07}_{-     0.10}$ & 
    0.943$^{+    0.251}_{-    0.181}$ & 
   0.0218$^{+   0.0035}_{-   0.0035}$ &  & 
 & 
   0.0005$^{+   0.0002}_{-   0.0002}$ &    381.79/496 \\
A3558 & 1T &  5.90$^{+ 0.10}_{- 0.10}$ & 
0.324$^{+0.015}_{-0.015}$ & 
0.0663$^{+0.0005}_{-0.0005}$ &  & 
 &  & 1282.65/1465 \\
 & 2T &      5.27 & 
    0.330 & 
   0.0440 &  & 
     7.31 &    0.0226 &   1282.28/1463 \\
 & T$+$IC &      5.90$^{+     0.11}_{-     0.10}$ & 
    0.350$^{+    0.016}_{-    0.025}$ & 
   0.0610$^{+   0.0040}_{-   0.0040}$ &  & 
 & 
   0.0017$^{+   0.0013}_{-   0.0013}$ &   1277.75/1464 \\
A3562 & 1T &  5.07$^{+ 0.61}_{- 0.55}$ & 
0.406$^{+0.150}_{-0.142}$ & 
0.0176$^{+0.0010}_{-0.0010}$ &  & 
 &  &   59.05/134 \\
 & 2T &      5.08 & 
    0.414 & 
   0.0085 &  & 
     5.00 &    0.0091 &     58.81/132 \\
 & T$+$IC &      5.11$^{+     0.62}_{-     0.54}$ & 
    0.398$^{+    0.155}_{-    0.141}$ & 
   0.0176$^{+   0.0011}_{-   0.0039}$ &  & 
 & 
$<$    0.0012 &     59.51/133 \\
A3571 & 1T &  7.20$^{+ 0.14}_{- 0.14}$ & 
0.372$^{+0.019}_{-0.019}$ & 
0.1105$^{+0.0008}_{-0.0008}$ &  & 
 &  & 1621.68/1883 \\
 & 2T &      7.22 & 
    0.372 & 
   0.0343 &  & 
     7.19 &    0.0762 &   1621.68/1881 \\
 & T$+$IC &      7.20$^{+     0.15}_{-     0.15}$ & 
    0.388$^{+    0.024}_{-    0.027}$ & 
   0.1065$^{+   0.0043}_{-   0.0038}$ &  & 
 & 
$<$    0.0026 &   1621.82/1882 \\
A1795 & 1T &  5.63$^{+ 0.08}_{- 0.08}$ & 
0.365$^{+0.013}_{-0.013}$ & 
0.0794$^{+0.0005}_{-0.0005}$ &  & 
 &  & 1741.93/1916 \\
 & 2T &      4.33 & 
    0.383 & 
   0.0325 &  & 
     6.63 &    0.0479 &   1671.68/1914 \\
 & T$+$IC &      5.62$^{+     0.08}_{-     0.09}$ & 
    0.390$^{+    0.014}_{-    0.014}$ & 
   0.0753$^{+   0.0030}_{-   0.0033}$ &  & 
 & 
   0.0014$^{+   0.0010}_{-   0.0010}$ &   1674.47/1915 \\
A3581 & 1T &  1.88$^{+ 0.04}_{- 0.04}$ & 
0.557$^{+0.059}_{-0.055}$ & 
0.0272$^{+0.0010}_{-0.0010}$ &  & 
 &  &  428.73/555 \\
 & 2T &      1.74 & 
    0.571 & 
   0.0265 &  & 
    17.23 &    0.0012 &    413.71/553 \\
 & T$+$IC &      1.74$^{+     0.06}_{-     0.07}$ & 
    0.625$^{+    0.079}_{-    0.071}$ & 
   0.0243$^{+   0.0016}_{-   0.0016}$ &  & 
 & 
   0.0008$^{+   0.0003}_{-   0.0003}$ &    412.74/554 \\
MKW8 & 1T &  3.35$^{+ 0.29}_{- 0.21}$ & 
0.354$^{+0.100}_{-0.093}$ & 
0.0134$^{+0.0007}_{-0.0007}$ &  & 
 &  &  160.18/230 \\
 & 2T &      2.42 & 
    0.370 & 
   0.0088 &  & 
     5.45 &    0.0050 &    157.94/228 \\
 & T$+$IC &      3.00$^{+     0.41}_{-     0.41}$ & 
    0.488$^{+    0.199}_{-    0.162}$ & 
   0.0103$^{+   0.0024}_{-   0.0030}$ &  & 
 & 
$<$    0.0017 &    157.59/229 \\
A2029 & 1T &  8.01$^{+ 0.21}_{- 0.21}$ & 
0.428$^{+0.029}_{-0.029}$ & 
0.0780$^{+0.0008}_{-0.0008}$ &  & 
 &  &  877.70/952 \\
 & 2T &      0.29 & 
    0.458 & 
   0.5676 &  & 
     8.48 &    0.0752 &    844.34/950 \\
 & T$+$IC &      8.13$^{+     0.24}_{-     0.23}$ & 
    0.501$^{+    0.052}_{-    0.046}$ & 
   0.0675$^{+   0.0044}_{-   0.0044}$ &  & 
 & 
   0.0034$^{+   0.0014}_{-   0.0014}$ &    861.38/951 \\
A2052 & 1T &  3.01$^{+ 0.05}_{- 0.05}$ & 
0.498$^{+0.029}_{-0.029}$ & 
0.0479$^{+0.0006}_{-0.0006}$ &  & 
 &  &  723.70/858 \\
 & 2T &      2.80 & 
    0.505 & 
   0.0250 &  & 
     3.24 &    0.0229 &    716.70/856 \\
 & T$+$IC &      2.81$^{+     0.11}_{-     0.10}$ & 
    0.610$^{+    0.066}_{-    0.065}$ & 
   0.0417$^{+   0.0030}_{-   0.0026}$ &  & 
 & 
   0.0017$^{+   0.0007}_{-   0.0008}$ &    704.61/857 \\
MKW3S & 1T &  3.36$^{+ 0.06}_{- 0.06}$ & 
0.385$^{+0.027}_{-0.026}$ & 
0.0396$^{+0.0006}_{-0.0006}$ &  & 
 &  &  722.14/847 \\
 & 2T &      3.95 & 
    0.389 & 
   0.0191 &  & 
     2.89 &    0.0208 &    720.11/845 \\
 & T$+$IC &      3.23$^{+     0.10}_{-     0.15}$ & 
    0.433$^{+    0.051}_{-    0.038}$ & 
   0.0361$^{+   0.0023}_{-   0.0033}$ &  & 
 & 
   0.0011$^{+   0.0010}_{-   0.0007}$ &    716.12/846 \\
A2065 & 1T &  6.46$^{+ 0.53}_{- 0.47}$ & 
0.261$^{+0.077}_{-0.076}$ & 
0.0292$^{+0.0009}_{-0.0009}$ &  & 
 &  &  166.22/258 \\
 & 2T &      4.73 & 
    0.268 & 
   0.0069 &  & 
     7.04 &    0.0224 &    166.19/256 \\
 & T$+$IC &      6.44$^{+     0.55}_{-     0.49}$ & 
    0.274$^{+    0.088}_{-    0.083}$ & 
   0.0278$^{+   0.0021}_{-   0.0038}$ &  & 
 & 
$<$    0.0016 &    166.41/257 \\
A2063 & 1T &  4.32$^{+ 0.15}_{- 0.12}$ & 
0.345$^{+0.033}_{-0.033}$ & 
0.0371$^{+0.0007}_{-0.0006}$ &  & 
 &  &  649.81/783 \\
 & 2T &      4.07 & 
    0.345 & 
   0.0199 &  & 
     4.62 &    0.0173 &    655.50/781 \\
 & T$+$IC &      4.21$^{+     0.20}_{-     0.15}$ & 
    0.380$^{+    0.055}_{-    0.063}$ & 
   0.0348$^{+   0.0027}_{-   0.0041}$ &  & 
 & 
$<$    0.0020 &    653.90/782 \\
A2142 & 1T & 10.26$^{+ 0.83}_{- 0.74}$ & 
0.202$^{+0.318}_{-0.202}$ & 
0.0629$^{+0.0042}_{-0.0042}$ &  & 
 &  &   47.02/98 \\
 & 2T &      9.10 & 
    0.224 & 
   0.0619 &  & 
    64.00 &    0.0032 &     42.08/96 \\
 & T$+$IC &      9.54$^{+     1.00}_{-     1.04}$ & 
$<$     0.615 & 
   0.0561$^{+   0.0068}_{-   0.0068}$ &  & 
 & 
   0.0025$^{+   0.0020}_{-   0.0020}$ &     42.63/97 \\
A2147 & 1T &  4.99$^{+ 0.66}_{- 0.53}$ & 
0.250$^{+0.126}_{-0.120}$ & 
0.0412$^{+0.0023}_{-0.0023}$ &  & 
 &  &  104.01/159 \\
 & 2T &      4.18 & 
    0.259 & 
   0.0261 &  & 
     6.84 &    0.0153 &    103.33/157 \\
 & T$+$IC &      5.02$^{+     0.68}_{-     0.53}$ & 
    0.246$^{+    0.120}_{-    0.118}$ & 
   0.0415$^{+   0.0024}_{-   0.0043}$ &  & 
 & 
$<$    0.0013 &    104.08/158 \\
A2199 & 1T &  4.45$^{+ 0.09}_{- 0.09}$ & 
0.363$^{+0.021}_{-0.020}$ & 
0.1019$^{+0.0011}_{-0.0011}$ &  & 
 &  &  918.41/1078 \\
 & 2T &      2.66 & 
    0.382 & 
   0.0235 &  & 
     5.01 &    0.0803 &    911.87/1076 \\
 & T$+$IC &      4.41$^{+     0.10}_{-     0.11}$ & 
    0.375$^{+    0.024}_{-    0.023}$ & 
   0.0986$^{+   0.0036}_{-   0.0040}$ &  & 
 & 
$<$    0.0024 &    916.19/1077 \\
A2204 & 1T &  7.10$^{+ 0.24}_{- 0.23}$ & 
0.397$^{+0.029}_{-0.028}$ & 
0.0467$^{+0.0006}_{-0.0006}$ &  & 
 &  &  628.73/781 \\
 & 2T &      4.21 & 
    0.494 & 
   0.0255 &  & 
    12.93 &    0.0232 &    608.41/779 \\
 & T$+$IC &      7.15$^{+     0.27}_{-     0.27}$ & 
    0.487$^{+    0.078}_{-    0.063}$ & 
   0.0380$^{+   0.0048}_{-   0.0047}$ &  & 
 & 
   0.0025$^{+   0.0013}_{-   0.0014}$ &    619.76/780 \\
A2256 & 1T &  6.99$^{+ 0.34}_{- 0.38}$ & 
0.301$^{+0.045}_{-0.044}$ & 
0.0526$^{+0.0011}_{-0.0009}$ &  & 
 &  &  341.67/443 \\
 & 2T &      0.40 & 
    0.326 & 
   0.1181 &  & 
     7.59 &    0.0504 &    326.25/441 \\
 & T$+$IC &      6.94$^{+     0.36}_{-     0.35}$ & 
    0.324$^{+    0.049}_{-    0.047}$ & 
   0.0488$^{+   0.0032}_{-   0.0032}$ &  & 
 & 
   0.0013$^{+   0.0011}_{-   0.0011}$ &    337.34/442 \\
A2255 & 1T &  7.43$^{+ 0.80}_{- 0.71}$ & 
0.269$^{+0.107}_{-0.104}$ & 
0.0237$^{+0.0008}_{-0.0008}$ &  & 
 &  &  104.15/193 \\
 & 2T &      6.84 & 
    0.268 & 
   0.0130 &  & 
     8.21 &    0.0107 &    104.16/191 \\
 & T$+$IC &      7.41$^{+     0.86}_{-     0.67}$ & 
    0.263$^{+    0.111}_{-    0.096}$ & 
   0.0238$^{+   0.0008}_{-   0.0018}$ &  & 
 & 
$<$    0.0005 &    103.80/192 \\
A3667 & 1T &  6.60$^{+ 0.11}_{- 0.11}$ & 
0.268$^{+0.015}_{-0.015}$ & 
0.0758$^{+0.0005}_{-0.0005}$ &  & 
 &  & 1497.45/1652 \\
 & 2T &      5.07 & 
    0.286 & 
   0.0425 &  & 
     9.35 &    0.0343 &   1485.95/1650 \\
 & T$+$IC &      6.70$^{+     0.12}_{-     0.13}$ & 
    0.304$^{+    0.023}_{-    0.026}$ & 
   0.0658$^{+   0.0046}_{-   0.0035}$ &  & 
 & 
   0.0032$^{+   0.0011}_{-   0.0015}$ &   1482.95/1651 \\
S1101 & 1T &  2.65$^{+ 0.06}_{- 0.06}$ & 
0.336$^{+0.038}_{-0.037}$ & 
0.0259$^{+0.0006}_{-0.0006}$ &  & 
 &  &  418.31/534 \\
 & 2T &      1.83 & 
    0.331 & 
   0.0101 &  & 
     3.12 &    0.0168 &    414.11/532 \\
 & T$+$IC &      2.56$^{+     0.12}_{-     0.12}$ & 
    0.363$^{+    0.052}_{-    0.048}$ & 
   0.0249$^{+   0.0013}_{-   0.0015}$ &  & 
 & 
$<$    0.0007 &    414.49/533 \\
A2589 & 1T &  3.69$^{+ 0.13}_{- 0.12}$ & 
0.543$^{+0.052}_{-0.050}$ & 
0.0205$^{+0.0004}_{-0.0004}$ &  & 
 &  &  335.38/446 \\
 & 2T &      0.35 & 
    0.554 & 
   0.0396 &  & 
     3.87 &    0.0195 &    326.42/444 \\
 & T$+$IC &      3.33$^{+     0.22}_{-     0.20}$ & 
    0.762$^{+    0.198}_{-    0.144}$ & 
   0.0154$^{+   0.0026}_{-   0.0025}$ &  & 
 & 
   0.0015$^{+   0.0007}_{-   0.0007}$ &    325.13/445 \\
A2597 & 1T &  3.36$^{+ 0.07}_{- 0.07}$ & 
0.329$^{+0.024}_{-0.024}$ & 
0.0274$^{+0.0004}_{-0.0004}$ &  & 
 &  &  626.72/721 \\
 & 2T &      2.29 & 
    0.360 & 
   0.0178 &  & 
     5.43 &    0.0109 &    600.23/719 \\
 & T$+$IC &      2.95$^{+     0.15}_{-     0.17}$ & 
    0.473$^{+    0.085}_{-    0.066}$ & 
   0.0211$^{+   0.0021}_{-   0.0022}$ &  & 
 & 
   0.0018$^{+   0.0005}_{-   0.0006}$ &    598.61/720 \\
A2634 & 1T &  4.50$^{+ 0.56}_{- 0.45}$ & 
0.292$^{+0.148}_{-0.140}$ & 
0.0182$^{+0.0011}_{-0.0010}$ &  & 
 &  &  108.16/140 \\
 & 2T &      4.55 & 
    0.292 & 
   0.0092 &  & 
     4.46 &    0.0091 &    108.16/138 \\
 & T$+$IC &      4.54$^{+     0.56}_{-     0.48}$ & 
    0.275$^{+    0.142}_{-    0.127}$ & 
   0.0184$^{+   0.0006}_{-   0.0032}$ &  & 
 & 
$<$    0.0010 &    107.58/139 \\
A2657 & 1T &  5.14$^{+ 0.30}_{- 0.28}$ & 
0.284$^{+0.065}_{-0.063}$ & 
0.0256$^{+0.0008}_{-0.0007}$ &  & 
 &  &  273.87/356 \\
 & 2T &      2.16 & 
    0.383 & 
   0.0112 &  & 
     7.83 &    0.0165 &    263.96/354 \\
 & T$+$IC &      5.15$^{+     0.30}_{-     0.29}$ & 
    0.284$^{+    0.066}_{-    0.063}$ & 
   0.0256$^{+   0.0008}_{-   0.0032}$ &  & 
 & 
$<$    0.0010 &    273.84/355 \\
A4038 & 1T &  3.17$^{+ 0.05}_{- 0.05}$ & 
0.371$^{+0.025}_{-0.024}$ & 
0.0593$^{+0.0008}_{-0.0008}$ &  & 
 &  &  870.11/1058 \\
 & 2T &      2.54 & 
    0.384 & 
   0.0416 &  & 
     4.83 &    0.0190 &    858.73/1056 \\
 & T$+$IC &      3.02$^{+     0.10}_{-     0.10}$ & 
    0.426$^{+    0.048}_{-    0.042}$ & 
   0.0538$^{+   0.0032}_{-   0.0032}$ &  & 
 & 
   0.0017$^{+   0.0009}_{-   0.0009}$ &    861.42/1057 \\
A4059 & 1T &  4.23$^{+ 0.13}_{- 0.12}$ & 
0.428$^{+0.036}_{-0.035}$ & 
0.0341$^{+0.0006}_{-0.0006}$ &  & 
 &  &  480.11/694 \\
 & 2T &      2.66 & 
    0.443 & 
   0.0072 &  & 
     4.71 &    0.0273 &    478.71/692 \\
 & T$+$IC &      4.23$^{+     0.13}_{-     0.14}$ & 
    0.428$^{+    0.050}_{-    0.035}$ & 
   0.0341$^{+   0.0006}_{-   0.0028}$ &  & 
 & 
$<$    0.0009 &    480.14/693

\enddata
\tablenotetext{a}{Parameters for the 2T model are unconstrained.}
\tablenotetext{b}{Normalization of the {\tt APEC} thermal spectrum,
which is given by $\{ 10^{-14} / [ 4 \pi (1+z)^2 D_A^2 ] \} \, \int n_e n_H
\, dV$, where $z$ is the redshift, $D_A$ is the angular diameter distance,
$n_e$ is the electron density, $n_H$ is the ionized hydrogen density,
and $V$ is the volume of the cluster.}
\tablenotetext{c}{Value is the normalization of the power-law component
for the T$+$IC model, which is the photon flux at a photon energy of
1 keV in units of photons cm$^{-2}$ s$^{-1}$ keV$^{-1}$.
For the 2T model, the value is the normalization of 
the second {\tt APEC} thermal
model in units of cm$^{-5}$.}
\end{deluxetable}

\begin{deluxetable}{lcccccccc}
\tablewidth{0pt}
\tablecaption{Joint Thermal and non-thermal Fits to the EPIC and 
BAT Spectra (3--195 keV)
\label{tab:bathi:joint2}}
\tablehead{
 & & \multicolumn{3}{c}{Component 1} && \multicolumn{2}{c}{Component 2} & \\
 \cline{3-5}
 \cline{7-8}
 & & $kT$ & abund & Norm.\tablenotemark{b} && $kT$ & 
Norm.\tablenotemark{c} & \\
Name & Model\tablenotemark{a} & (keV) & Z$_\odot$ & (cm$^{-5}$) & & (keV) &  & 
$\chi^2$/dof
}
\startdata
A0085 & 1T &  6.94$^{+ 0.31}_{- 0.26}$ & 
0.363$^{+0.034}_{-0.033}$ & 
0.0746$^{+0.0016}_{-0.0015}$ &  & 
 &  &  410.01/534 \\
 & 2T &      6.90 & 
    0.364 & 
   0.0449 &  & 
     7.01 &    0.0298 &    410.28/532 \\
 & T$+$IC &      6.91$^{+     0.30}_{-     0.27}$ & 
    0.362$^{+    0.037}_{-    0.030}$ & 
   0.0746$^{+   0.0015}_{-   0.0022}$ &  & 
 & 
$<$    0.0007 &    411.40/533 \\
A0119 & 1T &  6.71$^{+ 0.87}_{- 0.79}$ & 
0.243$^{+0.082}_{-0.080}$ & 
0.0285$^{+0.0020}_{-0.0015}$ &  & 
 &  &  115.75/162 \\
 & 2T &      6.78 & 
    0.246 & 
   0.0158 &  & 
     6.59 &    0.0127 &    115.65/160 \\
 & T$+$IC &      6.73$^{+     0.88}_{-     0.80}$ & 
    0.242$^{+    0.087}_{-    0.076}$ & 
   0.0286$^{+   0.0020}_{-   0.0028}$ &  & 
 & 
$<$    0.0009 &    115.69/161 \\
A0133 & 1T &  4.27$^{+ 0.34}_{- 0.27}$ & 
0.442$^{+0.053}_{-0.051}$ & 
0.0211$^{+0.0012}_{-0.0011}$ &  & 
 &  &  134.52/237 \\
 & 2T &      2.10 & 
    0.442 & 
   0.0003 &  & 
     4.34 &    0.0207 &    134.60/235 \\
 & T$+$IC &      4.03$^{+     0.59}_{-     0.35}$ & 
    0.462$^{+    0.134}_{-    0.068}$ & 
   0.0204$^{+   0.0019}_{-   0.0039}$ &  & 
 & 
$<$    0.0016 &    134.21/236 \\
NGC507 & 1T &  1.85$^{+ 0.35}_{- 0.27}$ & 
0.829$^{+0.757}_{-0.431}$ & 
0.0073$^{+0.0029}_{-0.0019}$ &  & 
 &  &   64.52/99 \\
 & 2T &      1.60 & 
    1.003 & 
   0.0069 &  & 
    64.00 &    0.0005 &     62.13/97 \\
 & T$+$IC &      1.61$^{+     0.44}_{-     0.40}$ & 
    1.434$^{+   -1.434}_{-    0.924}$ & 
   0.0052$^{+   0.0044}_{-   0.0036}$ &  & 
 & 
$<$    0.0006 &     62.87/98 \\
A0262 & 1T &  2.33$^{+ 0.09}_{- 0.08}$ & 
0.412$^{+0.059}_{-0.055}$ & 
0.0531$^{+0.0027}_{-0.0026}$ &  & 
 &  &  297.16/379 \\
 & 2T &      2.31 & 
    0.412 & 
   0.0480 &  & 
     2.54 &    0.0051 &    297.16/377 \\
 & T$+$IC &      2.33$^{+     0.10}_{-     0.09}$ & 
    0.403$^{+    0.065}_{-    0.048}$ & 
   0.0535$^{+   0.0025}_{-   0.0029}$ &  & 
 & 
$<$    0.0004 &    297.48/378 \\
A0400 & 1T &  2.34$^{+ 0.30}_{- 0.23}$ & 
0.352$^{+0.168}_{-0.143}$ & 
0.0195$^{+0.0031}_{-0.0027}$ &  & 
 &  &  138.40/159 \\
 & 2T &      2.49 & 
    0.352 & 
   0.0087 &  & 
     2.22 &    0.0107 &    138.42/157 \\
 & T$+$IC &      2.42$^{+     0.25}_{-     0.31}$ & 
    0.346$^{+    0.163}_{-    0.141}$ & 
   0.0196$^{+   0.0032}_{-   0.0027}$ &  & 
 & 
$<$    0.0005 &    137.80/158 \\
A0399 & 1T &  7.63$^{+ 0.71}_{- 0.67}$ & 
0.233$^{+0.058}_{-0.056}$ & 
0.0349$^{+0.0014}_{-0.0014}$ &  & 
 &  &  152.92/229 \\
 & 2T &      7.67 & 
    0.233 & 
   0.0105 &  & 
     7.62 &    0.0244 &    152.92/227 \\
 & T$+$IC &      7.69$^{+     0.63}_{-     0.76}$ & 
    0.235$^{+    0.057}_{-    0.057}$ & 
   0.0347$^{+   0.0016}_{-   0.0019}$ &  & 
 & 
$<$    0.0007 &    153.07/228 \\
A3112 & 1T &  5.29$^{+ 0.27}_{- 0.22}$ & 
0.455$^{+0.032}_{-0.031}$ & 
0.0339$^{+0.0011}_{-0.0011}$ &  & 
 &  &  374.66/448 \\
 & 2T &      4.03 & 
    0.550 & 
   0.0284 &  & 
    15.19 &    0.0073 &    364.14/446 \\
 & T$+$IC &      5.10$^{+     0.29}_{-     0.24}$ & 
    0.527$^{+    0.066}_{-    0.056}$ & 
   0.0296$^{+   0.0032}_{-   0.0028}$ &  & 
 & 
   0.0015$^{+   0.0009}_{-   0.0011}$ &    367.28/447 \\
Fornax & 1T &  2.01$^{+ 0.17}_{- 0.13}$ & 
0.253$^{+0.096}_{-0.085}$ & 
0.0192$^{+0.0022}_{-0.0021}$ &  & 
 &  &  471.90/503 \\
 & 2T &      1.60 & 
    0.254 & 
   0.0177 &  & 
     3.40 &    0.0038 &    470.55/501 \\
 & T$+$IC &      1.57$^{+     0.25}_{-     0.28}$ & 
    0.488$^{+    0.408}_{-    0.179}$ & 
   0.0176$^{+   0.0036}_{-   0.0044}$ &  & 
 & 
   0.0008$^{+   0.0005}_{-   0.0004}$ &    465.05/502 \\
2A0335 & 1T &  3.22$^{+ 0.13}_{- 0.12}$ & 
0.400$^{+0.036}_{-0.035}$ & 
0.0941$^{+0.0037}_{-0.0037}$ &  & 
 &  &  254.73/381 \\
 & 2T &      2.67 & 
    0.411 & 
   0.0553 &  & 
     3.93 &    0.0407 &    254.16/379 \\
 & T$+$IC &      3.22$^{+     0.13}_{-     0.13}$ & 
    0.399$^{+    0.040}_{-    0.032}$ & 
   0.0943$^{+   0.0036}_{-   0.0042}$ &  & 
 & 
$<$    0.0010 &    254.74/380 \\
IIIZw54 & 1T &  3.04$^{+ 0.28}_{- 0.23}$ & 
0.242$^{+0.068}_{-0.063}$ & 
0.0170$^{+0.0015}_{-0.0015}$ &  & 
 &  &  156.99/228 \\
 & 2T &      3.01 & 
    0.242 & 
   0.0089 &  & 
     3.07 &    0.0081 &    156.99/226 \\
 & T$+$IC &      2.81$^{+     0.50}_{-     0.41}$ & 
    0.280$^{+    0.130}_{-    0.101}$ & 
   0.0159$^{+   0.0025}_{-   0.0025}$ &  & 
 & 
$<$    0.0012 &    157.91/227 \\
A3158 & 1T &  6.34$^{+ 0.55}_{- 0.52}$ & 
0.348$^{+0.063}_{-0.061}$ & 
0.0388$^{+0.0019}_{-0.0019}$ &  & 
 &  &  154.76/217 \\
 & 2T &      6.83 & 
    0.348 & 
   0.0173 &  & 
     5.91 &    0.0214 &    154.52/215 \\
 & T$+$IC &      6.32$^{+     0.54}_{-     0.55}$ & 
    0.350$^{+    0.066}_{-    0.061}$ & 
   0.0387$^{+   0.0020}_{-   0.0030}$ &  & 
 & 
$<$    0.0010 &    154.81/216 \\
NGC1550 & 1T &  1.55$^{+ 0.19}_{- 0.14}$ & 
0.299$^{+0.212}_{-0.163}$ & 
0.0285$^{+0.0075}_{-0.0061}$ &  & 
 &  &   81.90/128 \\
 & 2T &      1.56 & 
    0.299 & 
   0.0145 &  & 
     1.53 &    0.0141 &     82.39/126 \\
 & T$+$IC &      1.31$^{+     0.36}_{-     0.30}$ & 
    0.390$^{+    0.442}_{-    0.223}$ & 
   0.0305$^{+   0.0151}_{-   0.0096}$ &  & 
 & 
$<$    0.0010 &     81.43/127 \\
EXO0422 & 1T &  3.21$^{+ 0.15}_{- 0.13}$ & 
0.336$^{+0.035}_{-0.034}$ & 
0.0289$^{+0.0013}_{-0.0013}$ &  & 
 &  &  330.67/446 \\
 & 2T &      2.80 & 
    0.340 & 
   0.0150 &  & 
     3.62 &    0.0142 &    330.65/444 \\
 & T$+$IC &      3.06$^{+     0.25}_{-     0.37}$ & 
    0.372$^{+    0.128}_{-    0.061}$ & 
   0.0276$^{+   0.0023}_{-   0.0036}$ &  & 
 & 
$<$    0.0017 &    329.39/445 \\
A3266 & 1T &  8.48$^{+ 0.45}_{- 0.35}$ & 
0.196$^{+0.032}_{-0.031}$ & 
0.0790$^{+0.0016}_{-0.0015}$ &  & 
 &  &  564.78/730 \\
 & 2T &      6.85 & 
    0.195 & 
   0.0151 &  & 
     8.95 &    0.0640 &    564.31/728 \\
 & T$+$IC &      8.41$^{+     0.44}_{-     0.36}$ & 
    0.199$^{+    0.032}_{-    0.032}$ & 
   0.0770$^{+   0.0033}_{-   0.0036}$ &  & 
 & 
$<$    0.0022 &    564.43/729 \\
A0496 & 1T &  4.59$^{+ 0.14}_{- 0.13}$ & 
0.388$^{+0.022}_{-0.022}$ & 
0.0805$^{+0.0016}_{-0.0016}$ &  & 
 &  &  631.47/766 \\
 & 2T &      4.35 & 
    0.388 & 
   0.0519 &  & 
     5.03 &    0.0287 &    631.48/764 \\
 & T$+$IC &      4.59$^{+     0.16}_{-     0.12}$ & 
    0.386$^{+    0.023}_{-    0.021}$ & 
   0.0806$^{+   0.0013}_{-   0.0023}$ &  & 
 & 
$<$    0.0006 &    630.73/765 \\
A3376 & 1T &  5.77$^{+ 1.12}_{- 0.95}$ & 
0.450$^{+0.146}_{-0.129}$ & 
0.0086$^{+0.0010}_{-0.0008}$ &  & 
 &  &   63.94/84 \\
 & 2T &      1.98 & 
    0.523 & 
   0.0025 &  & 
     6.83 &    0.0070 &     63.31/82 \\
 & T$+$IC &      5.25$^{+     1.33}_{-     1.31}$ & 
    0.630$^{+    0.661}_{-    0.217}$ & 
   0.0066$^{+   0.0028}_{-   0.0032}$ &  & 
 & 
$<$    0.0018 &     63.24/83 \\
A3391 & 1T &  6.85$^{+ 0.59}_{- 0.46}$ & 
0.314$^{+0.054}_{-0.052}$ & 
0.0200$^{+0.0008}_{-0.0007}$ &  & 
 &  &  202.73/303 \\
 & 2T &      6.83 & 
    0.314 & 
   0.0088 &  & 
     6.87 &    0.0112 &    202.75/301 \\
 & T$+$IC &      6.75$^{+     0.57}_{-     0.56}$ & 
    0.353$^{+    0.071}_{-    0.083}$ & 
   0.0178$^{+   0.0028}_{-   0.0027}$ &  & 
 & 
$<$    0.0018 &    203.22/302 \\
A3395s & 1T &  6.03$^{+ 1.29}_{- 1.09}$ & 
0.240$^{+0.109}_{-0.101}$ & 
0.0076$^{+0.0010}_{-0.0007}$ &  & 
 &  &   56.82/124 \\
 & 2T &      4.31 & 
    0.258 & 
   0.0039 &  & 
     7.64 &    0.0039 &     56.24/122 \\
 & T$+$IC &      5.88$^{+     1.48}_{-     0.69}$ & 
    0.269$^{+    0.154}_{-    0.134}$ & 
   0.0067$^{+   0.0019}_{-   0.0020}$ &  & 
 & 
$<$    0.0010 &     56.82/123 \\
R1504 & 1T &  8.32$^{+ 0.71}_{- 0.56}$ & 
0.407$^{+0.055}_{-0.051}$ & 
0.0495$^{+0.0020}_{-0.0019}$ &  & 
 &  &  940.16/788 \\
 & 2T &      6.93 & 
    0.412 & 
   0.0198 &  & 
     9.48 &    0.0299 &    939.83/786 \\
 & T$+$IC &     11.45$^{+     1.28}_{-     0.99}$ & 
    0.444$^{+    0.094}_{-    0.077}$ & 
   0.0473$^{+   0.0014}_{-   0.0043}$ &  & 
 & 
$<$    0.0013 &   1014.97/787 \\
A0576 & 1T &  4.37$^{+ 0.61}_{- 0.54}$ & 
0.363$^{+0.088}_{-0.083}$ & 
0.0231$^{+0.0026}_{-0.0019}$ &  & 
 &  &   91.85/129 \\
 & 2T &      2.72 & 
    0.430 & 
   0.0136 &  & 
     6.37 &    0.0113 &     90.65/127 \\
 & T$+$IC &      4.16$^{+     0.74}_{-     0.59}$ & 
    0.402$^{+    0.125}_{-    0.109}$ & 
   0.0216$^{+   0.0037}_{-   0.0039}$ &  & 
 & 
$<$    0.0018 &     90.96/128 \\
A0754 & 1T &  9.46$^{+ 0.43}_{- 0.42}$ & 
0.286$^{+0.034}_{-0.033}$ & 
0.0696$^{+0.0011}_{-0.0011}$ &  & 
 &  &  528.60/645 \\
 & 2T &      8.13 & 
    0.288 & 
   0.0564 &  & 
    16.61 &    0.0143 &    529.13/643 \\
 & T$+$IC &      9.40$^{+     0.43}_{-     0.42}$ & 
    0.284$^{+    0.038}_{-    0.033}$ & 
   0.0695$^{+   0.0011}_{-   0.0043}$ &  & 
 & 
$<$    0.0017 &    531.98/644 \\
HydraA & 1T &  4.39$^{+ 0.19}_{- 0.18}$ & 
0.282$^{+0.026}_{-0.026}$ & 
0.0414$^{+0.0014}_{-0.0013}$ &  & 
 &  &  330.74/443 \\
 & 2T &      3.89 & 
    0.297 & 
   0.0343 &  & 
     6.84 &    0.0079 &    330.58/441 \\
 & T$+$IC &      4.32$^{+     0.25}_{-     0.24}$ & 
    0.293$^{+    0.038}_{-    0.035}$ & 
   0.0399$^{+   0.0028}_{-   0.0031}$ &  & 
 & 
$<$    0.0016 &    330.55/442 \\
A1060 & 1T &  3.43$^{+ 0.09}_{- 0.09}$ & 
0.383$^{+0.025}_{-0.023}$ & 
0.0558$^{+0.0016}_{-0.0012}$ &  & 
 &  &  519.24/641 \\
 & 2T &      3.43 & 
    0.383 & 
   0.0290 &  & 
     3.42 &    0.0268 &    519.24/639 \\
 & T$+$IC &      3.42$^{+     0.09}_{-     0.10}$ & 
    0.384$^{+    0.029}_{-    0.022}$ & 
   0.0559$^{+   0.0015}_{-   0.0021}$ &  & 
 & 
$<$    0.0006 &    522.27/640 \\
A1367 & 1T &  4.17$^{+ 0.26}_{- 0.23}$ & 
0.298$^{+0.040}_{-0.039}$ & 
0.0297$^{+0.0015}_{-0.0014}$ &  & 
 &  &  259.15/344 \\
 & 2T &      1.05 & 
    0.302 & 
   0.0052 &  & 
     4.32 &    0.0283 &    258.74/342 \\
 & T$+$IC &      3.93$^{+     0.32}_{-     0.34}$ & 
    0.349$^{+    0.063}_{-    0.058}$ & 
   0.0262$^{+   0.0031}_{-   0.0032}$ &  & 
 & 
   0.0013$^{+   0.0011}_{-   0.0011}$ &    254.94/343 \\
MKW4 & 1T &  1.73$^{+ 0.28}_{- 0.26}$ & 
0.886$^{+1.045}_{-0.514}$ & 
0.0124$^{+0.0065}_{-0.0040}$ &  & 
 &  &   27.27/48 \\
 & 2T &      1.73 & 
    0.822 & 
   0.0126 &  & 
     0.71 &    0.0000 &     27.43/46 \\
 & T$+$IC &      1.67$^{+     0.33}_{-     0.42}$ & 
    0.984$^{+   -0.984}_{-    0.601}$ & 
   0.0119$^{+   0.0069}_{-   0.0080}$ &  & 
 & 
$<$    0.0007 &     27.21/47 \\
ZwCl1215 & 1T &  7.64$^{+ 0.51}_{- 0.50}$ & 
0.299$^{+0.043}_{-0.042}$ & 
0.0247$^{+0.0008}_{-0.0007}$ &  & 
 &  &  279.58/372 \\
 & 2T &      6.83 & 
    0.302 & 
   0.0044 &  & 
     7.83 &    0.0203 &    278.15/370 \\
 & T$+$IC &      7.63$^{+     0.54}_{-     0.48}$ & 
    0.300$^{+    0.049}_{-    0.040}$ & 
   0.0248$^{+   0.0007}_{-   0.0022}$ &  & 
 & 
$<$    0.0008 &    278.15/371 \\
NGC4636 & 1T &  2.30$^{+ 3.13}_{- 1.02}$ & 
0.197$^{+4.127}_{-0.197}$ & 
0.0019$^{+0.0033}_{-0.0019}$ &  & 
 &  &  101.60/154 \\
 & 2T &     17.23 & 
    1.654 & 
   0.0004 &  & 
     0.72 &    0.0036 &     99.60/152 \\
 & T$+$IC &      0.37$^{+     2.82}_{-     0.31}$ & 
$<$     0.000 & 
   0.0252$^{+ 173.0428}_{-   0.0243}$ &  & 
 & 
$<$    0.0003 &     99.45/153 \\
A3526 & 1T &  3.75$^{+ 0.13}_{- 0.04}$ & 
0.532$^{+0.013}_{-0.013}$ & 
0.1184$^{+0.0028}_{-0.0059}$ &  & 
 &  & 1487.12/1496 \\
 & 2T &      4.05 & 
    0.512 & 
   0.0562 &  & 
     3.78 &    0.0563 &   1486.08/1494 \\
 & T$+$IC &      3.93$^{+     0.11}_{-     0.05}$ & 
    0.511$^{+    0.010}_{-    0.017}$ & 
   0.1121$^{+   0.0016}_{-   0.0040}$ &  & 
 & 
$<$    0.0005 &   1493.02/1495 \\
A1644 & 1T &  5.74$^{+ 0.49}_{- 0.54}$ & 
0.301$^{+0.051}_{-0.049}$ & 
0.0411$^{+0.0024}_{-0.0018}$ &  & 
 &  &  231.49/305 \\
 & 2T &      4.36 & 
    0.314 & 
   0.0205 &  & 
     6.85 &    0.0219 &    229.61/303 \\
 & T$+$IC &      5.69$^{+     0.53}_{-     0.60}$ & 
    0.309$^{+    0.062}_{-    0.055}$ & 
   0.0400$^{+   0.0034}_{-   0.0048}$ &  & 
 & 
$<$    0.0022 &    231.32/304 \\
A1650 & 1T &  6.12$^{+ 0.25}_{- 0.25}$ & 
0.399$^{+0.029}_{-0.028}$ & 
0.0270$^{+0.0007}_{-0.0007}$ &  & 
 &  &  454.08/603 \\
 & 2T &      5.45 & 
    0.405 & 
   0.0189 &  & 
     7.87 &    0.0082 &    453.97/601 \\
 & T$+$IC &      5.96$^{+     0.40}_{-     0.27}$ & 
    0.411$^{+    0.096}_{-    0.040}$ & 
   0.0260$^{+   0.0017}_{-   0.0047}$ &  & 
 & 
$<$    0.0019 &    453.35/602 \\
A1651 & 1T &  6.94$^{+ 0.62}_{- 0.63}$ & 
0.411$^{+0.068}_{-0.064}$ & 
0.0331$^{+0.0019}_{-0.0013}$ &  & 
 &  &  133.74/199 \\
 & 2T &      5.40 & 
    0.414 & 
   0.0035 &  & 
     7.13 &    0.0297 &    133.53/197 \\
 & T$+$IC &      6.48$^{+     0.69}_{-     0.65}$ & 
    0.495$^{+    0.123}_{-    0.095}$ & 
   0.0273$^{+   0.0043}_{-   0.0044}$ &  & 
 & 
   0.0022$^{+   0.0015}_{-   0.0015}$ &    128.72/198 \\
Coma & 1T &  8.59$^{+ 0.17}_{- 0.14}$ & 
0.249$^{+0.015}_{-0.015}$ & 
0.2435$^{+0.0021}_{-0.0019}$ &  & 
 &  & 1454.06/1835 \\
 & 2T &      8.33 & 
    0.249 & 
   0.0923 &  & 
     8.77 &    0.1514 &   1453.89/1833 \\
 & T$+$IC &      8.58$^{+     0.18}_{-     0.14}$ & 
    0.248$^{+    0.015}_{-    0.015}$ & 
   0.2436$^{+   0.0021}_{-   0.0027}$ &  & 
 & 
$<$    0.0009 &   1454.15/1834 \\
NGC5044 & 1T &  1.44$^{+ 0.17}_{- 0.15}$ & 
0.685$^{+0.540}_{-0.328}$ & 
0.0184$^{+0.0067}_{-0.0048}$ &  & 
 &  &  179.51/238 \\
 & 2T &      1.43 & 
    0.688 & 
   0.0185 &  & 
     0.03 &    0.0000 &    179.48/236 \\
 & T$+$IC &      1.37$^{+     0.22}_{-     0.28}$ & 
    0.801$^{+    1.701}_{-    0.435}$ & 
   0.0174$^{+   0.0080}_{-   0.0073}$ &  & 
 & 
$<$    0.0006 &    180.08/237 \\
A3558 & 1T &  6.22$^{+ 0.15}_{- 0.15}$ & 
0.334$^{+0.016}_{-0.016}$ & 
0.0643$^{+0.0009}_{-0.0009}$ &  & 
 &  &  911.87/1133 \\
 & 2T &      5.97 & 
    0.334 & 
   0.0423 &  & 
     6.71 &    0.0220 &    911.85/1131 \\
 & T$+$IC &      6.23$^{+     0.14}_{-     0.16}$ & 
    0.335$^{+    0.016}_{-    0.016}$ & 
   0.0642$^{+   0.0010}_{-   0.0019}$ &  & 
 & 
$<$    0.0007 &    911.80/1132 \\
A3562 & 1T &  5.41$^{+ 1.22}_{- 0.97}$ & 
0.405$^{+0.158}_{-0.146}$ & 
0.0168$^{+0.0024}_{-0.0018}$ &  & 
 &  &   29.73/78 \\
 & 2T &      5.39 & 
    0.413 & 
   0.0081 &  & 
     5.34 &    0.0087 &     29.58/76 \\
 & T$+$IC &      5.24$^{+     1.29}_{-     0.88}$ & 
    0.415$^{+    0.182}_{-    0.146}$ & 
   0.0170$^{+   0.0022}_{-   0.0038}$ &  & 
 & 
$<$    0.0012 &     29.59/77 \\
A3571 & 1T &  7.41$^{+ 0.19}_{- 0.19}$ & 
0.381$^{+0.021}_{-0.020}$ & 
0.1089$^{+0.0013}_{-0.0013}$ &  & 
 &  & 1281.49/1551 \\
 & 2T &      7.39 & 
    0.381 & 
   0.0336 &  & 
     7.42 &    0.0753 &   1281.48/1549 \\
 & T$+$IC &      7.44$^{+     0.21}_{-     0.18}$ & 
    0.380$^{+    0.022}_{-    0.019}$ & 
   0.1091$^{+   0.0012}_{-   0.0023}$ &  & 
 & 
$<$    0.0008 &   1278.53/1550 \\
A1795 & 1T &  5.87$^{+ 0.12}_{- 0.12}$ & 
0.376$^{+0.014}_{-0.014}$ & 
0.0778$^{+0.0009}_{-0.0009}$ &  & 
 &  & 1344.31/1584 \\
 & 2T &      4.49 & 
    0.382 & 
   0.0252 &  & 
     6.49 &    0.0537 &   1345.49/1582 \\
 & T$+$IC &      5.85$^{+     0.12}_{-     0.11}$ & 
    0.376$^{+    0.017}_{-    0.014}$ & 
   0.0779$^{+   0.0009}_{-   0.0026}$ &  & 
 & 
$<$    0.0009 &   1346.85/1583 \\
A3581 & 1T &  2.00$^{+ 0.11}_{- 0.10}$ & 
0.419$^{+0.086}_{-0.078}$ & 
0.0272$^{+0.0023}_{-0.0022}$ &  & 
 &  &  200.26/287 \\
 & 2T &      1.83 & 
    0.473 & 
   0.0273 &  & 
    13.54 &    0.0009 &    198.23/285 \\
 & T$+$IC &      1.82$^{+     0.24}_{-     0.13}$ & 
    0.530$^{+    0.153}_{-    0.161}$ & 
   0.0255$^{+   0.0033}_{-   0.0027}$ &  & 
 & 
$<$    0.0009 &    198.62/286 \\
MKW8 & 1T &  3.82$^{+ 0.60}_{- 0.47}$ & 
0.321$^{+0.104}_{-0.097}$ & 
0.0121$^{+0.0014}_{-0.0013}$ &  & 
 &  &   80.29/129 \\
 & 2T &      3.47 & 
    0.319 & 
   0.0062 &  & 
     4.20 &    0.0059 &     80.27/127 \\
 & T$+$IC &      3.53$^{+     0.84}_{-     0.78}$ & 
    0.384$^{+    0.321}_{-    0.160}$ & 
   0.0105$^{+   0.0029}_{-   0.0031}$ &  & 
 & 
$<$    0.0016 &     80.21/128 \\
A2029 & 1T &  8.47$^{+ 0.37}_{- 0.28}$ & 
0.453$^{+0.034}_{-0.033}$ & 
0.0757$^{+0.0014}_{-0.0014}$ &  & 
 &  &  551.00/641 \\
 & 2T &      7.41 & 
    0.457 & 
   0.0338 &  & 
     9.55 &    0.0420 &    549.99/639 \\
 & T$+$IC &      8.39$^{+     0.33}_{-     0.30}$ & 
    0.482$^{+    0.046}_{-    0.047}$ & 
   0.0711$^{+   0.0051}_{-   0.0044}$ &  & 
 & 
$<$    0.0033 &    548.15/640 \\
A2052 & 1T &  3.21$^{+ 0.10}_{- 0.09}$ & 
0.473$^{+0.032}_{-0.031}$ & 
0.0451$^{+0.0014}_{-0.0014}$ &  & 
 &  &  430.61/532 \\
 & 2T &      3.22 & 
    0.473 & 
   0.0236 &  & 
     3.20 &    0.0216 &    430.61/530 \\
 & T$+$IC &      3.07$^{+     0.19}_{-     0.19}$ & 
    0.519$^{+    0.080}_{-    0.061}$ & 
   0.0429$^{+   0.0031}_{-   0.0034}$ &  & 
 & 
$<$    0.0019 &    441.27/531 \\
MKW3S & 1T &  3.64$^{+ 0.13}_{- 0.12}$ & 
0.381$^{+0.028}_{-0.028}$ & 
0.0368$^{+0.0011}_{-0.0011}$ &  & 
 &  &  420.99/524 \\
 & 2T &      3.81 & 
    0.381 & 
   0.0181 &  & 
     3.47 &    0.0187 &    420.98/522 \\
 & T$+$IC &      3.63$^{+     0.13}_{-     0.23}$ & 
    0.383$^{+    0.074}_{-    0.029}$ & 
   0.0366$^{+   0.0013}_{-   0.0043}$ &  & 
 & 
$<$    0.0014 &    420.98/523 \\
A2065 & 1T &  6.58$^{+ 0.90}_{- 0.75}$ & 
0.260$^{+0.080}_{-0.078}$ & 
0.0289$^{+0.0020}_{-0.0017}$ &  & 
 &  &  100.40/165 \\
 & 2T &      6.44 & 
    0.261 & 
   0.0067 &  & 
     6.61 &    0.0222 &    100.37/163 \\
 & T$+$IC &      6.53$^{+     0.92}_{-     0.81}$ & 
    0.269$^{+    0.091}_{-    0.083}$ & 
   0.0281$^{+   0.0027}_{-   0.0038}$ &  & 
 & 
$<$    0.0016 &    100.19/164 \\
A2063 & 1T &  4.50$^{+ 0.23}_{- 0.21}$ & 
0.339$^{+0.034}_{-0.033}$ & 
0.0360$^{+0.0012}_{-0.0012}$ &  & 
 &  &  437.10/519 \\
 & 2T &      4.42 & 
    0.344 & 
   0.0191 &  & 
     4.65 &    0.0168 &    429.73/517 \\
 & T$+$IC &      4.52$^{+     0.27}_{-     0.31}$ & 
    0.344$^{+    0.049}_{-    0.033}$ & 
   0.0360$^{+   0.0012}_{-   0.0038}$ &  & 
 & 
$<$    0.0013 &    429.76/518 \\
A2142 & 1T & 10.41$^{+ 1.03}_{- 0.84}$ & 
0.195$^{+0.348}_{-0.195}$ & 
0.0645$^{+0.0061}_{-0.0061}$ &  & 
 &  &   27.85/60 \\
 & 2T &      8.95 & 
    0.209 & 
   0.0634 &  & 
    64.00 &    0.0035 &     22.38/58 \\
 & T$+$IC &      9.23$^{+     1.21}_{-     1.24}$ & 
$<$     0.577 & 
   0.0576$^{+   0.0073}_{-   0.0073}$ &  & 
 & 
   0.0031$^{+   0.0020}_{-   0.0021}$ &     21.97/59 \\
A2147 & 1T &  5.62$^{+ 1.14}_{- 0.86}$ & 
0.239$^{+0.138}_{-0.128}$ & 
0.0381$^{+0.0044}_{-0.0037}$ &  & 
 &  &   56.38/99 \\
 & 2T &      5.55 & 
    0.238 & 
   0.0234 &  & 
     5.68 &    0.0147 &     56.43/97 \\
 & T$+$IC &      5.50$^{+     1.23}_{-     0.77}$ & 
    0.239$^{+    0.142}_{-    0.126}$ & 
   0.0383$^{+   0.0041}_{-   0.0043}$ &  & 
 & 
$<$    0.0011 &     56.49/98 \\
A2199 & 1T &  4.59$^{+ 0.14}_{- 0.13}$ & 
0.367$^{+0.022}_{-0.021}$ & 
0.0994$^{+0.0021}_{-0.0021}$ &  & 
 &  &  573.33/746 \\
 & 2T &      4.39 & 
    0.367 & 
   0.0498 &  & 
     4.80 &    0.0496 &    573.18/744 \\
 & T$+$IC &      4.57$^{+     0.16}_{-     0.14}$ & 
    0.372$^{+    0.025}_{-    0.025}$ & 
   0.0982$^{+   0.0031}_{-   0.0039}$ &  & 
 & 
$<$    0.0017 &    572.92/745 \\
A2204 & 1T &  7.46$^{+ 0.33}_{- 0.32}$ & 
0.414$^{+0.033}_{-0.032}$ & 
0.0453$^{+0.0011}_{-0.0010}$ &  & 
 &  &  375.74/507 \\
 & 2T &      5.44 & 
    0.458 & 
   0.0309 &  & 
    15.49 &    0.0159 &    366.89/505 \\
 & T$+$IC &      7.39$^{+     0.40}_{-     0.18}$ & 
    0.450$^{+    0.058}_{-    0.067}$ & 
   0.0442$^{+   0.0021}_{-   0.0067}$ &  & 
 & 
$<$    0.0025 &    375.45/506 \\
A2256 & 1T &  7.91$^{+ 0.48}_{- 0.46}$ & 
0.324$^{+0.053}_{-0.051}$ & 
0.0499$^{+0.0015}_{-0.0015}$ &  & 
 &  &  199.21/262 \\
 & 2T &      8.58 & 
    0.330 & 
   0.0213 &  & 
     7.25 &    0.0286 &    195.46/260 \\
 & T$+$IC &      7.70$^{+     0.47}_{-     0.46}$ & 
    0.328$^{+    0.055}_{-    0.049}$ & 
   0.0498$^{+   0.0014}_{-   0.0028}$ &  & 
 & 
$<$    0.0011 &    194.84/261 \\
A2255 & 1T &  7.18$^{+ 1.13}_{- 0.86}$ & 
0.256$^{+0.108}_{-0.101}$ & 
0.0242$^{+0.0017}_{-0.0016}$ &  & 
 &  &   55.90/119 \\
 & 2T &      6.85 & 
    0.251 & 
   0.0134 &  & 
     7.78 &    0.0110 &     55.55/117 \\
 & T$+$IC &      7.16$^{+     1.26}_{-     0.82}$ & 
    0.246$^{+    0.110}_{-    0.094}$ & 
   0.0245$^{+   0.0016}_{-   0.0021}$ &  & 
 & 
$<$    0.0006 &     55.81/118 \\
A3667 & 1T &  7.13$^{+ 0.20}_{- 0.20}$ & 
0.276$^{+0.017}_{-0.016}$ & 
0.0730$^{+0.0009}_{-0.0009}$ &  & 
 &  & 1148.25/1320 \\
 & 2T &      6.77 & 
    0.277 & 
   0.0399 &  & 
     7.57 &    0.0332 &   1148.23/1318 \\
 & T$+$IC &      7.16$^{+     0.21}_{-     0.19}$ & 
    0.271$^{+    0.018}_{-    0.016}$ & 
   0.0731$^{+   0.0003}_{-   0.0028}$ &  & 
 & 
$<$    0.0010 &   1156.48/1319 \\
S1101 & 1T &  2.85$^{+ 0.14}_{- 0.13}$ & 
0.335$^{+0.044}_{-0.042}$ & 
0.0236$^{+0.0012}_{-0.0012}$ &  & 
 &  &  213.81/283 \\
 & 2T &      2.16 & 
    0.353 & 
   0.0166 &  & 
     3.90 &    0.0088 &    212.44/281 \\
 & T$+$IC &      2.85$^{+     0.14}_{-     0.24}$ & 
    0.334$^{+    0.059}_{-    0.039}$ & 
   0.0237$^{+   0.0012}_{-   0.0015}$ &  & 
 & 
$<$    0.0005 &    214.42/282 \\
A2589 & 1T &  3.86$^{+ 0.23}_{- 0.21}$ & 
0.545$^{+0.055}_{-0.053}$ & 
0.0197$^{+0.0009}_{-0.0009}$ &  & 
 &  &  163.53/252 \\
 & 2T &      3.41 & 
    0.622 & 
   0.0186 &  & 
    52.96 &    0.0016 &    157.95/250 \\
 & T$+$IC &      3.43$^{+     0.40}_{-     0.36}$ & 
    0.754$^{+    0.288}_{-    0.179}$ & 
   0.0153$^{+   0.0033}_{-   0.0029}$ &  & 
 & 
   0.0014$^{+   0.0008}_{-   0.0010}$ &    158.69/251 \\
A2597 & 1T &  3.89$^{+ 0.17}_{- 0.16}$ & 
0.316$^{+0.025}_{-0.025}$ & 
0.0238$^{+0.0009}_{-0.0009}$ &  & 
 &  &  310.13/407 \\
 & 2T &      3.43 & 
    0.317 & 
   0.0124 &  & 
     4.39 &    0.0114 &    310.12/405 \\
 & T$+$IC &      3.78$^{+     0.29}_{-     0.31}$ & 
    0.337$^{+    0.062}_{-    0.047}$ & 
   0.0227$^{+   0.0020}_{-   0.0026}$ &  & 
 & 
$<$    0.0012 &    310.37/406 \\
A2634 & 1T &  4.81$^{+ 1.19}_{- 0.88}$ & 
0.266$^{+0.140}_{-0.132}$ & 
0.0180$^{+0.0029}_{-0.0023}$ &  & 
 &  &   62.69/79 \\
 & 2T &      0.44 & 
    0.297 & 
   0.4055 &  & 
     5.76 &    0.0153 &     61.04/77 \\
 & T$+$IC &      4.84$^{+     1.12}_{-     0.94}$ & 
    0.269$^{+    0.148}_{-    0.131}$ & 
   0.0179$^{+   0.0030}_{-   0.0027}$ &  & 
 & 
$<$    0.0008 &     62.73/78 \\
A2657 & 1T &  5.73$^{+ 0.64}_{- 0.60}$ & 
0.251$^{+0.068}_{-0.066}$ & 
0.0242$^{+0.0016}_{-0.0013}$ &  & 
 &  &  176.95/235 \\
 & 2T &      2.91 & 
    0.267 & 
   0.0054 &  & 
     6.45 &    0.0199 &    177.03/233 \\
 & T$+$IC &      5.77$^{+     0.60}_{-     0.64}$ & 
    0.252$^{+    0.067}_{-    0.068}$ & 
   0.0241$^{+   0.0017}_{-   0.0021}$ &  & 
 & 
$<$    0.0007 &    176.95/234 \\
A4038 & 1T &  3.39$^{+ 0.12}_{- 0.11}$ & 
0.345$^{+0.026}_{-0.025}$ & 
0.0559$^{+0.0019}_{-0.0017}$ &  & 
 &  &  579.52/726 \\
 & 2T &      3.07 & 
    0.350 & 
   0.0298 &  & 
     3.73 &    0.0267 &    579.38/724 \\
 & T$+$IC &      3.36$^{+     0.14}_{-     0.18}$ & 
    0.352$^{+    0.045}_{-    0.031}$ & 
   0.0554$^{+   0.0024}_{-   0.0032}$ &  & 
 & 
$<$    0.0013 &    579.42/725 \\
A4059 & 1T &  4.43$^{+ 0.23}_{- 0.21}$ & 
0.428$^{+0.037}_{-0.036}$ & 
0.0329$^{+0.0012}_{-0.0011}$ &  & 
 &  &  259.39/427 \\
 & 2T &      4.39 & 
    0.431 & 
   0.0162 &  & 
     4.39 &    0.0168 &    261.26/425 \\
 & T$+$IC &      4.44$^{+     0.22}_{-     0.22}$ & 
    0.425$^{+    0.040}_{-    0.033}$ & 
   0.0329$^{+   0.0013}_{-   0.0016}$ &  & 
 & 
$<$    0.0005 &    259.32/426

\enddata
\tablenotetext{a}{Parameters for the 2T model are unconstrained.}
\tablenotetext{b}{Normalization of the {\tt APEC} thermal spectrum,
which is given by $\{ 10^{-14} / [ 4 \pi (1+z)^2 D_A^2 ] \} \, \int n_e n_H
\, dV$, where $z$ is the redshift, $D_A$ is the angular diameter distance,
$n_e$ is the electron density, $n_H$ is the ionized hydrogen density,
and $V$ is the volume of the cluster.}
\tablenotetext{c}{Value is the normalization of the power-law component
for the T$+$IC model, which is the photon flux at a photon energy of
1 keV in units of photons cm$^{-2}$ s$^{-1}$ keV$^{-1}$.
For the 2T model, the value is the normalization of 
the second {\tt APEC} thermal
model in units of cm$^{-5}$.}
\end{deluxetable}

\end{document}